\documentclass[journal,10pt]{IEEEtran}
\usepackage{notation}
\usepackage{epsfig}
\usepackage{color}
\usepackage{amsmath}
\usepackage{amssymb}
\usepackage{enumerate}
\usepackage{multirow}
\usepackage{bbm}
\usepackage{algorithm}
\usepackage{algorithmic}
\usepackage{amsmath,multicol}
\ifCLASSINFOpdf
\else
\fi
\begin{document}
%
\title{Tree-Structure Expectation Propagation for LDPC Decoding over the BEC}
%
%
%

\author{Pablo M. Olmos,
        Juan Jos\'e Murillo-Fuentes,
        and~Fernando P\'erez-Cruz
\thanks{This work was partially funded by Spanish government (Ministerio de Educaci\'on y Ciencia, TEC2009-14504-C02-{01,02}, Consolider-Ingenio 2010 CSD2008-00010), Universidad Carlos III (CCG10-UC3M/TIC-5304) and European Union (FEDER).}        
\thanks{P. M. Olmos and F. P\'erez-Cruz are with Dept.~Teor\'\i a de la Se\~nal y Comunicaciones, Universidad Carlos III de Madrid
(Spain). E-mail: {\tt \{olmos,
fernando\}@tsc.uc3m.es}}
\thanks{J.J. Murillo-Fuentes is with the Dept.~Teor\'\i a de la Se\~nal y Comunicaciones,
Escuela T\'ecnica Superior de Ingenier\'ia, Universidad de Sevilla, Paseo de los
Descubrimientos~s/n, 41092 Sevilla, Spain. E-mail: {\tt 
murillo@us.es} }
}

%
%

\markboth{IEEE Transactions on Information Theory}%
{Olmos \MakeLowercase{\textit{et al.}}: Tree Expectation Propagation for LDPC Decoding over the BEC}
%



\maketitle

\begin{abstract}
We present the tree-structure expectation propagation (Tree-EP) algorithm to decode low-density parity-check (LDPC) codes over discrete memoryless channels (DMCs). EP generalizes belief propagation (BP) in two ways. First, it can be used with any exponential family distribution over the cliques in the graph. Second, it can impose additional constraints on the marginal distributions. We use this second property to impose pair-wise marginal constraints over pairs of variables connected to a check node of the LDPC code's Tanner graph. Thanks to these additional constraints, the Tree-EP marginal estimates for each variable in the graph are more accurate than those provided by BP. We also reformulate the Tree-EP algorithm for the binary erasure channel (BEC) as a peeling-type algorithm (\TEP\!\!) and we show that the algorithm has the same computational complexity as BP and it decodes a higher fraction of errors. We describe the \TEP decoding process by a set of differential equations that represents the expected residual graph evolution as a function of the code parameters. The solution of these equations is used to predict the TEP decoder performance in both the asymptotic regime and the finite-length regime over the BEC. While the asymptotic threshold of the TEP decoder is the same as the BP decoder for regular and optimized codes, we propose a scaling law (SL) for finite-length LDPC codes, which accurately approximates the TEP improved performance and facilitates its optimization.
%
%
%
\end{abstract}
 

%
\section{Introduction}

\IEEEPARstart{L}{}ow-density parity-check (LDPC) codes are well known channel capacity-approaching (c.a.) linear codes. In his PhD \cite{Gallager63}, Gallager proposed LDPC codes along with linear-time practical decoding methods, among which the belief propagation (BP) algorithm plays a fundamental role. BP was later redescribed and popularized in the articial intelligence community to perform approximate inference over graphical models, see for instance \cite{pearl88,Mackay96, Mackay99}. Given a factor graph that represents a joint probability density function (pdf) $\pdf(\V)$ of a set of discrete random variables \cite{Loeliger04}, BP estimates the marginal probability function for each variable. It uses a local message-passing algorithm between the nodes of the graph. The complexity of this algorithm is linear in the number of nodes \cite{pearl88}. For tree-like graphs, the BP solution is exact, but for graphs with cycles, BP is strictly suboptimal \cite{Frey01, Wiberg96, Aji00}.

Linear block codes can be represented using factor (Tanner) graphs \cite{Tanner81}, where the factor nodes enforce the parity check constraints. For LDPC codes, the presence of cycles in the Tanner graph quickly decays with the code length $\n$. For large block lengths, a channel decoder based on BP achieves an excellent performance, close to the bitwise maximum a posteriori (bit-MAP) decoding, in certain scenarios \cite{Mackay96,Urbanke01-2}. Nevertheless, the bit-MAP solution can only be achieved when the code length, code density and computational complexity go to infinity \cite{Urbanke08-2,Urbanke08,Kudekar10}. 

The analysis of the BP for LDPC decoding over independent and identically distributed channels is detailed in \cite{Urbanke02, Oswald02}, in which the limiting performance and code optimization are addressed. For the binary erasure channel (BEC), the BP decoder presents an alternative formulation, in which the known variable nodes (encoded bits) are removed from the graph after each iteration. The BP, under this interpretation, is referred to as the \emph{peeling decoder} (PD) \cite{Urbanke08-2}. In \cite{Luby01}, the authors investigate the PD  limiting performance by describing the expected LDPC graph evolution throughout the decoding process by a set of differential equations.  The asymptotic performance for the BP decoder is summarized in the computation of the so-called BP  \emph{threshold} \cite{Urbanke01-2,Urbanke08-2,Luby01,Brink04}, which  defines the limit of its decodable region for an LDPC code.

The analysis of BP decoding performance in the finite-length regime is based on the evaluation of the presence of \emph{stopping sets} (SSs) in the LDPC graph \cite{Urbanke02}, which can severely degrade the decoder performance. In \cite{Urbanke02,Orlitsky02}, the authors provide tools to compute the exact BP average performance. However, this task becomes computationally challenging if the degree of irregularity or block length increases \cite{Urbanke08-2}. Alternatively, we can separate the contributions to the error rate of large-size errors, which dominate in the waterfall region \cite{Urbanke09}, from small failures, which cause error floors \cite{Urbanke02}. \emph{Scaling laws} (SLs) were proposed in \cite{Urbanke09,Amraoui05}  to accurately predict the BP performance in the waterfall region. For the BEC, they are based on the PD covariance evolution for a given graph as a function of the code length. Covariance evolution was solved for any LDPC ensemble in \cite{Takayuki10}. On the other hand, the analysis of the error floor is addressed by determining the dominant terms of the code weight distribution \cite{Urbanke02,Orlitsky02}. Precise expressions for the asymptotic bit-MAP and BP error floor are derived in \cite{Gallager63, Di06, Orlitsky05}.

Expectation propagation (EP) \cite{Minka01} can be understood as a generalization of BP to construct tractable approximations 
of a joint pdf $p(\V)$. Consider the set of all possible probability distributions in a given exponential family that map over the same factor graph. EP minimizes within this family the inclusive Kullback-Leibler (KL) divergence \cite{Cover05} with respect to $p(\V)$. 
In \cite{Minka01,Yedidia00}, it is shown that BP can be reformulated as EP 
by considering a discrete family of pdfs that factorizes as the product of single-variable multinomial terms, i.e. $q(\V)= q_{1}(\Vi{1})q_{2}(\Vi{2})\ldots q_{\n}(\Vi{\n})$. 
EP generalizes BP in two ways: first, it is not restricted to discrete random variables. And second, EP naturally formulates to include more versatile approximating factorizations \cite{Minka03, Ghahramani97}. In this paper, we focus on EP to construct a Markov tree-structure to approximate the original graph. Conditional factors in the tree-structure  are able to capture pairwise interactions that single factors neglect.
We refer to this algorithm as \emph{tree-structure expectation propagation} (Tree-EP).  We borrow from the theoretical framework of the Tree-EP algorithm to design a new decoding approach to decode LDPC codes over discrete memoryless channels (DMCs) and we analyze the decoder performance for the BEC. 

For the erasure channel, we show that the Tree-EP can be reinterpreted as a peeling-type algorithm that formulates as an improved PD. We refer to this simplified algorithm as the TEP decoder. The TEP decoder was presented in \cite{Olmos10-2,Olmos11}, where we empirically observed a noticeable gain in performance compared to BP for both regular and irregular LDPC codes. We now explain, analyze and predict this gain in performance for any LDPC code. First, we extend to the TEP decoder the methodology proposed in \cite{Luby01} to evaluate the expected graph evolution of the LDPC's Tanner graph. As the block size increases, we show the conditions for which the TEP decoder could improve the BP decoder. Nevertheless, for typical LDPC ensembles the TEP decoder is not able to improve the BP solution. In the second part of the paper, we concentrate on practical finite-length codes and we explain the gain provided by the TEP decoder compared to BP. Based on empirical evidence, we propose a SL to predict the TEP  performance for any given LDPC ensemble in the waterfall region, which captures the gain in performance that the TEP achieves with respect to BP for finite-length LDPC codes. Furthermore, the SL can be used for TEP-oriented finite-length codes optimization. Finally, we also prove that the decoder complexity is of the same order than BP, i.e. linear in the number of variables, unlike other techniques proposed to improve BP at a higher computational cost. For instance, we can mention variable guessing algorithms \cite{Pishro04}, the Maxwell decoder \cite{Urbanke08} and pivoting algorithms for efficient Gaussian elimination \cite{Burshtein04,Liva09,Saejoon08}, whose complexity is not linear unless we impose additional restrictions that may alter/compromise their performance, such as bounding the number of guessed variables or pivots.

The rest of the paper is organized as follows. Section\SEC{EP} is devoted to introducing the Tree-EP algorithm for block decoding over DMCs. In Section\SEC{TEPsect}, we particularize the algorithm for the BEC, yielding the \TEP decoder. In Section\SEC{EG}, we derive the differential equations that describe the decoder behavior for a given LDPC graph and we investigate its asymptotic behavior as well as the algorithm's complexity. In Section\SEC{FLregimen}, we describe the scaling law proposed to approximate the TEP finite-length performance for a given LDPC ensemble in the waterfall region. We conclude the paper in Section\SEC{Conclusions}.

\section{Tree-EP for LDPC decoding LDPC over memoryless channels}\LABSEC{EP}

Consider an LDPC binary code $\code$ with parity check matrix $\Hs$, of dimensions
$\k\times\n$, where $\k=\n(1-\rate)$, $\n$ is the code length and $\rate$ the rate of the code. By definition, any vector $\v$  in $\mathbb{F}_{2}^{\n}$ belongs to the code $\code$ as long as  $\v\Hs^{\top}=\vect{0}$, where $\mathbb{F}_{2}^{\n}$ is the $\n$-dimensional binary Galois field. Each row of $\Hs$ therefore imposes an even parity constraint on a subset of variables:
\begin{equation}\LABEQ{Parity}
\pf{j}(\v)\doteq\parity\left[\sum_{i\in\setp{j}}\vi{i}\!\!\! \mod\, 2=0\right]\quad \forall j=1,\ldots,\k;
\end{equation} 
where $\vi{i}$ is the $i$-th component of $\v$, $\setp{j}$ is the set of positions where the $j$-th row of $\Hs$ is one and 
$\parity[\cdot]$ is a boolean operator, which takes value one if the condition in its argument is verified. Note that, given the definition in \EQ{Parity}, we can write $\parity[\v\in\code]=\pf{1}(\v)\pf{2}(\v)\ldots\pf{\k}(\v)$.

Assume that an unknown codeword is transmitted through a discrete memoryless channel \cite{Cover05} and let $\y\in\mathcal{A}(\y)$ be the observed channel output, where $\mathcal{A}(\y)$ is the channel output alphabet. A bit-MAP decoder \cite{Moon05} minimizes the bit error rate (BER) by estimating the transmitted vector $\hat{\v}=[\hvi{1}, \hvi{2}, \ldots, \hvi{\n}]$ as follows\footnote{In the following, we use lower case letters to denote a particular realization of a random variable or vector, e.g $\V=\v$ means that $\V$ takes the $\v$ value.}:
\begin{align}\LABEQ{EP0}
\hvi{u}&=\arg\max_{v\in\{0,1\}} p(\Vi{u}=v|\y)=\arg\max_{v\in\{0,1\}}\sum_{\V\in\code:\Vi{u}=v}\pdf\left(\V|\y\right) \nonumber\\
&=\arg\max_{v\in\{0,1\}}\sum_{\V\in\code:\Vi{u}=v}\pdf\left(\y|\V\right)p(\V)\nonumber\\
&=\arg\max_{v\in\{0,1\}}\sum_{\V:\Vi{u}=v}\prod_{i=1}^{\n}\pdf(y_{i}|\Vi{i})\prod_{j=1}^{\k}\pf{j}(\V)
\end{align}
for $u=1,\ldots,\n$, where we have assumed that the channel is memoryless and that all codewords are equally probable 
\begin{align}\LABEQ{}
p(\V)=\frac{\parity[\V\in\code]}{2^{\n\rate}}.
\end{align}

For most LDPC codes of interest, the factor graph associated to the product $\pdf\left(\y|\V\right)p(\V)$ in \EQ{EP0}
yields a graph with cycles \cite{Vardy99}. Hence, the exact computation of the marginals of $p(\Vi{u}=v|\y)$ grows exponentially with the number of coded bits \cite{Frey01}. Belief propagation \cite{Gallager63,pearl88,Mackay96} is nowadays the standard algorithm to efficiently solve this problem in coding applications, because accurate estimates for each marginal are obtained at linear cost with $\n$. Besides, BP can be cast as an approximation of $\pdf\left(\V|\y\right)$ in \EQ{EP0} by a complete disconnected factor graph, i.e.
\begin{align}\LABEQ{BP estimate}
p(\V|\y)&\propto\prod_{i=1}^{\n}\pdf(y_{i}|\Vi{i})\prod_{j=1}^{\k}\pf{j}(\V)\approx\prod_{i=1}^{\n}\widehat{q}_{i,\text{BP}}(\Vi{i}),
\end{align}
where $\widehat{q}_{i,\text{BP}}(\Vi{i})$ is the BP estimate for the $i$-th variable \cite{Frey01,Wainwright08,Bishop06}.

\subsection{Tree-EP algorithm for LDPC decoding}\LABSEC{TEPLDPC}

The Tree-EP algorithm \cite{Minka03,Minka01thesis} improves BP decoding because it approximates the posterior $\pdf\left(\V|\y\right)$ in \EQ{EP0} with a tree (or forest) Markov-structure between the variables, i.e.:
\begin{equation}\LABEQ{TEP1}
q(\V)= \prod_{i=1}^{\n} q_{i}(\Vi{i}|\Vi{p_i}),
\end{equation}
where the set of pairs $\{(i,p_i)_{i=1}^\n\}$ forms a tree graph over the original set of variables nodes and $q_{i}(\Vi{i}|\Vi{p_i})$ approximates the conditional probability
$p(\Vi{i}|\Vi{p_i},\y)$. For some variable nodes $\Vi{p_i}$ might be missing, i.e. $p_i$ is empty, and we use single-variable factors $\q_{i}(\Vi{i})$ to approximate them. 

Using the approximation in \EQ{TEP1}, the complexity of the marginalization in \EQ{EP0} is linear with the number of variables. In \cite{Minka01thesis}, it is shown that the approximation in \EQ{TEP1}  provides more accurate estimates for the marginals of $p(\Vi{u}|\y)$ than BP, since it captures information about the joint marginal of pairs of variables that are then propagated through the graph. Indeed, note that the factorization in \EQ{BP estimate} is just a particular case of \EQ{TEP1}, because EP converges to the same solution as the BP \cite{Minka01,Yedidia00}, when we consider a completely disconnected graph, i.e. $p_i={\emptyset}$ $\forall i$.

Consider a family of discrete probability distributions that factorize according to \EQ{TEP1}, denoted hereafter by $\mathcal{F}_{tree}$. The optimal choice for $q(\V)\in\mathcal{F}_{tree}$, denoted by $\widehat{q}(\V)$, is such that it minimizes the inclusive KL divergence\footnote{We refer to the inclusive KL divergence with respect to $p$ when we minimize $D_{KL}(p||q)$ and to the exclusive KL divergence when we minimize $D_{KL}(q||p)$, which is the one used for mean-field approximations, see \cite{Minka01thesis} for a discussion.} with $p(\V|\y)$:
\begin{align}\LABEQ{EP4}
\widehat{\q}(\V)&=\arg\min_{\displaystyle q(\V)\in\mathcal{F}_{tree}} D_{KL}\Big(\pdf(\V|\y)||\q(\V)\Big).
\end{align}

The next lemma, first proved in \cite{Lauritzen92}, states that the resolution of the problem in \EQ{EP4} is as complex as the bit-MAP decoding problem in \EQ{EP0}. In both cases we have to perform exact marginalization over the posterior distribution $\pdf(\V|\y)$.
\lemma{\LABLEM{MomentM}
(Moment matching for inclusive KL-divergence minimization). Consider a set of 
discrete random variables $\V$ with joint pdf $\pdf(\V)$.  
Let $\mathcal{F}_{\ast}$ be a family of probability distributions that share a common factorization:
\begin{align}\LABEQ{expfamily}
q(\V)\in\mathcal{F}_{\ast}\Leftrightarrow q(\V)=\prod_{i=1}^{L}q_{i}(\V_{i}),
\end{align}
for some normalized functions $q_{i}(\V_{i})$, where $\V_{i}$ is a subset of $\V$, for $i=1,\ldots,L$.
Under these conditions, the function $\widehat{q}(\V)$ in $\mathcal{F}_{\ast}$ such that
\begin{align}\LABEQ{KLdivergence2}
\widehat{q}(\V)=\arg\min_{\displaystyle q(\V)\in\mathcal{F}_{\ast}} D_{KL}\Big(\pdf(\V)||q(\V)\Big)
\end{align}
is constructed as follows:
\begin{align}\LABEQ{Sol}
\widehat{q}_{i}(\V_{i})&=\sum_{\V\sim\V_{i}}\pdf(\V),\quad i=1,\ldots,L \\ \LABEQ{Sol2}
\widehat{q}(\V)&=\prod_{i=1}^{L}\widehat{q}_{i}(\V_{i}),
\end{align}
where $\V\sim\V_{i}$ denotes all the variables in $\V$ except those in $\V_{i}$.
}
\proof{
The proof of this lemma can be found in \cite{Lauritzen92,Boyen98}. $\blacksquare$}

Lemma\LEM{MomentM} can be directly applied to the problem in \EQ{EP4} and the optimum Markov-tree $\hat{\q}(\V)$ is such that
\begin{align}\LABEQ{Sol}
\widehat{q}_{i}(\Vi{i}|\Vi{p_i})&=\frac{\displaystyle\sum_{\V\sim\{\Vi{i},\Vi{p_i}\}}p(\V|\y)}{\displaystyle\sum_{\V\sim\Vi{p_i}}p(\V|\y)}=p(\Vi{i}|\Vi{p_i},\y)
\end{align}
for $i=1,\ldots,\n$ and
\begin{align}
\widehat{q}(\V)&=\prod_{i=1}^{\n} \widehat{q}_{i}(\Vi{i}|\Vi{p_i}).
\end{align} 
The marginal computation in \EQ{Sol} is of the same complexity order as \EQ{EP0}. Lemma\LEM{MomentM} only  provides the conditions to find the distribution $\widehat{q}(\V)$ in the family $\mathcal{F}_{tree}$ that is optimum in the sense of \EQ{EP4}. A different problem arises from the optimization of the family $\mathcal{F}_{tree}$ itself, namely how we choose the parent variables $\Vi{p_i}$, to achieve the highest accuracy for the least cost. As discussed in \cite{Minka01thesis}, 
while determining the cost of a given choice is straightforward, estimating the accuracy for each case is a non-trivial problem that highly depends on the distribution $p(\V)$ and the application at hand. In our case, the analysis of the Tree-EP decoder for the erasure channel in Section\SEC{TEPsect} provides the intuition to construct the family $\mathcal{F}_{tree}$ for any DMC. To describe the implementation of the Tree-EP algorithm, in the following we assume fixed the family $\mathcal{F}_{tree}$.

The Tree-EP algorithm overcomes this problem by iteratively approaching $\hat{\q}(\V)$ as follows. Define $\q^{\ell}(\V)$ as the EP approximation to $\hat{\q}(\V)$ at the end of the $\ell$-th iteration and let $\q^{\infty}(\V)\approx\hat{\q}(\V)$ be the Tree-EP solution after convergence\footnote{EP as BP might not converge for loopy graphs \cite{Minka01thesis}.}. Let $\qij{z}{j}{\ell}(\Vi{z},\Vi{p_z})$ be non-negative real functions, with $z=1,\ldots,\n$ and $j=1,\ldots,\k$, that are updated at each iteration so that 
\begin{align}\LABEQ{parityfacW}
\pf{j}(\V)\approx\cep{j}{\ell}(\V)=\prod_{z=1}^{\n}\qij{z}{j}{\ell}(\Vi{z},\Vi{p_z}),
\end{align}
Thus, the Tree-EP approximation to the posterior $p(\V|\y)$ at iteration $\ell$, i.e $\q^{\ell}(\V)$, is constructed as follows:
\begin{align}\LABEQ{qTEP}
q^{\ell}(\V)&=\prod_{z=1}^{\n} q_{z}^{\ell}(\Vi{z}|\Vi{p_z})\nonumber\\
&\doteq\prod_{z=1}^{\n}\pdf(y_{z}|\Vi{z})\prod_{j=1}^{\k}\cep{j}{\ell}(\V)\nonumber\\
&=\prod_{z=1}^{\n}\left(\pdf(y_{z}|\Vi{z})\prod_{j=1}^{\k}\qij{z}{j}{\ell}(\Vi{z},\Vi{p_z})\right).
\end{align}
 
The Tree-EP is described in Algorithm \ref{TEPAlg}. At iteration $\ell$,  only the $m$-th factor, $\cep{m}{\ell}(\V)$, is refined. In Step 7 we replace $\cep{m}{\ell-1}(\V)$ by the true value $\pf{m}(\V)$ in the $q^{\ell-1}(\V)$ function. The resulting function is denoted by $f(\V,\ell,m)$. Then, by Lemma\LEM{MomentM},  in Steps 8 to 10 we compute $q^{\ell}(\V)$ as the solution of the following problem
\begin{align}\LABEQ{problem}
q^{\ell}(\V)=\arg\min_{\displaystyle q(\V)\in\mathcal{F}_{tree}} D_{KL}\Big(f(\V,\ell,m)||q(\V)\Big).
\end{align}
At this point, the TEP solution becomes suboptimal with respect to \EQ{EP4} but tractable: the computation of the marginals $q(\Vi{i}|\Vi{p_i})$ for $i=1,\ldots,\n$ over $f(\V,\ell,m)$ in \EQ{margfTEP} and \EQ{margfTEP2} can be performed efficiently. Let us first express the factorization of $f(\V,\ell,m)$ in a more convenient way:
\begin{align}\LABEQ{ftreeTEP}
f(\V,\ell,m)&=\pf{m}(\V)\frac{\q^{\ell-1}(\V)}{\cep{m}{\ell-1}(\V)}\nonumber\\&=\pf{m}(\V)\prod_{z=1}^{\n}p(y_{z}|\Vi{z})\prod_{\substack{j=1\\j\neq m}}^{\k}\cep{j}{\ell-1}(\V)\nonumber\\
&=\pf{m}(\V)\prod_{z=1}^{\n}g_z^{\ell-1,m}(\Vi{z},\Vi{p_{z}}),
\end{align}
where we have introduced the following auxiliary functions
\begin{align}\LABEQ{factors}
g_z^{\ell-1,m}(\Vi{z},\Vi{p_{z}})\doteq\pdf(y_{z}|\Vi{z})\prod_{\substack{j=1\\j\neq m}}^{\k}\qij{z}{j}{\ell-1}(\Vi{z},\Vi{p_{z}}).
\end{align}

Therefore, the marginalization of \EQ{ftreeTEP} in \EQ{margfTEP} yields
\begin{align}\LABEQ{eqmargn}
q(\Vi{i},\Vi{p_i})=\sum_{\V\sim\{\Vi{i},\Vi{p_i}\}}\pf{m}(\V)\prod_{z=1}^{\n}g_z^{\ell-1,m}(\Vi{z},\Vi{p_{z}}).
\end{align}
As in \EQ{parityfacW}, the product $\prod_{z=1}^{\n}g_z^{\ell-1,m}(\Vi{z},\Vi{p_{z}})$ maps over the same factor graph than the tree structure chosen in \EQ{TEP1}. Therefore, the presence of cycles in the factor graph of $f(\V,\ell,m)$ is due to the parity factor $\pf{m}(\V)$. The graph is cycle-free as long as, among the variables connected to the parity check node $\pf{m}(\V)$, none of them are linked by a conditional term $q(\Vi{i}|\Vi{p_i})$ in \EQ{TEP1}, as illustrated in Fig. \FIG{GraphsTEP}(a). Otherwise the graph presents cycles, as shown in Fig. \FIG{GraphsTEP}(b). These cycles play a crucial role in understanding why the Tree-EP algorithm outperforms the BP solution. In the first case, the marginal computation in \EQ{margfTEP} is  solved at linear cost by message-passing. For the latter, where the graph is not completely cycle-free, we can compute the pairwise marginals using Pearl's cutset conditioning algorithm \cite{Minka03}, \cite{Becker99}.

Pearl's algorithm proceeds by breaking each cycle assuming a set of the variables involved as known, e.g. $\Vi{o}$ in Fig. \FIG{GraphsTEP}(b). Then, the marginals of the remaining variables can be computed at low-cost by message-passing. The overall complexity of this method is exponential with the number of assume-observed variables. However, we prove that for the BEC the complexity of the Tree-EP algorithm is linear with the number of variables, i.e. of the same order as BP. 


 \begin{algorithm}\label{TEPAlg}
\begin{algorithmic}[1]
\STATE $\ell=0$.
\vspace{0.05cm}
\STATE Initialize $\cep{j}{0}(\V)=1$ for $j=1,\ldots,\k$.
\vspace{0.05cm}
\STATE Initialize $\q^{0}(\V)=\displaystyle\prod_{i=1}^{\n}\pdf(y_{i}|\Vi{i})$.
\vspace{0.05cm}
\REPEAT 
\STATE $\ell=\ell+1$.
\vspace{0.05cm}
\STATE{Chose a $\cep{m}{\ell}(\V)$ to refine: $m=\text{mod}(\ell,\k)$.}
\vspace{0.05cm}
\STATE{Remove $\cep{m}{\ell-1}(\V)$  from $\q^{\ell-1}(\V)$ and include the \emph{true term} $\pf{m}(\V)$:
\begin{align}\LABEQ{fTEP}
f(\V,\ell,m)&\doteq\pf{m}(\V)\frac{\q^{\ell-1}(\V)}{\cep{m}{\ell-1}(\V)}.
\end{align}
}
\STATE{Compute $q_{i}^{\ell}(\Vi{i},\Vi{p_i})$ and $q_{i}^{\ell}(\Vi{i})$ for $i=1,\ldots,\n$:
\begin{align}\LABEQ{margfTEP}
q_{i}^{\ell}(\Vi{i},\Vi{p_i})&\propto\sum_{\V\sim\{\Vi{i},\Vi{p_i}\}}f(\V,\ell,m),\\\LABEQ{margfTEP2}
q_{i}^{\ell}(\Vi{i})&=\sum_{\Vi{p_{i}}}q_{i}^{\ell}(\Vi{i},\Vi{p_i}).
\end{align}
}
\vspace{0.05cm}
\STATE{Compute $q_{i}^{\ell}(\Vi{i}|\Vi{p_i})=\frac{\displaystyle q_{i}^{\ell}(\Vi{i},\Vi{p_i})}{\displaystyle q_{i}^{\ell}(\Vi{p_i})}$ for $i=1,\ldots,\n$.}
\vspace{0.05cm}
\STATE $q^{\ell}(\V)=\displaystyle\prod_{i=1}^{\n}q_{i}^{\ell}(\Vi{i}|\Vi{p_i})$.
\vspace{0.05cm}
\STATE{Compute $\qij{i}{m}{\ell}(\Vi{i},\Vi{p_i})$ from $q^{\ell}(\V)$. By \EQ{qTEP},
\begin{align}\LABEQ{recompute}
\qij{i}{m}{\ell}(\Vi{i},\Vi{p_i})=\frac{q_{i}^{\ell}(\Vi{i}|\Vi{p_i})}{p(y_{i}|\Vi{i})\displaystyle\prod_{\substack{j=1\\j\neq m}}^{\k}\qij{i}{j}{\ell-1}(\Vi{i},\Vi{p_i})}
\end{align}
for $i=1,\ldots,\n$.
}
\vspace{0.05cm}
\STATE{$\cep{m}{\ell}(\V)=\displaystyle\prod_{i=1}^{\n}\qij{i}{m}{\ell}(\Vi{i},\Vi{p_i})$.}
\vspace{0.05cm}
\UNTIL all $\cep{j}{\ell}(\V)$, $j=1,\ldots,\k$, converge.
\vspace{0.05cm}
\STATE{$\widehat{q}(\V)\approx q^{\infty}(\V)$}.
\end{algorithmic}
\caption{Tree-EP algorithm for a predefined tree structure.}
\end{algorithm}

\begin{figure}[h]\LABFIG{GraphsTEP}
\centering
\begin{tabular}{c}
\includegraphics[scale=0.4]{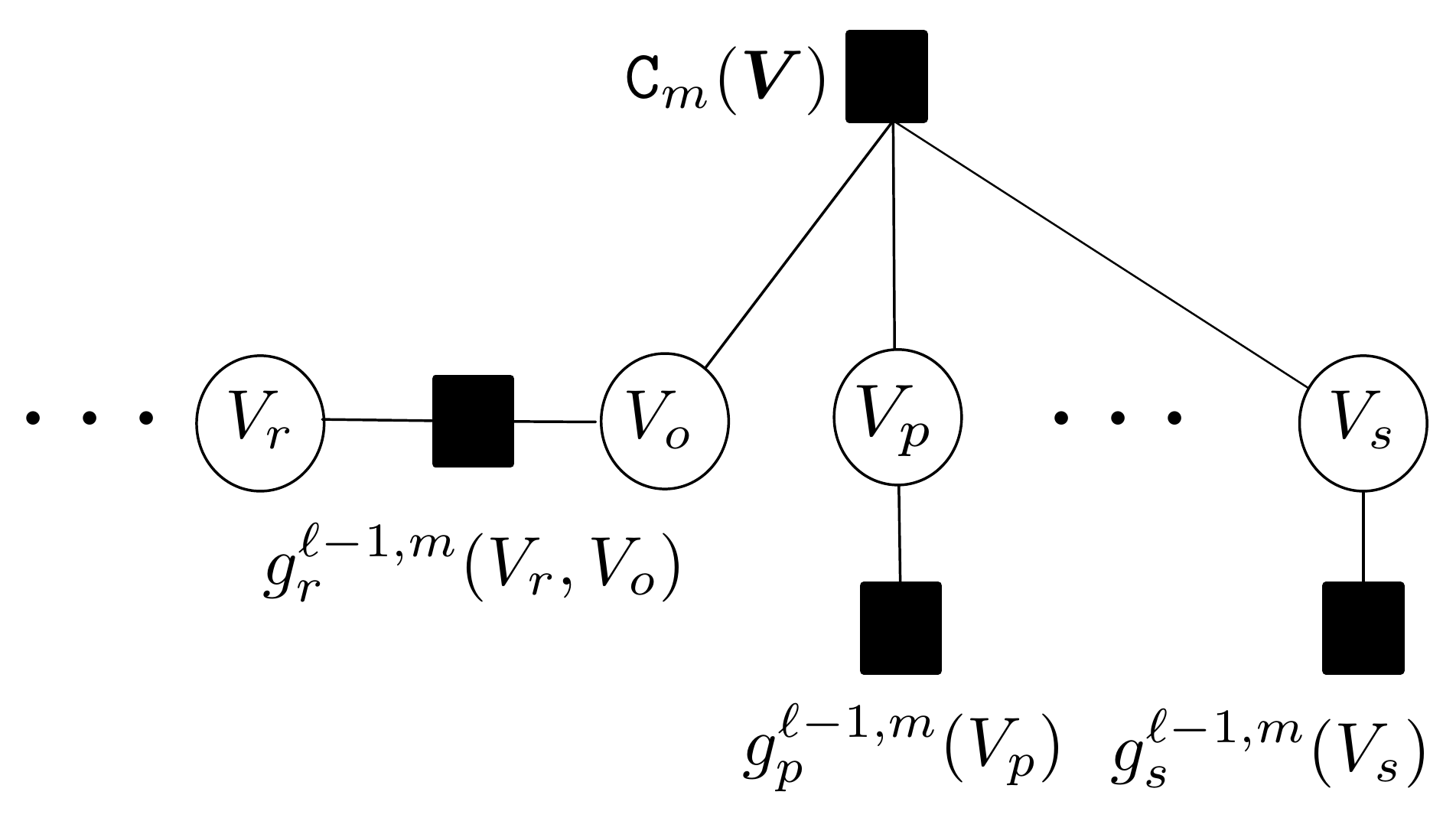}\\\\(a)\\\\
\includegraphics[scale=0.4]{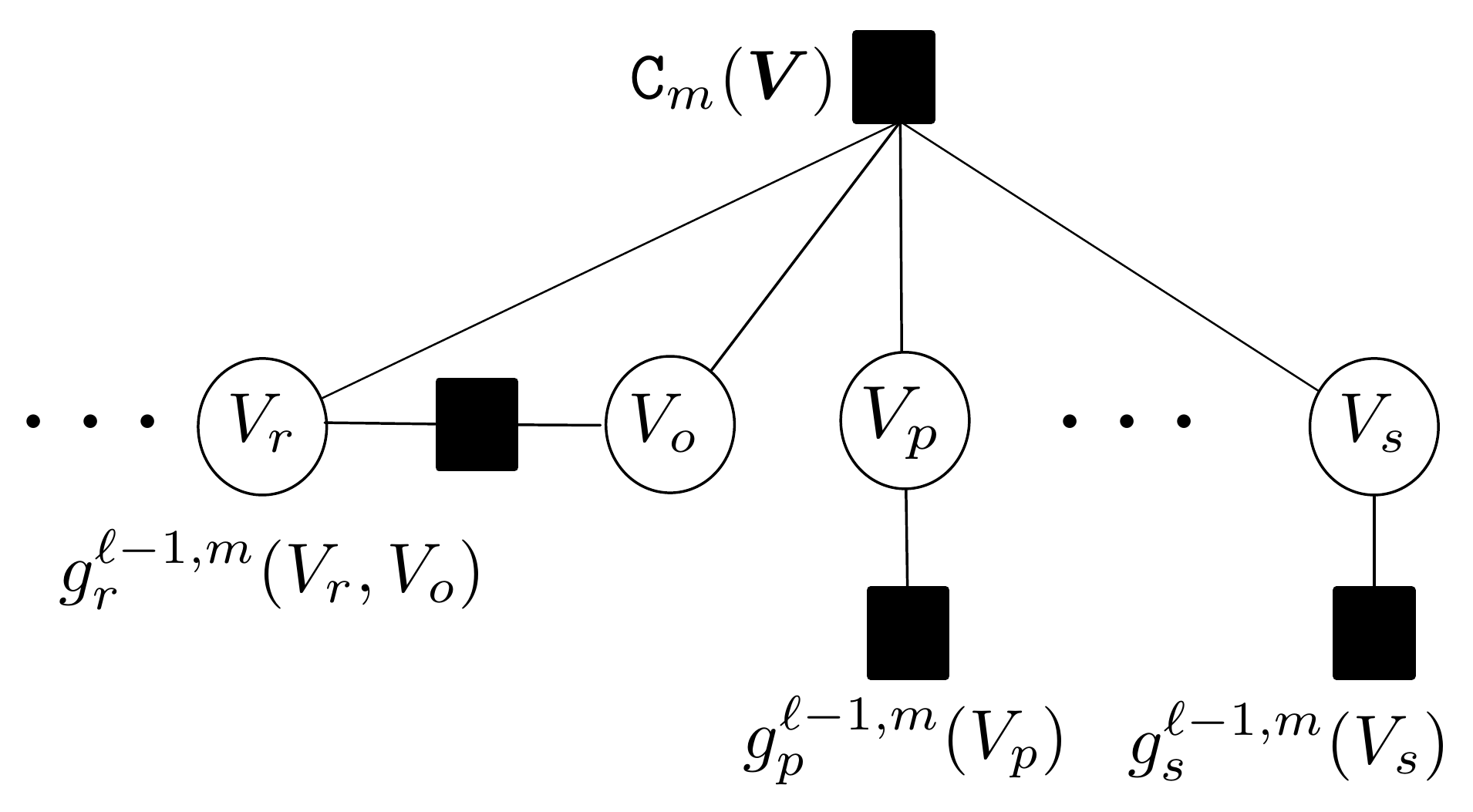}\\\\(b)
\end{tabular}\caption{Example of factor graph associated to $f(\V,\ell,m)$. In (a), the graph is cycle free since $\pf{m}$ does not connect any variable $\Vi{i}$ with its parent $\Vi{p_i}$. In (b), the graph has a cycle since both $\Vi{r}$ and $\Vi{p_r}=\Vi{o}$ belong to $\setp{m}$.} 
\end{figure}

\section{Tree-EP decoding for the BEC}\LABSEC{TEPsect}


For the BEC, the likelihood function for a particular variable $\Vi{i}$, $p(y_{i}|\Vi{i})$, provides a complete description about its value when it has not been erased:
\begin{align}\LABEQ{l1}
&p(y_{i}=1|\Vi{i}=1)=1-\pe, \quad p(y_{i}=1|\Vi{i}=0)=0,\\\LABEQ{l2}
&p(y_{i}=0|\Vi{i}=1)=0, \quad p(y_{i}=0|\Vi{i}=1)=1-\pe.
\end{align}
In this case, we say that the variable is known. Otherwise, when the variable is erased, the likelihood function for this variable is constant and the uncertainty about its value is complete: 
\begin{align}\LABEQ{l3}
p(y_{i}=\;?|\Vi{i}=1)&=p(y_{i}=\;?|\Vi{i}=0)=\pe.
\end{align}
This two-state behavior of the likelihood function makes the pairwise marginal functions $\q_{i}^{\ell}(\Vi{i},\Vi{p_i})$ in \EQ{eqmargn} and \EQ{margfTEP} alternate between just four states, depending on what we know about these variables at the end of the $\ell$-th iteration. Furthermore, we can describe a finite set of  scenarios for which $\q_{i}^{\ell}(\Vi{i},\Vi{p_i})$ might alternate between these states.   
This result is used later to propose a simplified reformulation of the decoding algorithm. Let us detail the possible outcomes and cases of interest by running the algorithm at different times, after initialization.

\vspace{0.2cm}
\emph{\underline{Iteration $\ell=1$}.} 
In Step 7, we compute
\begin{align}\LABEQ{q10}
f(\V,\ell=1,m=1)=\pf{1}(\V)\prod_{z=1}^{\n}\pdf(y_{z}|\Vi{z}),
\end{align}
where the likelihood terms $\pdf(y_{z}|\Vi{z})$ for $z=1,\ldots,\n$ take the form \EQ{l1}-\EQ{l3}. 
 Over \EQ{q10}, we compute the marginals of $\{(\Vi{i},\Vi{p_i})_{i=1}^\n\}$. For any pair $(\Vi{i},\Vi{p_i})$, we observe the following scenarios:\vspace{0.2cm}
\begin{itemize}
\item If $\Vi{i}$ and $\Vi{p_i}$ have not been erased,  then marginalization yields:
\begin{align}\LABEQ{q11}
\q_{i}^{1}(\Vi{i},\Vi{p_i})&\propto\pdf(y_{i}|\Vi{i})\pdf(y_{p_i}|\Vi{p_i})\nonumber\\
&=\parity\left[\Vi{i}=v_1,\Vi{p_i}=v_2\right],
\end{align}
where $v_1$ and $v_2$ are, respectively, the values of $\Vi{i}$ and $\Vi{p_i}$.
\item If $\Vi{i}$ is erased and $\Vi{p_i}$ is not, then the marginalization in \EQ{q10} might reveal the value of $\Vi{i}$. This scenario only happens if $\Vi{i}$ is connected to the check $\pf{1}(\V)$ and, in addition, the rest of variables connected to this check are known. In this way, the parity restriction fixes the value of the variable and the marginalization yields the same result than in \EQ{q11}.
Otherwise $\Vi{i}$ remains unknown:
\begin{align}\LABEQ{q122}
\q_{i}^{1}(\Vi{i},\Vi{p_i})=\frac{1}{2}\parity\left[\Vi{p_i}=v_2\right],
\end{align}
where $v_2$ is the value of $\Vi{p_i}$. For instance, assume that, in Fig. \FIG{GraphsTEP} (a), $\Vi{o}$ and $\Vi{p}$ are known. When we compute the marginal for $\Vi{s}$, this variable gets revealed.


\item The case where $\Vi{p_i}$ is erased and $\Vi{i}$ is not is symmetric to the previous scenario.

\item If both $\Vi{i}$ and $\Vi{p_i}$ are erased, we compute 
\begin{align}\LABEQ{q14}
\q_{i}^{1}(\Vi{i},\Vi{p_i})\propto\sum_{\V\sim\{\Vi{i},\Vi{p_i}\}}\pf{1}(\V)\prod_{\substack{z=1\\z\neq \{i,p_i\}}}^{\n}\pdf(y_{z}|\Vi{z})=1/4
\end{align}
for any $(\Vi{i},\Vi{p_i})$ pair unless these two variables are the only unknown variables  connected to the check $\pf{1}(\V)$. In this case, due to the parity constraint, only one value of $\Vi{i}+\Vi{p_i}$ makes $\q_{i}^{1}(\Vi{i},\Vi{p_i})$ non-zero. In other words, an equality or inequality relationship between both variables is found and $\q_{i}^{1}(\Vi{i},\Vi{p_i})$ either takes the form of
\begin{align}\LABEQ{q143}
\q_{i}^{1}(\Vi{i},\Vi{p_i})=\frac{1}{2}\parity[\Vi{i}=\Vi{p_i}]
\end{align}
or
\begin{align}\LABEQ{q144}
\q_{i}^{1}(\Vi{i},\Vi{p_i})=\frac{1}{2}\parity[\Vi{i}\neq\Vi{p_i}].
\end{align}
For instance, assume that, in Fig. \FIG{GraphsTEP}(b),  $\Vi{p}$ and $\Vi{s}$ are known. Then, we learn that $\Vi{r}$ is equal or opposite to $\Vi{o}$.
\end{itemize}

The latter is of importance to explain the advantage of the TEP decoder over BP, because we only obtain the result in \EQ{q143} or \EQ{q144} if we compute pairwise marginals. The BP algorithm  uses only a disconnected approximation for the factors and from them we cannot derive the (in)equality constraints in \EQ{q143} and \EQ{q144}.

It can be readily check for the BEC that when we compute the functions $q_{i}^{\ell}(\Vi{i}|\Vi{p_i})$ and $\qij{i}{m}{\ell}(\Vi{i},\Vi{p_i})$ in Steps 9, 10 and 11 of the algorithm, they are proportional to the pairwise marginal computed above\footnote{For instance, if we assume that $q_{i}^{\ell}(\Vi{i},\Vi{p_i})$ is of the form of \EQ{q11}, we first compute $q(\Vi{p_i})$:
\begin{align}
q(\Vi{p_i})=\sum_{\Vi{i}}q(\Vi{i},\Vi{p_i})=\parity[\Vi{p_i}=v_1]
\end{align}
and, therefore, 
\begin{align}
q_{i}^{\ell}(\Vi{i}|\Vi{p_i})=\frac{\parity\left[\Vi{i}=v_1,\Vi{p_i}=v_2\right]}{\parity[\Vi{p_i}=v_1]}=\parity\left[\Vi{i}=v_1,\Vi{p_i}=v_2\right],
\end{align}
where we take $q_{i}^{\ell}(\Vi{i}|\Vi{p_i})=0$ when $q(\Vi{p_i})=0$.}:
\begin{align}\LABEQ{q1funcs}
q_{i}^{\ell}(\Vi{i}|\Vi{p_i})\propto\q_{i}^{\ell}(\Vi{i},\Vi{p_i}),\\\LABEQ{q1funcs2}
\qij{i}{m}{\ell}(\Vi{i},\Vi{p_i})\propto\q_{i}^{\ell}(\Vi{i},\Vi{p_i}).
\end{align}

\emph{\underline{Iteration $\ell$}.} We follow a induction procedure to analyze the result of the $\ell$-th iteration.
Given the factorization of the function $f(\V,\ell,m)$ in \EQ{ftreeTEP}, the current information of each variable is contained in the $g_z^{\ell-1,m}(\Vi{z},\Vi{p_{z}})$ functions for $z=1,\ldots,\n$ in \EQ{factors}. By replacing \EQ{q1funcs2} into \EQ{factors}, we observe that these functions are also proportional to $\q_{z}^{\ell-1}(\Vi{z},\Vi{p_z})$. Therefore,
\begin{align}\LABEQ{adfasa}
f(\V,\ell,m)\propto\pf{m}(\V)\prod_{z=1}^{\n}\q_{z}^{\ell-1}(\Vi{z},\Vi{p_z}).
\end{align}
By enumerating  the possible outcomes of the marginal $\q_{i}^{\ell}(\Vi{i},\Vi{p_i})$ over  $f(\V,\ell,m)$  in \EQ{adfasa}, we conclude that the discussion for $\ell=1$ extends for this case almost literally. By induction, we have proved that the functions $\q_{i}^{\ell}(\Vi{i},\Vi{p_i})$ only belong to one of the four states described in \EQ{q11}-\EQ{q144}.

Nevertheless, for any $\ell> 1$, we reveal a new scenario for which a variable can be de-erased, thanks to the imposed (in)equality pairwise condition in \EQ{q143} and \EQ{q144}. Assume that at iteration $\ell-1$, we learn that $V_r$ and $V_o$ are equal and at iteration $\ell$ we process the check node $\pf{m}(\mathbf{V})$ depicted Fig. 1 (b). If $V_s=0$, then the erased variable $V_p$ should be zero to fulfill the parity constraint. This scenario is not possible if we cannot capture equality relationships, which are void for the BP decoder. Therefore, the tree structured approximation in \EQ{TEP1} provides with respect to the BP procedure an extra case in which a variable can be de-erased. The key aspect is to find (in)equality conditions between variables that can be used to decode other variables related to them by parity functions. This depends on the family $\mathcal{F}_{tree}$ that it is used to decode the received word.

The Tree-EP procedure for BEC presented next can be cast as the sequential search of three particular scenarios to perform the inference (e.g. build $\mathcal{F}_{tree}$), where the only thing that matters is the number of unrevealed variables in the processed check node. We can further simplify and reformulate this procedure as a peeling-type decoder. The key idea is to simplify the LDPC Tanner graph according to the information that we sequentially obtain from the encoded bits. We rename this reformulated algorithm as the TEP decoder \cite{Olmos10-2,Olmos11}.

\subsection{The TEP decoder}\LABSEC{TEPBEC}

Before detailing the TEP decoder algorithm, let us introduce some basic definitions about Tanner graphs. The Tanner graph of an LDPC code is induced by the parity-check matrix $\Hs$ as detailed in \cite{Tanner81,Moon05}. The graph has $\n$ variable nodes $\Vi{1},\ldots,\Vi{\n}$
and $\k=\n(1-\rate)$ parity check nodes $\CN{1},\ldots,\CN{\k}$. The degree of a check node, denoted as $\cd{j}$, is the number of variable nodes connected to it in the graph. Similarly, the degree of a variable node, denoted as $\vd{i}$, is the number of check nodes connected to that variable node. In the Tanner graph, we associate a zero parity value to each check node in the graph. As we iterate, the parity of a certain check node $\CN{j}$, which is denoted hereafter as $[\CN{j}]$, might change. 

\begin{algorithm}
\begin{algorithmic}[1]
\vspace{0.05cm}
\STATE Let $\y$ be the received codeword: $\y\in\{0,?,1\}^{\n}$.
\vspace{0.05cm}
\STATE Construct the index set $\overline{\xi}:\forall s\in \overline{\xi}$ then $y_{s}\neq?$.
\vspace{0.05cm}
\FORALL {$s\in\overline{\xi}$}
\vspace{0.05cm}
	\STATE Remove from the graph variable node $\Vi{s}$.
\vspace{0.05cm}
	\STATE If $y_{s}=1$, flip the parity of the check nodes connected to $\Vi{s}$.
\vspace{0.05cm}
\ENDFOR
\vspace{0.05cm}
\REPEAT
\vspace{0.05cm}
\STATE Look for a check node of degree one or two.
\vspace{0.05cm}
\IF {\textbf{$\CN{j}$ is found with degree one, connected to $\Vi{s}$,}.}
\vspace{0.05cm}
\STATE $\Vi{s}$ is decoded with value $[\CN{j}]$.
\vspace{0.05cm}
\STATE Remove both $\Vi{s}$ and $\CN{j}$ from the graph.
\vspace{0.05cm}
\STATE If $[\CN{j}]=1$, flip the parity of the check nodes connected to $\Vi{s}$.
\vspace{0.05cm}
\ELSIF {\textbf{$\CN{j}$ is found with degree two, connected to $\Vi{o}$ and $\Vi{r}$,}}
\vspace{0.05cm}
\STATE Remove $\CN{j}$ and $\Vi{o}$ from the graph. 
\STATE If $[\CN{j}]=1$, flip the parity of the check nodes connected to $\Vi{o}$.
\vspace{0.05cm}
\STATE Reconnect to $\Vi{r}$ the check nodes connected to $\Vi{o}$.
\vspace{0.05cm}
\ENDIF
\UNTIL {the graph is empty or there are no degree-one or two check nodes in the graph.}
\end{algorithmic}
\caption{TEP algorithm for LDPC decoding over BEC.}\label{AlgTEPBEC}
\end{algorithm}

The TEP decoder is detailed in  Algorithm \ref{AlgTEPBEC}. It is based on the sequential procedure of degree-one and two check nodes in the graph.
Processing a degree-one check node in Steps 10-12 of Algorithm \ref{AlgTEPBEC} is  equivalent to find an erased variable connected to a check node where the rest of variables are known. Since the BP solution is restricted to this case, the description of the BP as a peeling-type algorithm is obtained if we do not consider Steps 13-16 in Algorithm \ref{AlgTEPBEC} \cite{Luby01,Luby97}. In this sense, the TEP decoder emerges as an improved PD. Besides, we claim that the complexity of both decoders is of the same order, i.e. $\order(\n)$. We intentionally leave to Section\SEC{Complex} a detailed analysis of the TEP complexity.

The removal of a degree-two check node in Steps 13-16 of Algorithm \ref{AlgTEPBEC} represents the inference of an equality or inequality relationship. This process 
is sketched in Fig. \FIG{Fig1}. The variable $\Vi{1}$ heirs the connections of $\Vi{2}$ (solid lines) in Fig. \FIG{Fig1}(b). Finally, the check $\CN{1}$ and the variable $\Vi{2}$ can be removed, as shown in Fig. \FIG{Fig1}(c), because they have no further implication in the decoding process. $\Vi{2}$ is de-erased once $\Vi{1}$ is de-erased. Note that, when we remove a check node of degree two, we usually create a variable node with a higher degree while the degree of the check nodes remain unaltered. 

\begin{figure}[h]
\centering
\begin{tabular}{c}
\includegraphics[width=3 cm]{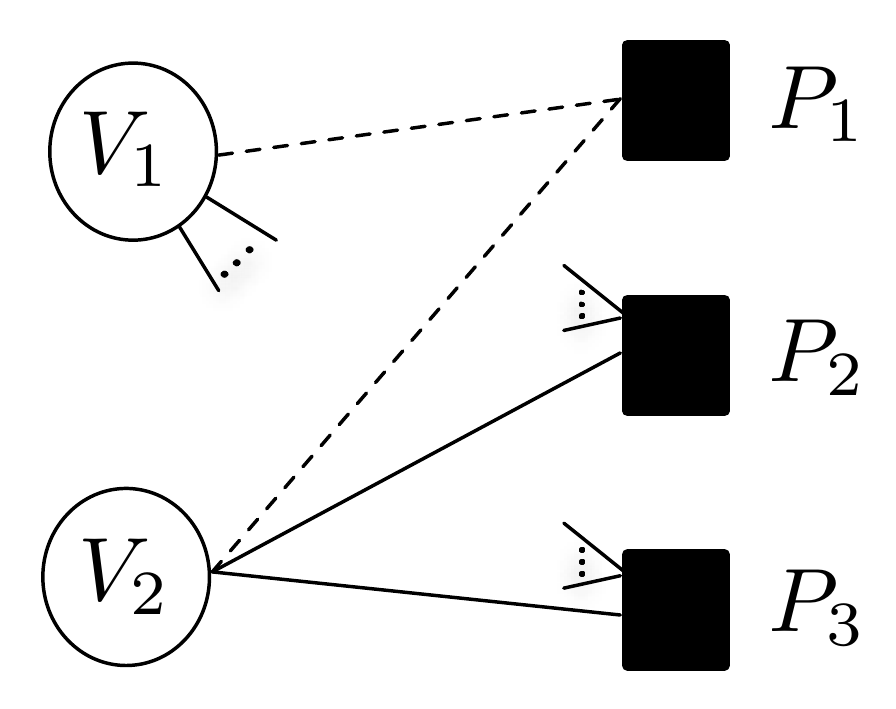}\\(a)\\  \includegraphics[width=5 cm]{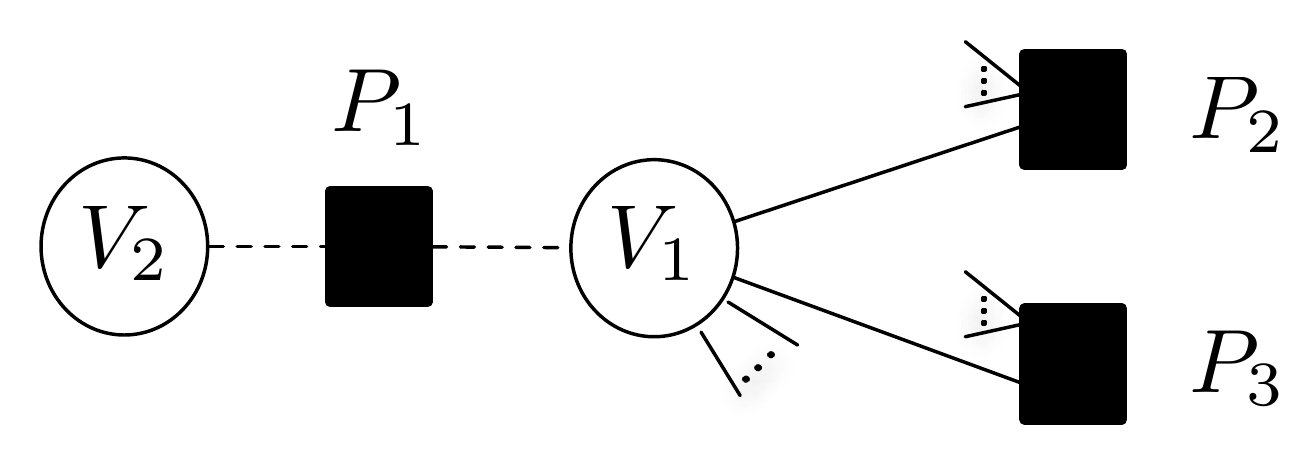}\\(b)\\\includegraphics[width=3 cm]{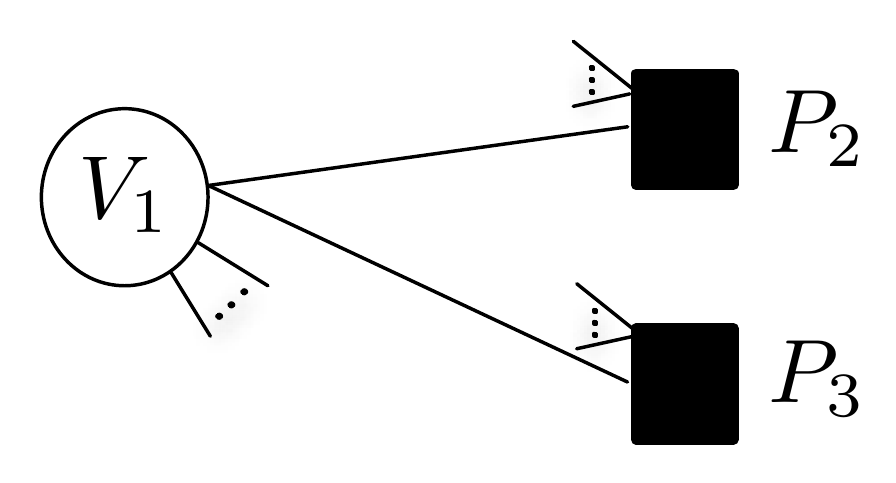}\\
(c)
\end{tabular}\caption{In (a) we show two variable nodes, $\Vi{1}$ and $\Vi{2}$, that share a check node of degree two, $\CN{1}$. In (b), $\Vi{1}$ heirs the connections of $\Vi{2}$ (solid lines). In (c), we show the graph once $\CN{1}$ and $\Vi{2}$ have been removed. If $\CN{1}$ is parity one, the parities of $\CN{2}$, $\CN{3}$ are reversed.}\LABFIG{Fig1}
\end{figure}

The TEP decoder eventually creates additional check nodes of degree one when we find a scenario equivalent to the one depicted in Fig. \FIG{FigTEP2}. Consider  two variable nodes connected to a check node of degree two that also share another check node with degree three, as illustrated in Fig. \FIG{FigTEP2}(a). After removing the check
node $\CN{3}$ and the variable node $\Vi{2}$, the check node $\CN{4}$ is now degree one, as illustrated in Fig. \FIG{FigTEP2}(b). At the beginning of the decoding algorithm, this scenario is very unlikely. However, as we remove variable and check nodes, the probability of this event grows, as we are reducing the graph and increasing the degree of the remaining variable nodes. 

\begin{figure}[h]
\centering
\begin{tabular}{cc}
\includegraphics[width=4 cm]{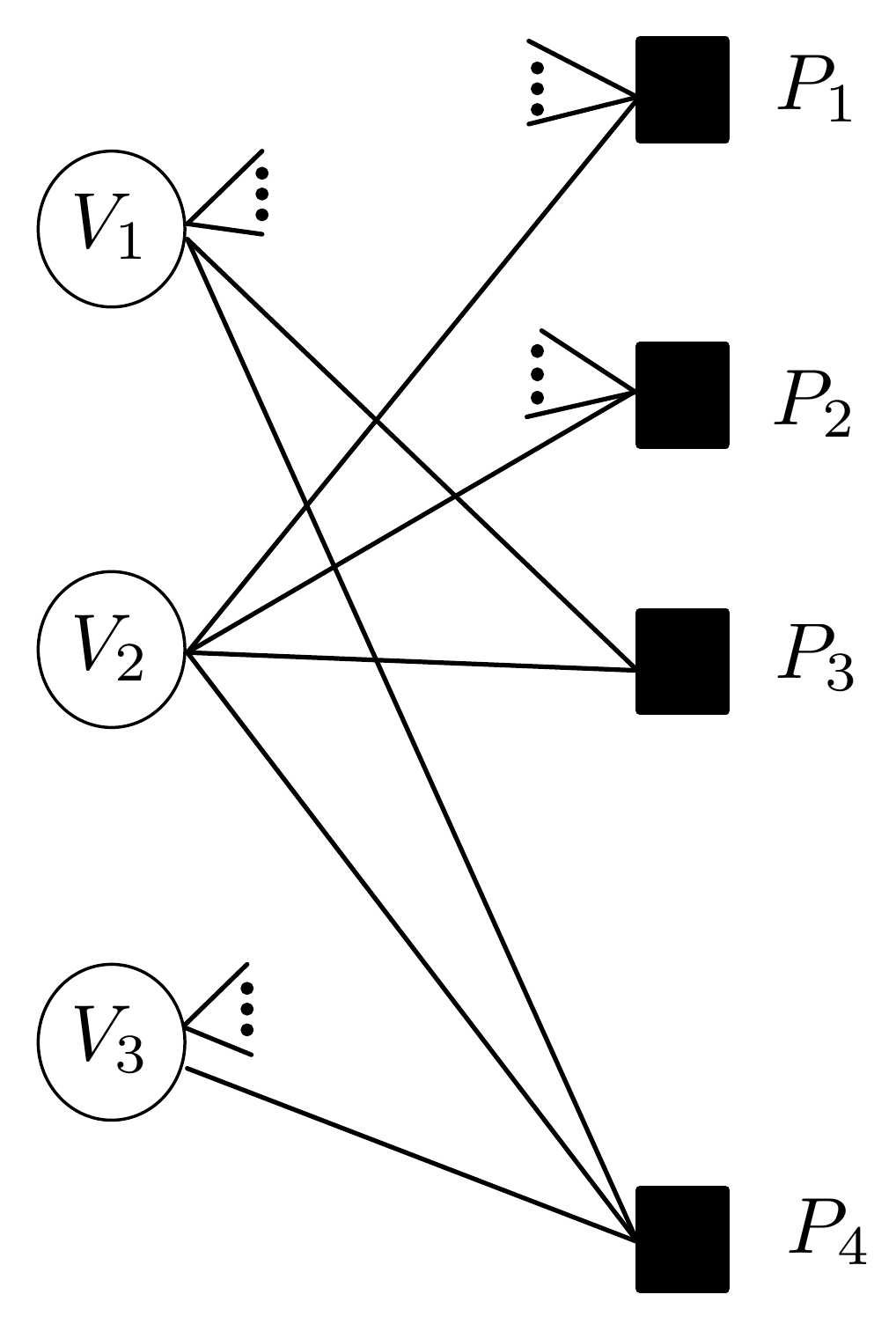} & \includegraphics[width=4 cm]{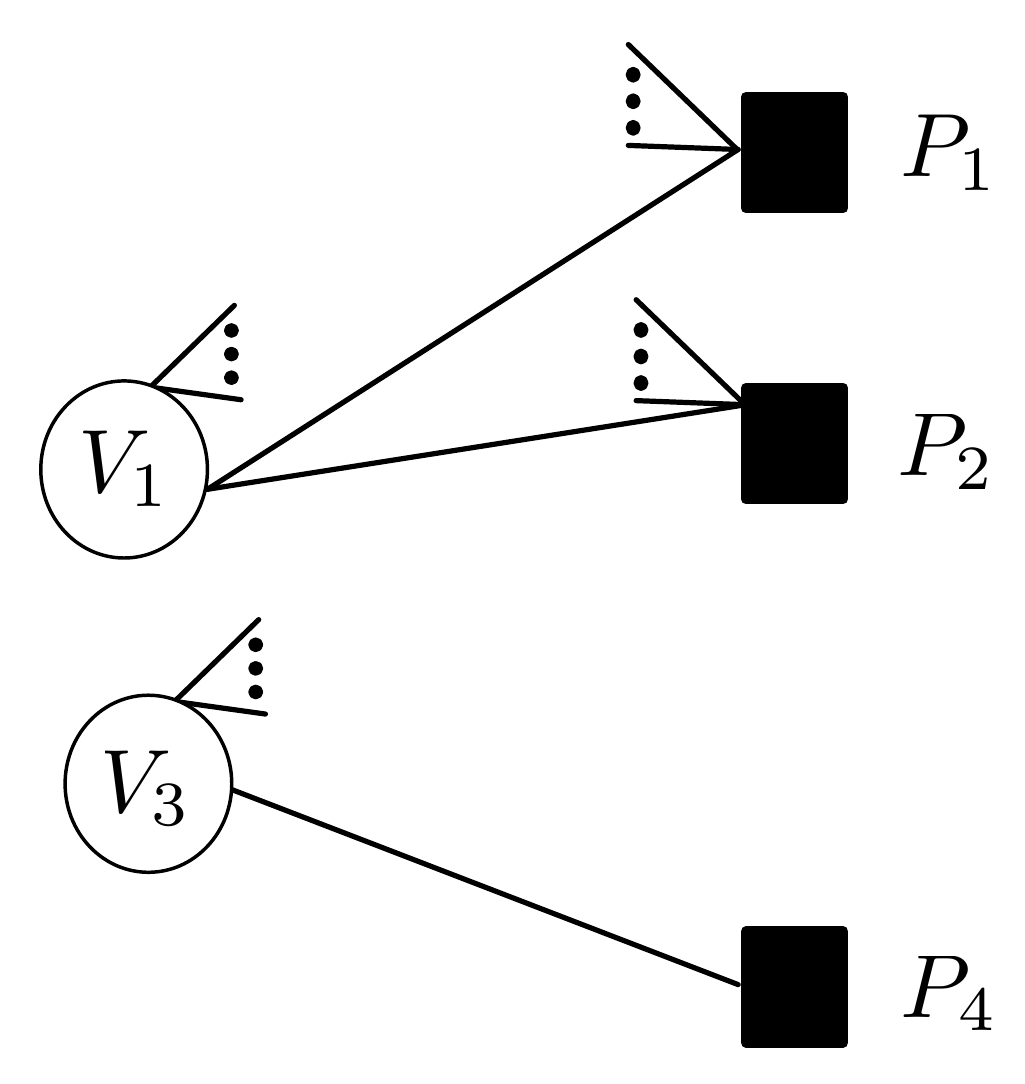}\\
(a) & (b)
\end{tabular}\caption{In (a), the variables $\Vi{1}$ and $\Vi{2}$ are connected to a two-degree check node, $\CN{3}$, and they also share a check node of degree three, $\CN{4}$.
In (b) we show the graph once the TEP has removed $\CN{3}$ and $\Vi{2}$.}\LABFIG{FigTEP2}
\end{figure}

Another important result of the TEP algorithm is that it applies the Tree-EP algorithm with no initial pairwise relations. The tree structure is not  fixed a priori. It dynamically includes a new pairwise relation in the tree structure whenever it processes a degree-two check node,  i.e. it updates $\mathcal{F}_{tree}$ on the fly. In Appendix\SEC{A1}, we show that the TEP decoding is independent of the ordering in which the variables are processed, as different processing orderings yield equivalent trees in the graph.

\subsection{Connection to previous works}

A similar procedure for removing degree-two check nodes was considered in the analysis of accumulate-repeat-accumulate (ARA) LDPC codes 
\cite{Abbasfar07,Urbanke08-2}. ARA codes were proposed to achieve channel capacity under BP decoding at bounded complexity. Roughly speaking, ARA codes are formed by the concatenation of an accumulate binary encoder, an irregular LDPC code and another accumulate binary encoder. In \cite{Sason07,Pfister05}, Pfister and Sason showed that ARA codes can be described for BEC as an equivalent irregular LDPC ensemble and hence they were able to compute the ARA BP threshold using standard techniques \cite{Urbanke01-2}. To obtain such equivalent LDPC ensemble, parity checks of degree two, corresponding to the accumulate encoding, were processed similarly to how the TEP decoder processes degree-two check nodes. Once they obtained the equivalent irregular LDPC ensemble, they just consider BP decoding. Nevertheless, we want to emphasize that the novelty of our proposal is twofold: first, we consider, describe and measure the effect of the removal of degree-two check nodes to improve the BP solution for any block code. And second, we have shown that the idea of propagating pairwise relationships can be extended for any binary DMC using the EP framework.

\section{\TEP decoder expected graph evolution}\LABSEC{EG}

Both the PD and the TEP decoder sequentially reduce the LDPC Tanner graph by either removing check nodes of degree one, or degree one and two. As a consequence, the decoding process yields a sequence of residual graphs. In \cite{Luby97,Luby01},  it is shown that if we apply the PD to elements of an LDPC ensemble, then the sequence of \emph{residual graphs} follows a typical path or expected evolution \cite{Urbanke08-2}. The authors described this path as the solution of a set of differential equations and characterized the typical deviation from it. Their analysis is based on a result on the evolution of (martingale) Markov processes due to Wormald \cite{Wormald}.  

In this section, we first introduce the Wormald's theorem and then particularize it to compute the expected evolution of the residual graphs for the TEP, which is used in the following to evaluate the decoder performance.

\subsection{Wormald's theorem}\LABSEC{Worlmald}

Consider a discrete-time Markov random process $\Zm{\parmW}{t}$ with finite $\Zdim$-dimension state space $\calA(Z)^{\Zdim}$ that depends on some parameter $\parmW>1$. Let $\Zmi{\parmW}{i}{t}$ denote the $i$-th component of $\Zm{\parmW}{t}$ for $i\in\{1,\ldots,\Zdim\}$.
 Let $\D$ be a subset of $\mathbb{R}^{d+1}$ containing those vectors $\left[0, \zi{1}, \ldots, \zi{d} \right]$ such that:

\begin{equation}\LABEQ{D}
p\left(\frac{\Zmi{\parmW}{i}{t=0}}{\parmW}=\zi{i}, 1\leq i\leq d\right)>0,
\end{equation}
 We define the stopping time $t_{\D}$ to be the minimum time so that $(t/\parmW,\Zmi{\parmW}{1}{t}/\parmW,\ldots,\Zmi{\parmW}{\Zdim}{t}/\parmW)\notin\D$. Furthermore, let $f_i(\cdot)$ for $1\leq i\leq d$, be functions from $\mathbb{R}^{d+1}\rightarrow\mathbb{R}$ such that the following conditions are fulfilled:

\begin{enumerate}
\item \label{c1} (Boundedness). There exists a constant $\Omega$ such  that for all $i\leq d$ and $\parmW>1$,
\begin{equation}\LABEQ{C1}
\left|\Zmi{\parmW}{i}{t+1}-\Zmi{\parmW}{i}{t}\right|\leq\Omega,
\end{equation}
for all $0\leq t\leq t_{\D}$.
\item \label{c2} (Trend functions). For all $1\leq i\leq d$ and $\parmW>1$,
\begin{align}\LABEQ{C2}
\E&\left[\Zmi{\parmW}{i}{t+1}-\Zmi{\parmW}{i}{t}\cond\Zm{\parmW}{t}\right]\nonumber\\&=f_{i}\left(\tau,\Zmi{\parmW}{1}{t}/\parmW,\ldots,\Zmi{\parmW}{d}{t}/\parmW\right)
\end{align}
for all $0\leq t\leq t_{\D}$, where $\tau=\frac{t}{\parmW}$.
\item \label{c3} (Lipschitz continuity). For each $i\leq d$, the function $f_i(\cdot)$ is Lipschitz continuous on the intersection of $\D$ with the half space $\left\{\left[t, \zi{1}, \ldots, \zi{d} \right]: t\geq0\right\}$, i.e., if $\boldsymbol{b}, \boldsymbol{c}\in\mathbb{R}^{d+1}$ belong to such intersection, then there exists a constant $\L$ such that
\begin{equation}\LABEQ{C3}
\left|f_i(\boldsymbol{b})-f_i(\boldsymbol{c})\right|\leq\L\sum_{j=1}^{d+1}\left|b_{j}-c_{j}\right|.
\end{equation}
\end{enumerate}

Under these conditions, the following holds:
\begin{itemize}
\item For $\left[0, b_{1}, \ldots, b_{d} \right]\in\D$, the system of differential equations
\begin{equation}\LABEQ{H1}
\frac{\partial\zi{i}}{\partial\tau}=f_i\left(\tau,  \zi{1}, \ldots, \zi{d}\right), \;\; i=1,\ldots,\Zdim,
\end{equation}
has a unique solution in $\D$ for $\zi{i}(\tau): \mathbb{R}\rightarrow\mathbb{R}$ with  $\zi{i}(0)=b_{i}$, $1\leq i\leq d$.

\item There exists a constant $\ca$ such that
\begin{align}\LABEQ{H2}
p&\left(|\Zmi{\parmW}{i}{t}/\parmW-\zi{i}(t/\parmW)|>\ca\parmW^{-\frac{1}{6}}\right)<\Zdim\parmW^{\frac{2}{3}}\exp(-\parmW^{\frac{1}{3}}/2),
\end{align}
for $i=1,\ldots,d$ and for $0\leq t \leq  t_{\D}$, where $\zi{i}(\tau)$ is the solution given by equation \EQ{H1} for 
\begin{align}\LABEQ{}
\zi{i}(0)=\E[\Zmi{\parmW}{i}{t=0}/\parmW].
\end{align}
\end{itemize}

The result in \EQ{H2} states that each realization of the process $\Zmi{\parmW}{i}{t}$ 
has a deviation from the solution of $\EQ{H1}$ smaller than $\order(\parmW^{-1/6})$ 
if $\parmW$ is large enough. Our goal is to show that this theorem is suitable to describe the LDPC graph evolution during the \TEP decoding process in certain scenarios.

\subsection{LDPC ensembles and residual graphs}\LABSEC{LDPC}

In this subsection, we introduce some basic notation about LDPC ensembles. An ensemble of codes, denoted by $\LDPC$, is defined by the code length $\n$ and the edge degree distribution (DD) pair $(\lamb{x},\rh{x})$ \cite{Urbanke01-2}:
\begin{align}\LABEQ{ensemble}
&\lambfull,\\
&\rhofull,
\end{align}
where $\lambi{i}$ represents the fraction of edges with left degree $i$ in the graph and $\rhi{j}$ is the fraction of edges with right degree $j$. The left
(right) degree of an edge is the degree of the variable (check) node it is connected to. The graph is specified in terms of fractions of edges, and not nodes, of each degree; this form is more convenient to analyze the convergence properties of the decoder. If the total number of edges in the graph
is denoted by $\Edges$, it can be readily checked that
\begin{equation}
\Edges=\frac{\n}{\sum_{i}\lambi{i}/i}.
\end{equation}

The design rate of the LDPC ensemble $\rate=\rate(\lamb{x},\rh{x})$ is set  as follows \cite{Urbanke08-2}:
\begin{align}\LABEQ{rate}
\rate=1-\frac{\Lambda_{\text{avg}}}{\Theta_{\text{avg}}},
\end{align}
where $\Lambda_{\text{avg}}$  and $\Theta_{\text{avg}}$ are, respectively, the average variable degree and the average check node degree in the graph. They can be computed from the graph DD:
\begin{align}\LABEQ{nodeavg}
\Lambda_{\text{avg}}&=\frac{1}{\int_{0}^{1}\lamb{\nu}\text{d}\nu},\\\LABEQ{nodeavg2}
\Theta_{\text{avg}}&=\frac{1}{\int_{0}^{1}\rh{\nu}\text{d}\nu}.
\end{align}

To analyze the expected graph evolution, each time step corresponds to each step of the decoder. $\Li{i}{t}$ and $\Ri{j}{t}$ are, respectively, the number of edges with left degree $i$ and right degree $j$ in the residual graph at time $t$ and we define $\li{i}{t}=\Li{i}{t}/\Edges$ and $\ri{i}{t}=\Ri{i}{t}/\Edges$. We denote by $E(t)$ the number of edges in the graph at time $t$ and define $\eremain{t}=E(t)/\Edges$. Hence, $\li{i}{t}/\eremain{t}$ for $i=1,\ldots,\Tlmax{t}$ and $\ri{j}{t}/\eremain{t}$ for $j=1,\ldots,\rhimax$ are the coefficients of the DD pair that defines the graph at time $t$. As we show in the next subsection, an small fraction of degree-one variable nodes might appear during the decoding process. Note that we have included an explicit dependency with time in $\Tlmax{t}$. As we described in Section\SEC{TEPBEC}, the removal of degree-two check nodes tends to increase the variable degree and, consequently, we expect the maximum left  degree to grow.
 
The remaining graph at time $t+1$ only depends on the graph at time $t$ and, hence, the sequence of graphs $\left(\Li{}{t}, \Ri{}{t}\right)$ along the time is a discrete time Markov process, where $L(t)=\{L_{i}(t)\}_{i=1}^{\Tlmax{t}}$, $R(t)=\{R_{j}(t)\}_{j=1}^{\rhimax}$. It can be shown that the DD sequence of the residual graphs constitutes a sufficient statistic \cite{Luby01,Urbanke09} for this Markov process and, therefore, it suffices to analyze their evolution.  

For a given LDPC ensemble and a BEC with parameter $\pe$, the TEP decoder performance is analyzed and predicted using the expected evolution of $\ri{1}{t}$ and $\ri{2}{t}$ along time. In this section, we first identify their dependence with the rest of the components of the DD in the graph. Then, we show the conditions in which they can be estimated along time using Worlmald's theorem.

\subsection{Expected graph evolution in one TEP iteration} \LABSEC{TEPiter}

We analyze the average evolution of the DD pair $\left(\Li{}{t}, \Ri{}{t}\right)$ in one iteration of the TEP decoder, i.e.
\begin{align}\LABEQ{Expected1}
&\E\left[\Li{}{t+1}-\Li{}{t}|\Li{}{t},\Ri{}{t}\right],\\\LABEQ{Expected2}
&\E\left[\Ri{}{t+1}-\Ri{}{t}|\Li{}{t},\Ri{}{t}\right].
\end{align}
At time $t$, the \TEP looks for a check node of degree one or two to remove it.  With probability  
\begin{equation}\LABEQ{PC}
\pc=\frac{\Ri{1}{t}}{\Ri{1}{t}+\Ri{2}{t}/2}
\end{equation}
the decoder selects a check node of degree one, which is denoted as Scenario $\cone$.
Alternatively, in Scenario $\ctwo$, a check node of degree two is removed with probability $\pco\doteq1-\pc$. The expected change in  the process $\left(\Li{}{t}, \Ri{}{t}\right)$ between $t$ and $t+1$ can be expressed as follows:
\begin{align}\LABEQ{ExL}
\E&\Big[\Li{i}{t+1}-\Li{i}{t}\cond\Li{}{t},\Ri{}{t}\Big]
=\\&\qquad\quad\pc\underbrace{\E\Big[\Li{i}{t+1}-\Li{i}{t}\cond\Li{}{t},\Ri{}{t}, \cone\Big]}_{I}\nonumber\\
&\qquad+\pco\underbrace{\E\Big[\Li{i}{t+1}-\Li{i}{t}\cond\Li{}{t},\Ri{}{t}, \ctwo\Big]}_{II}\nonumber,
\end{align}
for $i=1,\ldots,\Tlmax{t}$ and
\begin{align}\LABEQ{ExR}
\E&\Big[\Ri{j}{t+1}-\Ri{j}{t}\cond\Li{}{t},\Ri{}{t}\Big]
=\\&\quad\qquad\pc\underbrace{\E\Big[\Ri{j}{t+1}-\Ri{j}{t}\cond\Li{}{t},\Ri{}{t}, \cone\Big]}_{III}\nonumber\\
&\qquad+\pco\underbrace{\E\Big[\Ri{j}{t+1}-\Ri{j}{t}\cond\Li{}{t},\Ri{}{t}, \ctwo\Big]}_{IV}\nonumber,
\end{align}
for $j=1,\ldots,\rhimax$. In the following, we omit the pair $\{\Li{}{t},\Ri{}{t}\}$ in the expectations to keep the notation uncluttered. 

The terms $I$ in \EQ{ExL} and $III$ in \EQ{ExR} correspond to one iteration of the BP decoder and they were already computed in \cite{Luby01}. We include them for completeness:
\begin{align}\LABEQ{LBP}
\E\Big[\Li{i}{t+1}-\Li{i}{t}\cond \cone\Big]=-i\frac{\li{i}{t}}{\eremain{t}}\;\;,
\end{align}
for $i=1,\ldots,\Tlmax{t}$,
\begin{align}\LABEQ{RBP}
\E\Big[\Ri{j}{t+1}-&\Ri{j}{t}\cond\cone\Big]=\\
&j\left(\frac{\ri{j+1}{t}}{\eremain{t}}-\frac{\ri{j}{t}}{\eremain{t}}\right)(\lav{t}-1)-\delta(j-1)\nonumber
\end{align}
for $j=1,\ldots,\rhimax$, where $\delta(j)$ is the Kronecker's delta function and
\begin{equation}\LABEQ{l_avg}
\lav{t}=\sum_{i=1}^{\Tlmax{t}}i\,\li{i}{t}/\eremain{t}
\end{equation}
is the average edge left degree at time $t$. We now compute the terms $II$ and $IV$ in \EQ{ExL} and \EQ{ExR}, respectively. When a check node of degree two is removed, e.g., $\CN{j}$ in Fig. \FIG{Cases}, there are two possible subscenarios:
\begin{itemize}
\item $\ctwonoshare$: The variable nodes $\Vione$ and $\Vitwo$ connected with the check node $\CN{j}$ do not share another check node, depicted in Fig. \FIG{Cases}(a).
\item $\ctwoshare$: The variable nodes $\Vione$ and $\Vitwo$ connected with the check node $\CN{j}$ share at least another check node, depicted in Fig. \FIG{Cases}(b).
\end{itemize}

\begin{figure}
\centering
\begin{tabular}{cc}
  \includegraphics[width=4 cm]{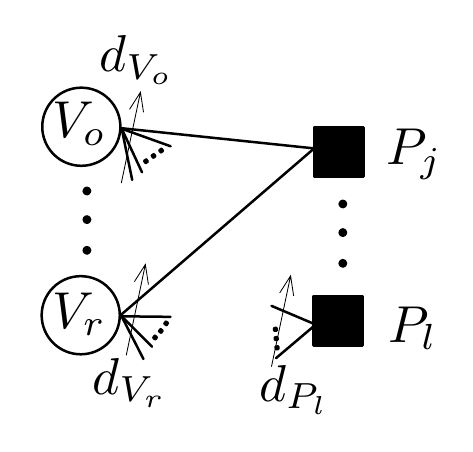} & \includegraphics[width=4 cm]{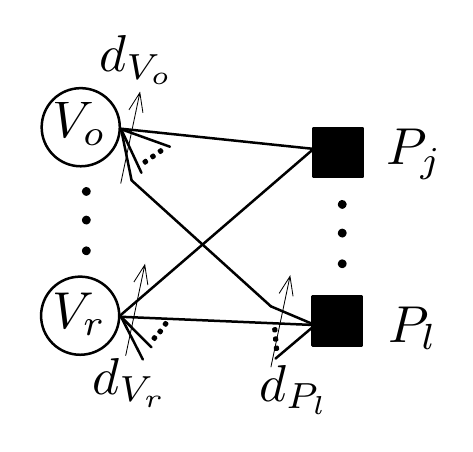} \\
  (a) & (b)
\end{tabular}
\caption{Examples of Subscenarios $\ctwonoshare$ in (a), given the check node $\CN{j}$ the variable nodes $\Vione$ and $\Vitwo$ only share the check node $\CN{j}$,  and $\ctwoshare$ in (b), in which they share another check node of degree $\cd{l}$.}\LABFIG{Cases}
\end{figure}

Let $\pA{t}$ be the probability of scenario $\ctwoshare$, i.e., $\pA{t}=\P\left(\ctwoshare|\ctwo\right)$. In Appendix\SEC{A3}, we show that this probability is given by:
\begin{equation}\LABEQ{pA}
\pA{t}=\frac{(\lav{t}-1)^{2}(\rav{t}-1)}{\eremain{t}\Edges},
\end{equation}
where 
\begin{equation}\LABEQ{r_avg}
\rav{t}=\sum_{j=1}^{\rhimax}j\,\ri{j}{t}/\eremain{t}
\end{equation}
is the average edge right degree. In Appendix\SEC{A3}, we also show that the scenario $\ctwoshare$ is dominated by the case in which the two variables connected to a check node of degree two only share another check node, as illustrated in Fig. \FIG{Cases}(b). To compute $II$ and $IV$ in \EQ{ExL} and \EQ{ExR}, we first evaluate the graph expected change for $\ctwonoshare$ and  $\ctwoshare$ and then average them using $\pA{t}$:
\begin{align}
&\E\big[\Li{i}{t+1}-\Li{i}{t}|\ctwo\big]=
\pA{t}\E\left[\Li{i}{t+1}-\Li{i}{t}| \ctwoshare\right]\nonumber\\\LABEQ{ExLS2}
&\qquad+\pAo{t}\E\left[\Li{i}{t+1}-\Li{i}{t}| \ctwonoshare\right],\\\nonumber\\ 
&\E\big[\Ri{j}{t+1}-\Ri{j}{t}|\ctwo\big]=
\pA{t}\E\left[\Ri{j}{t+1}-\Ri{j}{t}|\ctwoshare\right]\nonumber\\\LABEQ{ExRS2}
&\qquad+\pAo{t}\E\left[\Ri{j}{t+1}-\Ri{j}{t}|\ctwonoshare\right],
\end{align}
where $\pAo{t}=1-\pA{t}$. Now we evaluate each term in \EQ{ExLS2} and \EQ{ExRS2}.

\subsubsection{Expected change in the graph assuming $\ctwonoshare$}\LABSEC{S2A}

At time $t$, we remove the check node $\CN{j}$ in Fig. \FIG{Cases}(a), which is connected to
$\Vione$ and $\Vitwo$. If $\Vione$ is the remaining variable, its degree becomes $\vd{\indone}+\vd{\indtwo}-2$. 
From the edge perspective, the graph losses $\vd{\indone}$ edges with left degree $\vd{\indone}$ and $\vd{\indtwo}$ edges with left degree
 $\vd{\indtwo}$, and gains $\vd{\indone}+\vd{\indtwo}-2$ edges with left degree $\vd{\indone}+\vd{\indtwo}-2$. 
 Note also that we have the same result if $\Vi{q}$ is the remaining variable. The node degrees $\vd{\indone}$ and $\vd{\indtwo}$ are
 asymptotically pairwise independent \cite{Montanari09} and, thus, $\Vione$ and $\Vitwo$ are degree $i$ with probability $\li{i}{t}/\eremain{t}$ for $i=1,\ldots,\Tlmax{t}$.

We first focus on the evolution in the number of edges with left degree $i\geq3$ at time $t+1$, i.e. $\Li{i}{t+1}$. The sample 
space of $\Li{i}{t+1}-\Li{i}{t}$ is given by:
 \begin{equation}\LABEQ{Liset}
\Li{i}{t+1}-\Li{i}{t}\in\{-i,-2i,+i,0\}\qquad i\geq3,
 \end{equation}
where  
\begin{itemize}
\item $\Li{i}{t+1}-\Li{i}{t}=-i$, if $(\vd{\indone}=i$ XOR  $\vd{\indtwo}=i)$
AND $\vd{\indone}+\vd{\indtwo}-2\neq i$. 
\item $\Li{i}{t+1}-\Li{i}{t}=-2i$, if $\vd{\indone}=\vd{\indtwo}=i$. 
\item $\Li{i}{t+1}-\Li{i}{t}=+i$, if $\vd{\indone}\neq i, \vd{\indtwo}\neq i$ AND $ \vd{\indone}+\vd{\indtwo}-2=i$. 
\item $\Li{i}{t+1}-\Li{i}{t}=0$, otherwise.
\end{itemize}

The probability associated to each case can be easily evaluated. Finally, the expected change in the number of edges with left degree yields:
\begin{align}\LABEQ{LiNo}
&\E\Big[\Li{i}{t+1}-\Li{i}{t}\cond,\ctwonoshare\Big]=\nonumber\\
&\qquad-i\frac{\li{i}{t}}{\eremain{t}}2\left(1-\frac{\li{i}{t}}{\eremain{t}}-\frac{\li{2}{t}}{\eremain{t}}\right)-2i\left(\frac{\li{i}{t}}{\eremain{t}}\right)^{2}\nonumber\\&\qquad+i\convfull{2}{i}{1}. 
\end{align}

For $i=2$, the sample space reduces to $\Li{2}{t+1}-\Li{2}{t}\in\{-2,+2,0\}$
and, it is straightforward to show that:
\begin{align}\LABEQ{LiNo2}
\E&\left[\Li{2}{t+1}-\Li{2}{t}\cond\ctwonoshare\right]=\nonumber\\
&-2\frac{\li{2}{t}}{\eremain{t}}2\left(1-\frac{\li{2}{t}}{\eremain{t}}\right)-2\left(\frac{\li{2}{t}}{\eremain{t}}\right)^{2}+2\left(\frac{\li{1}{t}}{\eremain{t}}\frac{\li{3}{t}}{\eremain{t}}\right).
\end{align}
For the case $i=1$, we have
\begin{align}\LABEQ{LiNo1}
\E&\left[\Li{1}{t+1}-\Li{1}{t}\cond,\ctwonoshare\right]\nonumber\\
&=-2\frac{\li{1}{t}}{\eremain{t}}\left(1-\frac{\li{2}{t}}{\eremain{t}}\right)-2\left(\frac{\li{1}{t}}{\eremain{t}}\right)^{2}.
\end{align}

Note that degree-one variable nodes are not created in both scenarios $\cone$ and $\ctwonoshare$. Regarding the edge right degree distribution, only two edges of right degree two are lost and, hence,
\begin{eqnarray}\LABEQ{RiNo}
&&\E\left[\Ri{2}{t+1}-\Ri{2}{t}\cond\ctwonoshare\right]=-2,\\
&&\E\left[\Ri{j}{t+1}-\Ri{j}{t}\cond\ctwonoshare\right]=0,\;j\neq2.\qquad
\end{eqnarray}


\subsubsection{Expected graph evolution assuming $\ctwoshare$}\LABSEC{S2B}

We study now the scenario depicted at Fig. \FIG{Cases}(b), where the variables
 $\Vione$ and $\Vitwo$ are also linked to another check node $\CN{l}$ of degree $\cd{l}$. In this case, the degree of the remaining variable is $\vd{\indone}+\vd{\indtwo}-4$ and
 the check node $\CN{l}$ losses two edges and its degree reduces to $\cd{l}-2$. On the left side, the graph losses $\vd{\indone}$ edges with left degree $\vd{\indone}$ and $\vd{\indtwo}$ edges with left degree
 $\vd{\indtwo}$, and gains $\vd{\indone}+\vd{\indtwo}-4$ edges with left degree $\vd{\indone}+\vd{\indtwo}-4$. 
 
 For a given degree $i\neq4$, $\Li{i}{t+1}-\Li{i}{t}$ takes value in the set in \EQ{Liset}. The possible combinations of $\vd{\indone},\vd{\indtwo}$ and the associated $\Li{i}{t+1}-\Li{i}{t}$ values are now as follows:
 \vspace{0.3cm}
\begin{itemize}
\item $\Li{i}{t+1}-\Li{i}{t}=-i$, if $(\vd{\indone}=i$ XOR  $\vd{\indtwo}=i)$
AND $\vd{\indone}+\vd{\indtwo}-4\neq i$. 
\item $\Li{i}{t+1}-\Li{i}{t}=-2i$, if $\vd{\indone}=\vd{\indtwo}=i$.
\item $\Li{i}{t+1}-\Li{i}{t}=+i$, if $\vd{\indone}\neq i, \vd{\indtwo}\neq i$ AND $ \vd{\indone}+\vd{\indtwo}-4=i$.
\item $\Li{i}{t+1}-\Li{i}{t}=0$, otherwise.
\end{itemize}

The expected value of $\Li{i}{t+1}-\Li{i}{t}$ for $i\neq4$ is given by:
\begin{align}\LABEQ{LiS}
&\E\Big[\Li{i}{t+1}-\Li{i}{t}\cond,\ctwoshare\Big]=\nonumber\\
&\qquad-i\frac{\li{i}{t}}{\eremain{t}}2\left(1-\frac{\li{i}{t}}{\eremain{t}}-\frac{\li{4}{t}}{\eremain{t}}\right)-2i\left(\frac{\li{i}{t}}{\eremain{t}}\right)^{2}\nonumber\\&\qquad+i\convfull{4}{i}{3},
\end{align}
and, for the case $i=4$ we obtain:
\begin{align}\LABEQ{LiS4}
&\E\Big[\Li{4}{t+1}-\Li{4}{t}\cond\ctwoshare\Big]=\nonumber\\
&\qquad-4\frac{\li{4}{t}}{\eremain{t}}2\left(1-\frac{\li{4}{t}}{\eremain{t}}\right)-4\left(\frac{\li{4}{t}}{\eremain{t}}\right)^{2}\nonumber\\&\qquad+4 \convfull{4}{i}{3}.
\end{align}


The values of $i$ for which the number of edges involved in \EQ{LiNo} and \EQ{LiS} is larger than the number of edges in the graph are not allowed. We set a zero probability for them. We do not enumerate the complete list of these combinations for the sake of the readability of the section.

The importance of scenario $\ctwoshare$ lies on the fact that the check node $\CN{l}$ losses two edges and its degree reduces to $\cd{l}-2$. Therefore, check nodes with right degree one can be created. Since the check node has degree $\cd{l}=l$ with probability $\ri{l}{t}/\eremain{t}$, it can be shown that:
\begin{align}\LABEQ{Ri}
\E&\Big[\Ri{j}{t+1}-\Ri{j}{t}\cond\ctwoshare\Big]=j\left(\frac{\ri{j+2}{t}}{\eremain{t}}-\frac{\ri{j}{t}}{\eremain{t}}\right),
\end{align}
for $j>2$,
\begin{align}\LABEQ{Ri2}
&\E\Big[\Ri{2}{t+1}-\Ri{2}{t}\cond\ctwoshare\Big]=2\left(\frac{\ri{4}{t}}{\eremain{t}}-\frac{\ri{2}{t}}{\eremain{t}}\right)-2,
\end{align}
and
\begin{align}\LABEQ{Ri1}
\E\Big[\Ri{1}{t+1}-\Ri{1}{t}\cond\ctwoshare\Big]=\frac{\ri{3}{t}}{\eremain{t}}.
\end{align}

With the results in \EQ{LiNo}-\EQ{Ri1} and the probability $\pA{t}$ in \EQ{pA}, we are able to compute the terms $II$ and $IV$ in \EQ{ExL} and \EQ{ExR}, obtaining the expected graph evolution in one iteration of the \TEP decoder. It is important for the following analysis to note that, in any possible scenario, $\Ri{1}{t}$ and $\Ri{2}{t}$ only depend on the left DD through $\lav{t}$.

\subsection{Analysis of $\lav{t}$ in the asymptotic limit}\LABSEC{LAV}

The application of Wormald's Theorem to guarantee the concentration around the TEP expected  graph evolution derived in the previous subsection is not formally possible in the limit $\n\rightarrow\infty$. First, the maximum left degree $\Tlmax{t}$ is not bounded and the boundedness condition in \EQ{C1} might not hold. And second, note that the dimension of the Markov process $\{L(t),R(t)\}$ is not bounded either. However, the asymptotic limit performance can be studied by observing the evolution of the average left degree $\lav{t}$, which  
measures how likely is the creation of degree-one check nodes by removing degree-two check nodes, see Fig. \FIG{FigTEP2}. 

\begin{figure*}[]
\centering
\begin{tabular}{cc}
\includegraphics[width=8 cm]{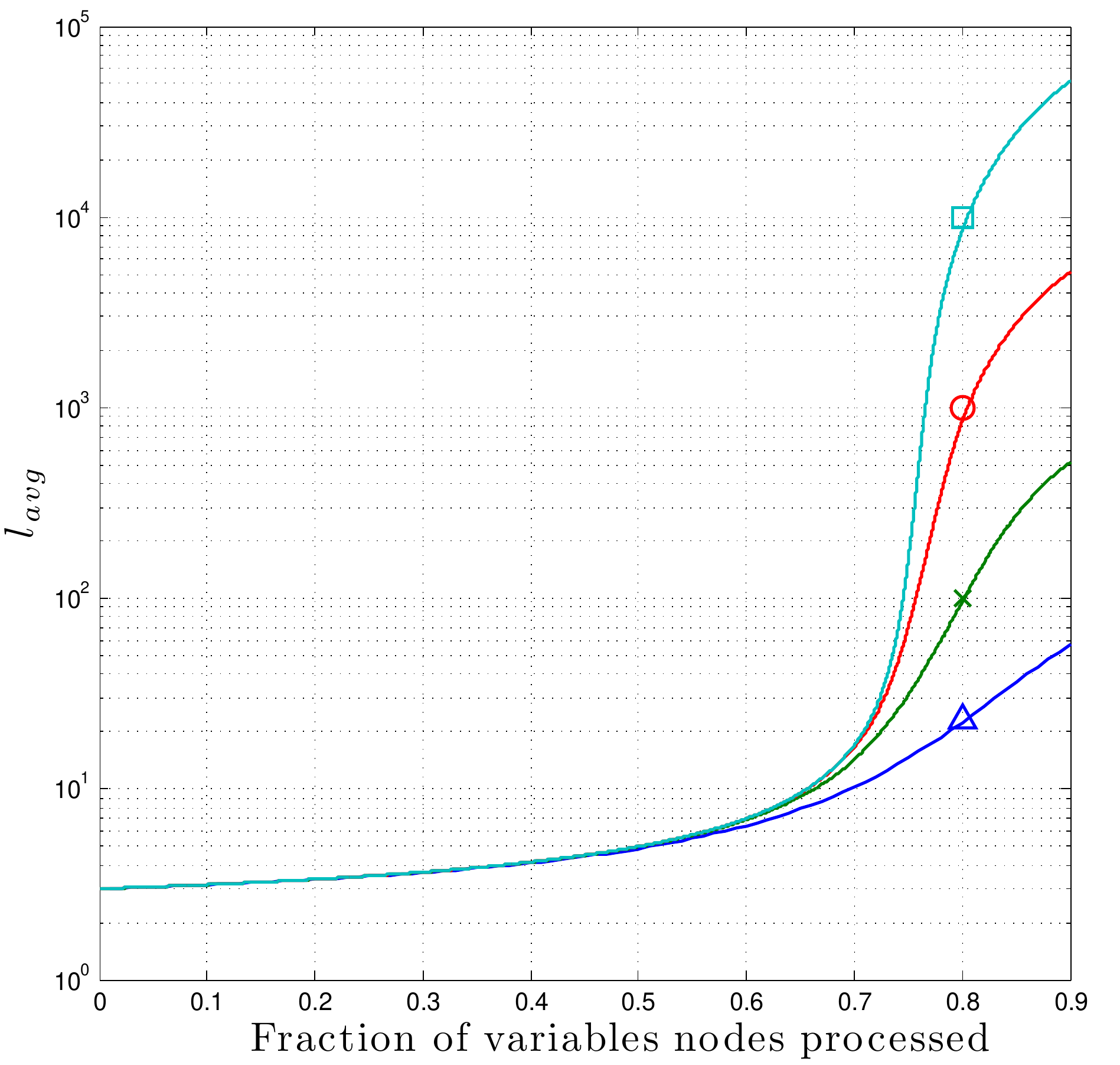} & \includegraphics[width=8 cm]{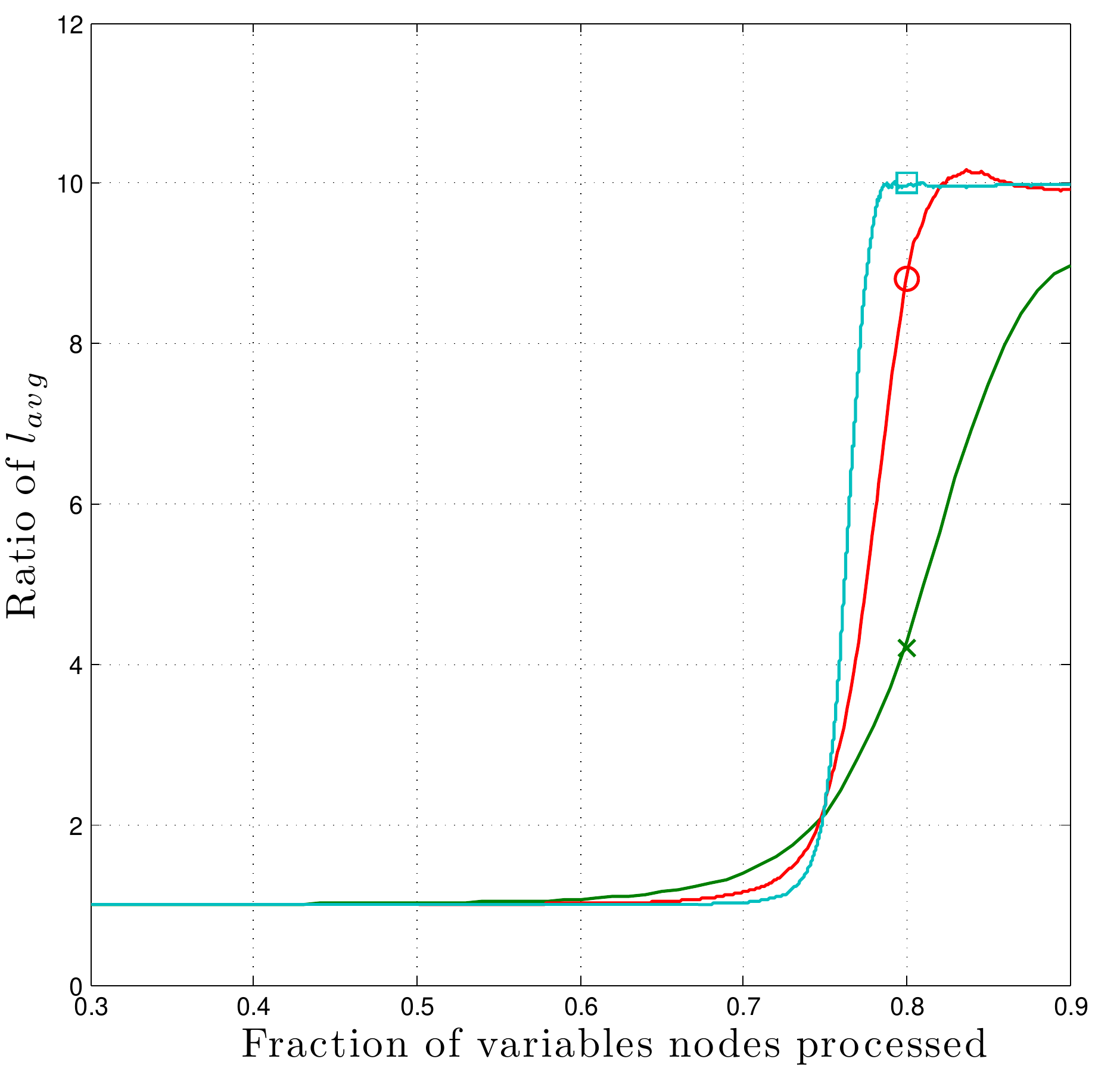}\\ (a) & (b)
\end{tabular}\caption{ In Fig. \FIG{Phasetran}(a), we plot the average left degree $\lav{t}$ computed using the urn model for $\n'=100$ $(\bigtriangleup)$, $1000$ $(\times)$, $10^{4}$ $(\circ)$, $10^{5}$ $(\square)$, as a function of the fraction of processed variables. In Fig. \FIG{Phasetran}(b) we plot the ratio between $\lav{t}$ for $\n'=10^{j+1}$ and $\n'=10^j$ for $j=2$ $(\times)$, $3$ $(\circ)$ and $4$ $(\square)$.}\LABFIG{Phasetran}
\end{figure*}

If we assume an $\LDPC$ ensemble with bounded complexity, i.e. finite $\lambda_{\text{avg}},\lambimax$ and $\rhimax$ values, then in the limit $\n\rightarrow\infty$ we get $\pA{t=0}=0$. As long as $\pA{t}=0$, the TEP and BP solutions are equivalent.
Therefore, the BP decoding threshold is only improved for those LDPC ensembles for which, as the TEP decoder runs, there exists some $t_{0}\leq t_{\D}$ such that
\begin{align}\LABEQ{limit capacity}
&\lim_{t\rightarrow t_{0}}\lim_{\n\rightarrow\infty}\pA{t}\nonumber\\&=\lim_{t\rightarrow t_{0}}\lim_{\n\rightarrow\infty}\frac{(\lav{t}-1)^{2}(\rav{t}-1)\sum_{i}\lambi{i}/i}{\eremain{t}\n}>0
\end{align}
or, equivalently, if  there exists some $t_{0}\leq t_{\D}$ such that
\begin{align}\LABEQ{lavinfty}
\lim_{t\rightarrow t_{0}}\lav{t}=\infty,
\end{align}
since the rest of the terms in \EQ{limit capacity} stay bounded during the TEP procedure if $\pe\geq\peBP$. If $\lav{t}$ becomes infinite, the asymptotic decoding threshold for the TEP might be higher than the decoding threshold of the BP decoder. However, we cannot rely on Wormald's Theorem to find the TEP threshold $\peTEP$. In this section, we analyze the conditions for $\lav{t}$ to go to infinity, which opens the possibility for $\peTEP>\peBP$. Although, we have not found so far any LDPC ensemble with a finite degree distribution meeting these conditions and, besides, the strategies followed to maximize this effect yield ensembles that lack practical interest, as shown later. We leave as an interesting open problem the search for practical LDPC ensembles for which $\peTEP>\peBP$ and, as well as, the exact computation of such threshold.

Assume an LDPC ensemble of infinite length. As proven in Appendix\SEC{A1}, the processing order is irrelevant and, hence, we can always run the BP first. 
The removal of degree-one check nodes does not increase the left average degree, i.e. $\lav{t}$ is finite \cite{Urbanke08-2}. Over the BP residual graph, processing a degree-two check node can be explained with a Polya urn model. Consider $\Tlmax{t}$ urns labeled $1,2,\ldots,\Tlmax{t}$. In urn $i$ we place a ball for each variable of degree $i$ in the graph. In each iteration, we take two balls from the urns. The urns are chosen independently with probability proportional to the number of balls per urn times its label. One ball is thrown away and the other one is placed in the urn labeled by the sum of the labels of the picked balls minus 2, introducing a new urn if it did not previously exist. For example, if we pick a ball with label ``3'' and a ball with label ``4'', we put one ball in the urn labeled ``5''. We repeat this process $c_2$ times, the number of degree-two check nodes in the BP residual graph. The resulting $\lav{t}$ is the sum of the number of balls in all urns multiplied by their labels.

It is straightforward to conclude that, as we process check nodes of degree two, $\lav{t}$ increases, but does it becomes infinite? On the one hand, if we start with $\n'$ balls and $c_2=\n'-1$, $\lav{t}$ becomes infinite as $\n'\rightarrow\infty$. On the other hand, if $c_2$ is small enough such that we do not pick twice the same ball, $\lav{t}$ stays finite. There must be an intermediate value for $c_2$ for which $\lav{t}$ becomes infinite. 

Let us illustrate this result with the $(3,6)$ regular LDPC code. We run the urn model described above to numerically compute $\lav{t}$. In Fig. \FIG{Phasetran}(a), we plot  $\lav{t}$ for $\n'=10^2$ $(\bigtriangleup)$, $10^3$ $(\times)$, $10^{4}$ $(\circ)$, $10^{5}$ $(\square)$, as a function of the fraction of processed variables $c_2/\n'$. In Fig. \FIG{Phasetran}(b) we plot the ratio between $\lav{t}$ for $\n'=10^{j+1}$ and $\n'=10^j$ for $j=2$ $(\times)$, $3$ $(\circ)$ and $4$ $(\square)$. Although the urn model is only valid asymptotically and as long as $\pA{t}=0$, we can see there is a clear phase-transition. If we process less than $74\%$ of the variables nodes we should not expect $\lav{t}$ to go to infinity. 
For a BEC channel with erasure probability just above the BP threshold ($\epsilon=0.4295$), the ratio between degree-two check nodes and variables in the BP residual graph is $0.625$ \cite{Luby01}. Therefore, for this ensemble the TEP decoder performs asymptotically as the BP decoder, i.e. $\peTEP=\peBP$, because $\lav{t}$ stays finite.


In order to maximize the fraction $c_2$ with respect to the number of variables in the BP residual graph there are two basic strategies. We can either design LDPC ensembles to minimize the size of the residual BP graph or we can design the ensemble to increase the presence of degree-two check nodes. Given the  known analytical expressions for the BP residual graph \cite{Luby01}, it is easy to prove that in the first case the solution yields standard irregular ensembles for which $\peBP\rightarrow\peMAP$, i.e. codes for which the threshold is given by the stability condition \cite{Oswald02,Luby01,Di06}. In this case, there is no margin and $\peBP=\peTEP=\pe_{\rm MAP}$. In the second case, the ensemble presents severe limitations: for instance,  by \EQ{rate}, any ensemble where all the check nodes are of degree two, i.e. $\rh{x}=x$,  has a rate
\begin{align}\LABEQ{}
\rate(\lamb{x},x)=1-\frac{2}{\int_{0}^{1}\lamb{\nu}\text{d}{\nu}}\in[0,1]\Rightarrow\lamb{x}=a+bx,
\end{align}
where $a,b\in\mathbb{R}$ are such that $a+b=1$. Either the code presents minimum distance of only one bit if $a>0$ or zero rate if $a=0$. We approach this undesired behavior as we increase the fraction of degree-two check nodes $\rhi{2}$. 

\subsection{TEP decoder complexity}\LABSEC{Complex}

In the next two lemmas we prove that, for most LDPC codes, namely those for which $\peTEP=\peBP$, the TEP complexity linearly scales with the code length $\n$:


\lemma{\LABLEM{finiteLAVG} Consider a transmission over a BEC of parameter $\pe$ 
using a code $\code$ sampled at random from $\LDPC$, where the polynomials $\lambda(x)$ and $\rho(x)$ are of  finite order and the code length $\n\rightarrow\infty$. Let $\Lambda_{\text{avg}}(t,\y,\code)$ and $\Theta_{\text{avg}}(t,\y,\code)$ be the evolution under TEP decoding of the average variable and check node degrees when the channel realization is $\y$. For any vector $\y$,  $\Lambda_{\text{avg}}(t,\y,\code)$ and $\Theta_{\text{avg}}(t,\y,\code)$ are bounded during the whole decoding process.}

\begin{proof}
See Appendix\SEC{A4}.
\end{proof}

The boundedness property of the variable degree evolution under TEP decoding, proved in Lemma\LEM{finiteLAVG}, has a significant impact in the decoding complexity.

\lemma{\LABLEM{Complexity}
 The complexity per iteration of the TEP algorithm for decoding LDPC codes of positive rate and finite maximum variable and check node degrees remains constant for any code for which $\peTEP=\peBP$. Compared to the BP complexity,  the TEP complexity differs at most in a constant complexity per iteration. 
 
The complexity per iteration can grow as a function of $\n$ only for $\pe\geq\peBP$ and for those ensembles for which 
the limiting condition in \EQ{limit capacity} is fulfilled.

\begin{proof}
See Appendix\SEC{A5}.
\end{proof}

\color{black}
%
%
%

\subsection{\TEP decoder differential equations} \LABSEC{TEPdiffeqs}

Unlike the asymptotic case, for finite-length LDPC ensembles such that $\peTEP=\peBP$, Wormald's Theorem can be applied to estimate the expected graph evolution along the decoding process. In particular, we are interested in the evolution of $\ri{1}{t}$ and $\ri{2}{t}$, which is basic to predict the finite-length performance as we later discuss  in Section\SEC{FLregimen}. Consider the TEP decoding of an arbitrarily large $\LDPC$ ensemble with finite code length.
 First, note that expressions  \EQ{ExL} and \EQ{ExR} play the role of the trend functions in Wormald's theorem:
\begin{align}
\E&\Big[\Ri{j}{t+1}-\Ri{j}{t}\Big]\LABEQ{Rig}\\\nonumber
&=\g{j}\left(\frac{t}{\Edges},\frac{\Li{\text{avg}}{t}}{\Edges},\frac{\Ri{1}{t}}{\Edges},\ldots,\frac{\Ri{\rhimax}{t}}{\Edges} \right),\\
\E&\Big[\Li{i}{t+1}-\Li{i}{t}\Big]\LABEQ{Lif}\\\nonumber&=\f{i}\left(\frac{t}{\Edges},\frac{\Li{}{t}}{\Edges},\frac{\Ri{1}{t}}{\Edges},\frac{\Ri{2}{t}}{\Edges} \right),
\end{align}
for $i=1,\ldots,\Edges$ and $j=1,\ldots,\rhimax$, where $L_{\text{avg}}(t)=\sum_ii\Li{i}{t}$.
Regarding the bounding condition in $\EQ{C1}$, for finite ensembles, any individual realization $|\Ri{j}{t+1}-\Ri{j}{t}|$ and $|\Li{i}{t+1}-\Li{i}{t}|$ are bounded by $\Edges$ at any time. This condition also ensures the Lipschitz continuity of the functions $\f{i}(\boldsymbol{\boldsymbol{b}})$ and $\g{j}(\boldsymbol{b})$ in \EQ{Rig} and \EQ{Lif} for all $\boldsymbol{b}$ contained in the set $\D$ defined in \EQ{D}. In this case, $\D$ is simply the region of possible initial conditions for the DD, i.e. the region within the  hypercube of unit length and dimension $\Edges+\rhimax$.


%

Given the former discussion, Wormald's Theorem ensures that when we solve the differential equations:
\begin{align}\LABEQ{System1}
&\frac{\partial\ri{j}{\tau}}{\partial\tau}=\g{j}\left(\tau,\lav{\tau},\ri{1}{\tau},\ldots,\ri{\rhimax}{\tau} \right),\\\nonumber\\\LABEQ{System2}
&\frac{\partial\lav{\tau}}{\partial\tau}=\sum_{i}i\f{i}\left(\tau,\li{2}{\tau},\ldots,\li{\Edges}{\tau},\ri{1}{t},\ri{2}{\tau} \right),\\\LABEQ{System3}
&\frac{\partial\li{i}{\tau}}{\partial\tau}=\f{i}\left(\tau,\li{2}{t},\ldots,\li{\Edges}{t},\ri{1}{t},\ldots,\ri{\rhimax}{t} \right),\;\;\;
\end{align}
for $j=1,\ldots,\rhimax$ and $i=1,\ldots,\Edges$ with initial conditions $\li{i}{0}=\E\left[\Li{i}{0}\right]/\Edges$ and $\ri{j}{0}=\E\left[\Ri{j}{0}\right]/\Edges$, then the solution for $\ri{1}{\tau}$ and $\ri{2}{\tau}$
is unique and, with high probability, by \EQ{H2}, does not differ more than $\order(E^{-1/6})$ from any particular realization of $\Ri{j}{t}/\Edges$ for $j=1$ and $2$. Because of the expected graph evolution equations 
derived in Section\SEC{TEPiter} assumed independency, the deviation that we can expect with respect to the solution given by \EQ{System1}-\EQ{System3} might rise up to $\order(E^{-1/6}+E^{-1})$. As we show in Appendix\SEC{A3}, any cycle involving $m$ variables decays as $E^{-m}$. The most dominant component, i.e. $E^{-1}$, is not significant compared to maximum deviation guaranteed by Wormald's theorem.

Note that the initial conditions, $\li{i}{0}$ and $\ri{j}{0}$,  contain the information about the ensemble $\LDPC$ and the channel \cite{Luby01,Urbanke08-2}:
\begin{align}\LABEQ{initialcond}
\li{i}{0}&=\pe\lambi{i},\quad i\leq\lambimax,\\\LABEQ{initialcond2}
\ri{j}{0}&=\sum_{m\geq j}\rhi{j}\binom{m-1}{j-1}\pe^{j}(1-\pe)^{m-j}, j=1,\ldots,\rhimax.
\end{align}

It is important to remark that, as described in Section\SEC{Worlmald}, the solution of \EQ{System1}-\EQ{System2} only holds for all $t< t_{\D}$. In our scenario, the stopping time $ t_{\D}$ is given by either the time instant at which the decoder stops because there are no degree-one or degree-two check nodes or the time when $e(\tau)$ cancels, denoting that all variables in the graph have been decoded.

\begin{figure*}[]
\centering
\begin{tabular}{cc}
\includegraphics[width=7.5 cm]{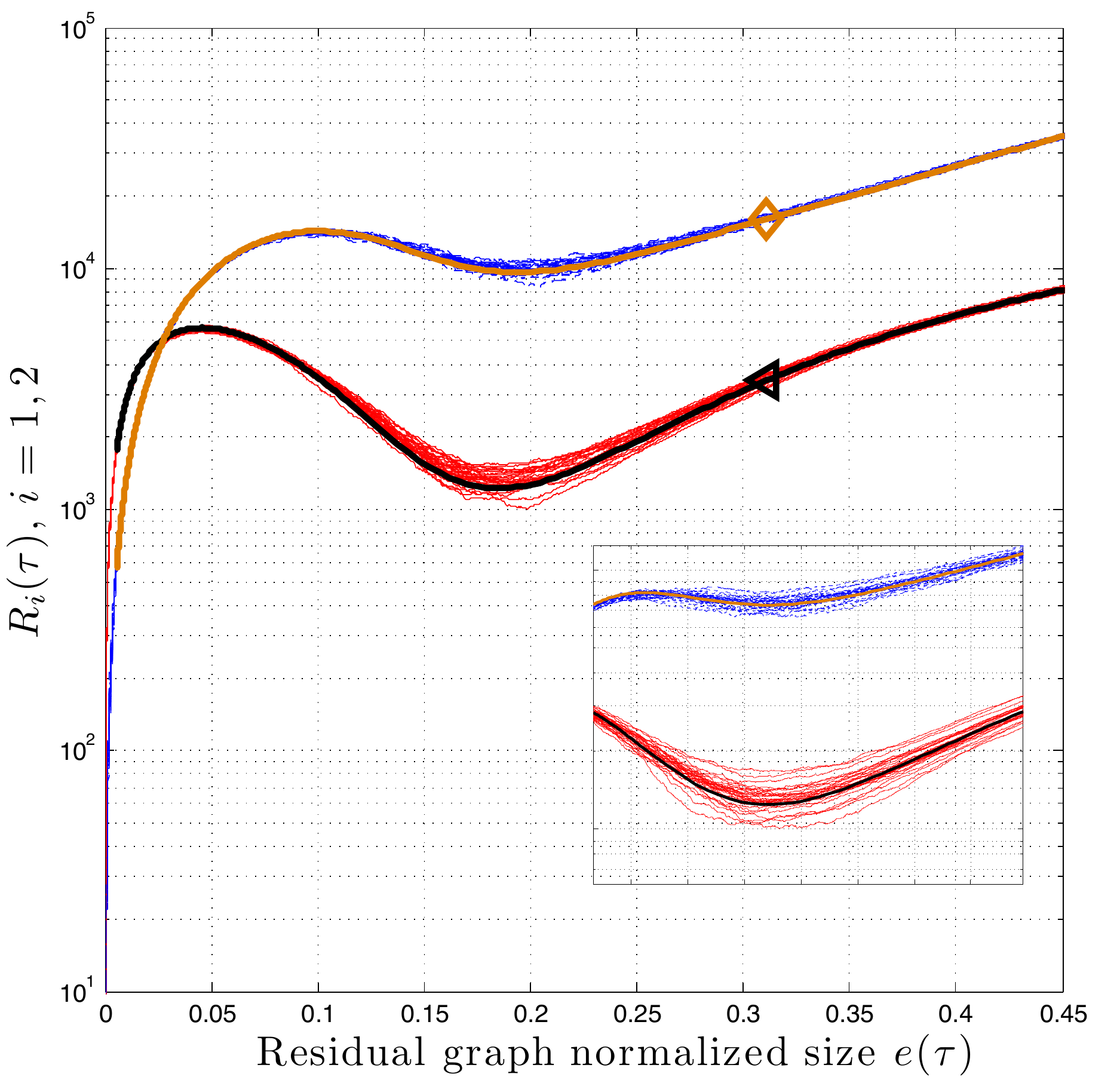} & \includegraphics[width=7.5 cm]{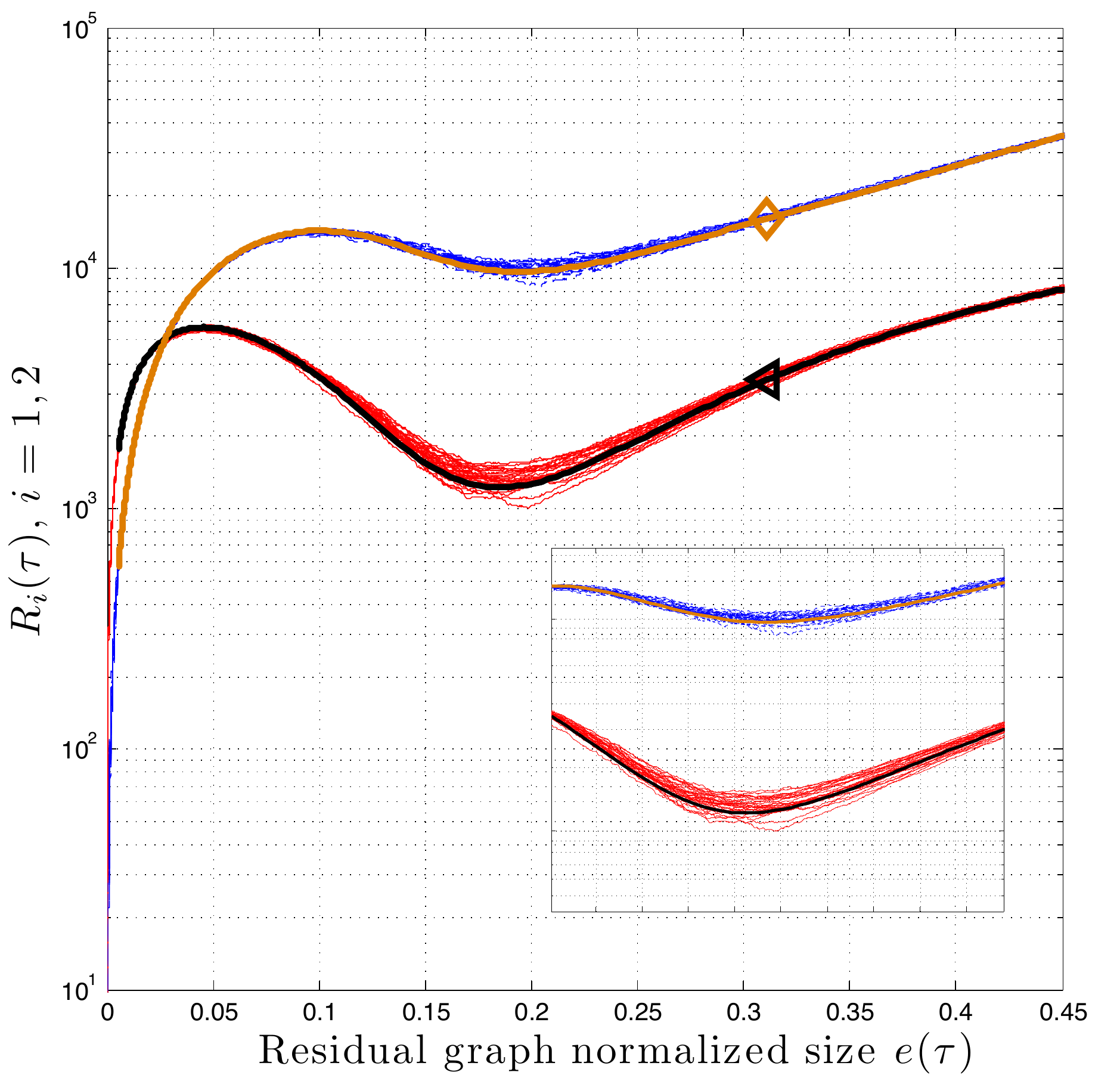} \\ \includegraphics[width=7.5 cm]{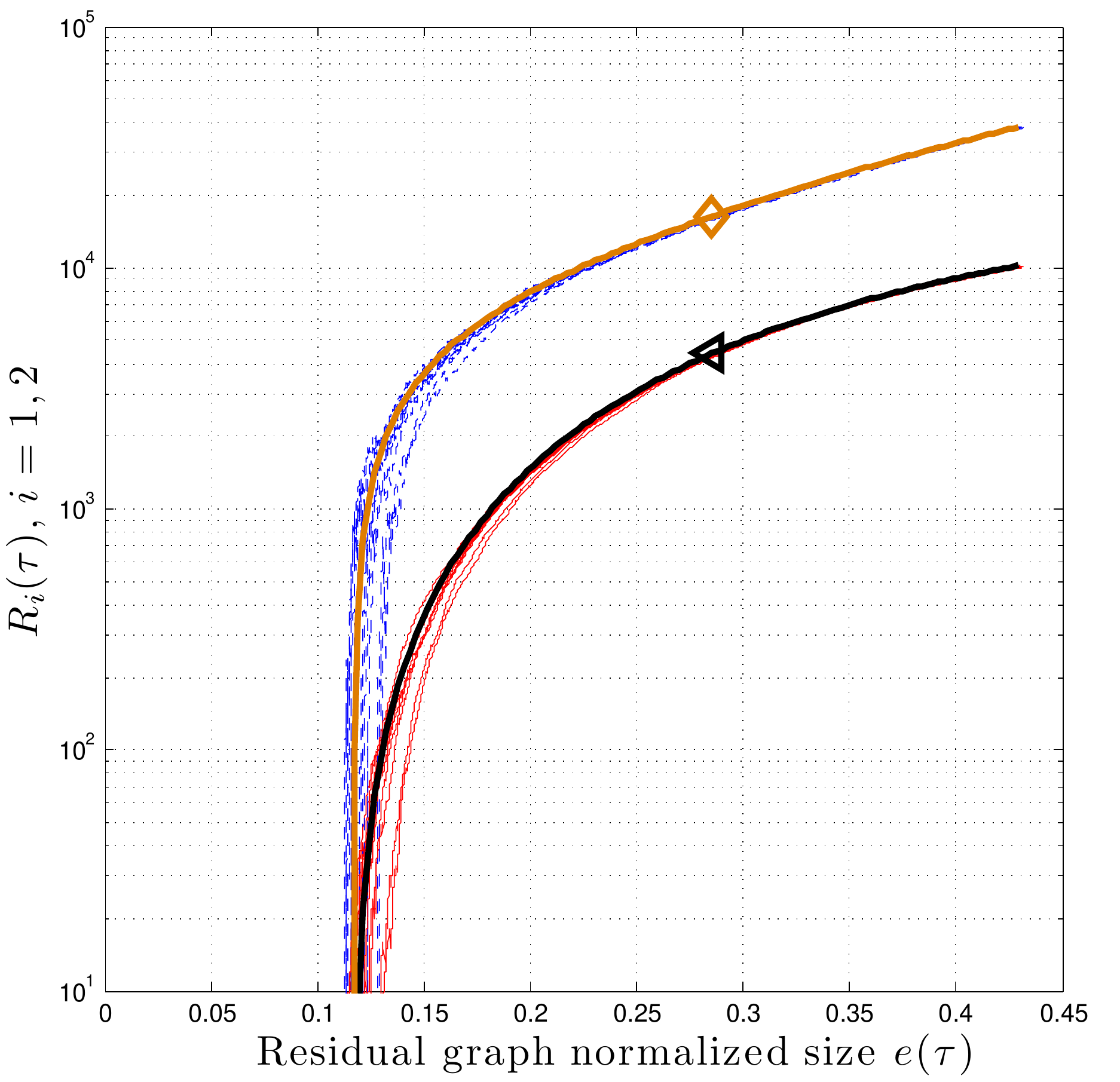}
& \includegraphics[width=7.5 cm]{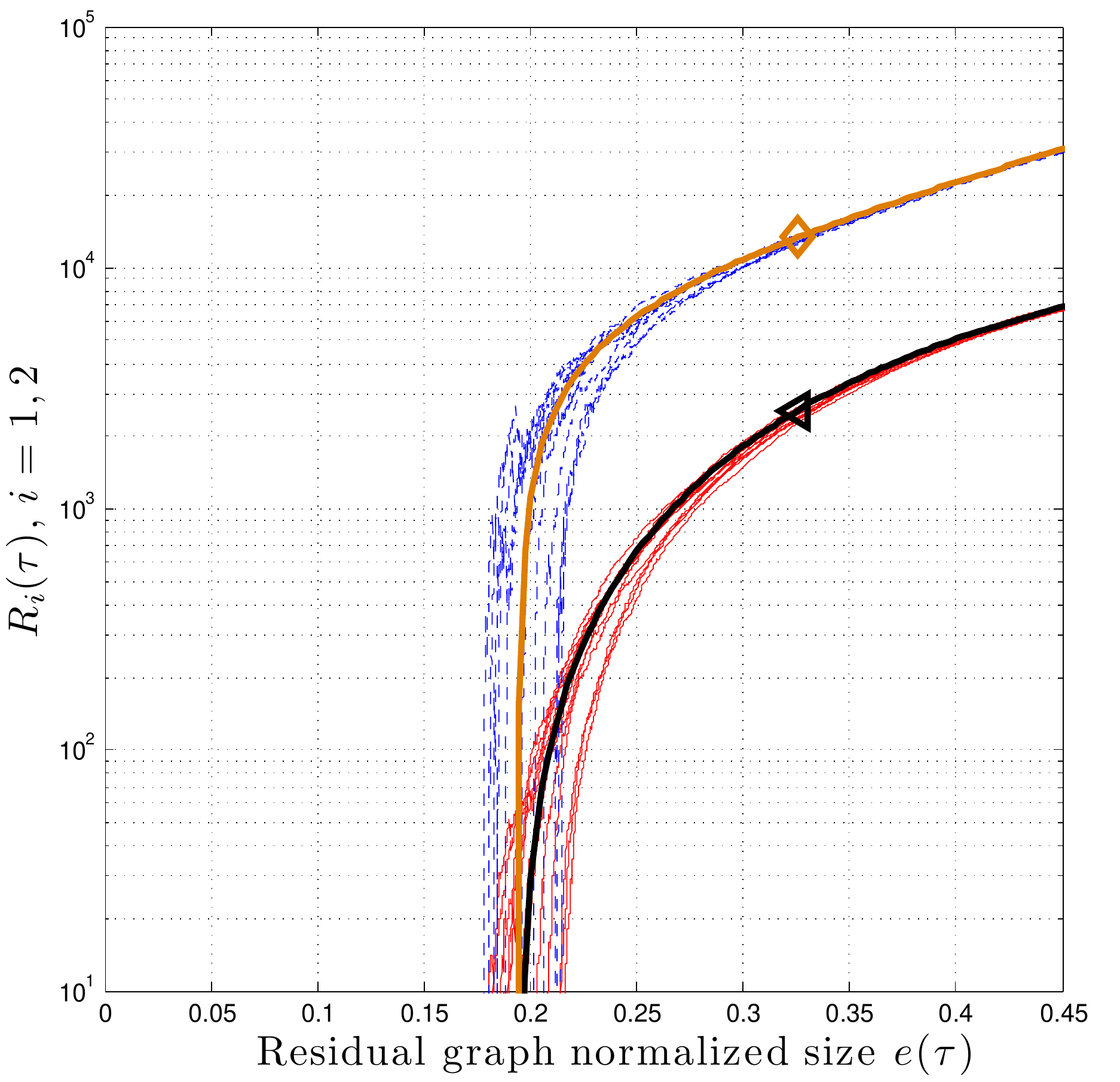}\\
(a) & (b)
\end{tabular}
\caption{In (a), we include the predicted evolution of $\Ri{1}{\tau}$ $(\lhd)$ and $\Ri{2}{\tau}$  $(\diamondsuit)$ for a regular $(3,6)$ code with $\n=2^{17}$ and an erasure probability $\pe=0.415<\peBP$ (top) and $\pe=0.43>\peBP$ (bottom), where $\peBP=0.4294$. In (b), we include the same result for the irregular DD defined in \EQ{IrrUrb} and \EQ{IrrUrb2} and an erasure probability $\pe=0.47<\peBP$ (top) and $\pe=0.483>\peBP$ (bottom), where $\peBP=0.4828$. We include in thin lines a set of $20$ individual decoding trajectories chosen at random.}
\LABFIG{AboveBP}
\end{figure*}

Let us illustrate the accuracy of this model to analyze the TEP decoder properties for a very large code length, $\n=2^{17}$. In Fig. \FIG{AboveBP}(a), we compare the solution of the system of differential equations in \EQ{System1} and \EQ{System2} for $R_{1}(\tau)=\ri{1}{\tau}\Edges$ and $R_{2}(\tau)=\ri{2}{\tau}\Edges$ for a regular $(3,6)$ code with 20 particular decoding trajectories obtained through simulation. We consider two cases: below ($\pe=0.415$) and above ($\pe=0.43$)  the BP threshold, i.e. $\peBP=0.4294$. We depict their evolution along the residual graph at each time, i.e. $e(\tau)$. We plot in thick lines the solution of our model for $R_{1}(\tau)$ $(\lhd)$ and $R_{2}(\tau)$ $(\diamond)$, and in thin lines the simulated trajectories,  $R_{1}(\tau)$ in solid and $R_{2}(\tau)$ in dashed line. In Fig. \FIG{AboveBP}(b), we reproduce the same experiment for the following irregular LDPC ensemble, 
\begin{align}\LABEQ{IrrUrb}
\lambda(x)&=\frac{1}{6}x+\frac{5}{6}x^3,\\\LABEQ{IrrUrb2}
\rho(x)&=x^{5},
\end{align}
where the BP threshold for this code is $\peBP=0.4828$. For this code, by running the urn model procedure described previously in Section\SEC{LAV}, we also find that $\lav{t}$ does not scale with $\n$.

In both cases, when we are below the BP threshold both the expected evolution curves and the empirical trajectories represent successful decoding since degree-one check nodes do not vanish until $e(\tau)$ tends to zero. Besides, note also that the longest deviation happens around $\tau\approx0.12$ and $\tau\approx0.18$, when the predicted curves for both  $R_{1}(\tau)$ and $R_{2}(\tau)$ have a relative minimum. This point is known as critical point and plays a fundamental role in the derivation of scaling laws to predict the performance in the finite-length regime \cite{Urbanke09,Urbanke08-2}, as explained in Section\SEC{FLregimen}. In \cite{Urbanke09,Amraoui05}, the authors show that the graph initial DD are Gaussian distributed around the mean in \EQ{initialcond} and \EQ{initialcond2}. Furthermore, they observe that as the PD performs, individual realizations are also Gaussian distributed around the mean computed in \cite{Luby01} using Wormald's theorem. And they show that the standard deviation is $\order(1/\sqrt{E})$, lower than the one warranted by Wormald's theorem. This result is a consequence of the channel properties.  In Fig. \FIG{AboveBP} we observe the same results for the TEP decoder.  

\section{TEP decoding of LDPC ensembles in the finite-length regime}\LABSEC{FLregimen}

\subsection{Motivation}

In \cite{Olmos10-2,Olmos11}, we empirically observed the gain in performance obtained when the TEP decoder is applied to decode some finite-length LDPC ensembles over the BEC. The gain of the proposed algorithm for practical codes can be analyzed from a different perspective based on the work by Polyanskiy \emph{et al.} \cite{Verdu10}. They present bounds on the maximum achievable coding rate for binary memoryless channels in the finite-length regime. These bounds can be regarded as the extension of the Shannon coding rate limit when the number of channel uses, i.e. code length, is fixed. For a given channel, a target probability of error $P_{\text{W}}$ and a desired code rate $\rate$, we can compute the minimum code length $\n_{\min}$ for which there exists a code that satisfies these requirements. 
\begin{figure}[h]
\centering
\includegraphics[width=9cm]{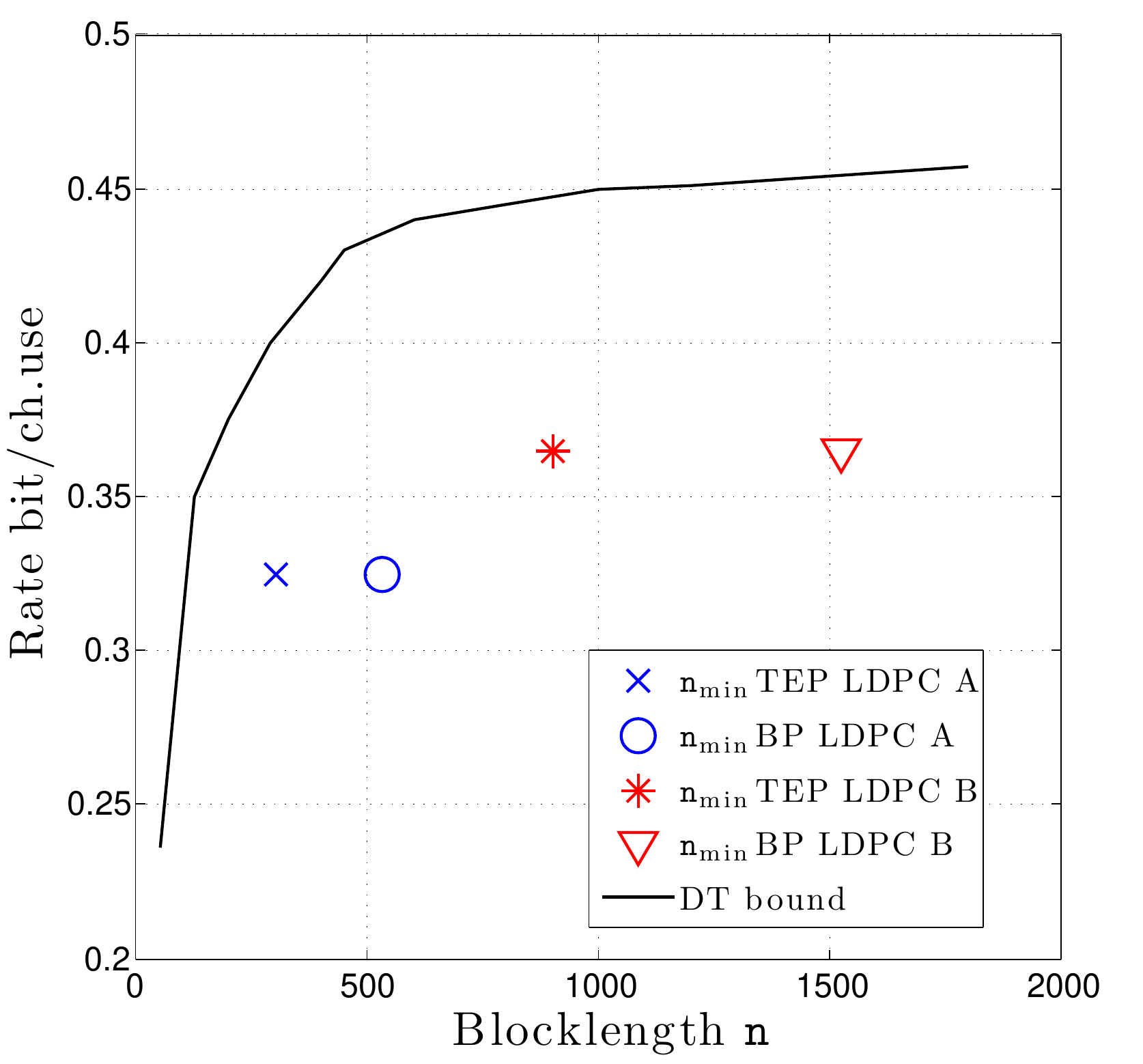}
\caption{Dependency-testing (DT) lower bound to the maximal achievable rate with supreme block error rate $P_{\text{W}}\leq10^{-3}$ over a BEC of parameter $\pe=0.5$ \cite{Verdu10}.  We include the minimum code length computed to obtain a valid performance for both the TEP and the BP and two LDPC ensembles $A$ and $B$ of rates $\rate_{A}=0.325$ and $\rate_{B}=0.365$.}\LABFIG{polyansky_TEP}
\end{figure}

In Fig. \FIG{polyansky_TEP}, we plot the dependency-testing (DT) lower bound in \cite{Verdu10} to the maximal achievable rate for a target block error rate of $P_{\text{W}}=10^{-3}$ over a BEC of parameter $\pe=0.5$. In \cite{Verdu10}, this lower bound is shown to be tight and much simpler to compute than the non-asymptotic maximum achievable rate. We considered two LDPC ensembles, referred to as code A and B. In both codes, $\lambda(x)=x^{2}$, while $\rho_B(x)=0.25x^{3}+0.75x^{4}$ and $\rho_A(x)=0.5x^{3}+0.5x^{4}$. The rate of each code is, respectively, $\rate_{A}=0.325$ and $\rate_{B}=0.365$. In Fig. \FIG{polyansky_TEP}, we depict the minimum code length for both the TEP and the BP decoders to empirically obtain a performance below $10^{-3}$ for each one of the two LDPC ensembles proposed. We observe that the use of the TEP decoder reduces by roughly half the block length to the optimum case given by the DT bound. Since the complexity of both algorithms is of the same order for these ensembles, the TEP decoder emerges as a powerful method to decode practical finite-length LDPC codes. 

In the light of these results, we focus on a theoretical description of the TEP gain with respect to BP for finite-length codes. We show that the TEP differential equations proposed in Section\SEC{TEPdiffeqs} are the key to measure and predict such gain. This result is used to extend to the TEP decoder the closed-form expressions proposed in \cite{Urbanke09,Amraoui05} to estimate the BP performance for some LDPC ensembles of regular or quasi-regular DD. They are referred to as BP scaling laws (SLs). For the TEP, we propose a simple SL, in which all parameters are analytically known as a function of the DD. We start by reviewing some important steps in the analysis of the BP performance for finite-length LDPC ensembles that are need for the \TEP finite-length analysis.

\subsection{BP decoder in the finite-length regime}\LABSEC{BPfinite}

In \cite{Urbanke09,Amraoui05}, the authors proved that the BP performance for finite-length LDPC codes can be predicted by analyzing the statistical presence of degree-one check nodes at a finite set of time instants during the whole decoding process. These points are referred to as \emph{critical points}.

\definition{BP-critical point of an LDPC ensemble. For a given LDPC ensemble with BP threshold $\peBP$, let $r_{1}^{\text{BP}}(\tau)$ be the expected evolution of the fraction of degree-one check nodes under BP decoding at $\pe$ in the limit $\n\rightarrow\infty$. We say that $\tau^{*}$ is a BP-critical point of the ensemble if 
\begin{align}\LABEQ{}
\lim_{\pe\rightarrow\peBP^{-}} r_{1}^{\text{BP}}(\tau^{*})=0.
\end{align}
}

In \cite{Luby01,Luby97}, the authors analytically compute  $r_{1}^{\text{BP}}(\tau)$ as a function of the LDPC degree distribution and the channel parameter $\pe$:
\begin{align}\LABEQ{R1PDLuby}
r_{1}^{\text{BP}}(u)=\epsilon\lambda(u)\bigg(u-1+\rho\left(1-\pe\lambda(u)\right)\bigg),
\end{align}
where $\frac{\partial u}{u}=\frac{-\partial \tau}{\eremain{\tau}}$.
The decoding process starts at $u=1$ and finishes at $u=0$. 
Let $X_{r_{1}}^{\text{BP}}(\tau,\n)$ be the random process that represents the evolution of the fraction of degree-one check nodes along the decoding process for a given ensemble of code length $\n$. Note that any realization $x_{r_{1}}^{\text{BP}}(\tau,\n)$ of such process represents a successful decoding as long as it is positive for any $\tau\in[0,\n/\Edges)$.  The process $X_{r_{1}}^{\text{BP}}(\tau,\n)$ presents some important properties \cite{Urbanke09, Amraoui05}:
\begin{enumerate}
\item $\E[X_{r_{1}}^{\text{BP}}(\tau,\n)]$ closely follows $r_{1}^{\text{BP}}(\tau)$ in \EQ{R1PDLuby}. For moderately-sized codes the mean of the process is essentially independent of $\n$.
\item The variance is of order $\order(\n^{-1})$. We denote it as $\delta^{\text{BP}}_{r_{1},r_{1}}(\tau^{})/\n$.
\item For any $\tau$, the distribution of $X_{r_{1}}^{\text{BP}}(\tau,\n)$ tends with $\n$ to a (truncated) Gaussian pdf.
\end{enumerate}

 \begin{figure}[h]
\centering
\begin{tabular}{c}
\includegraphics[width=7cm]{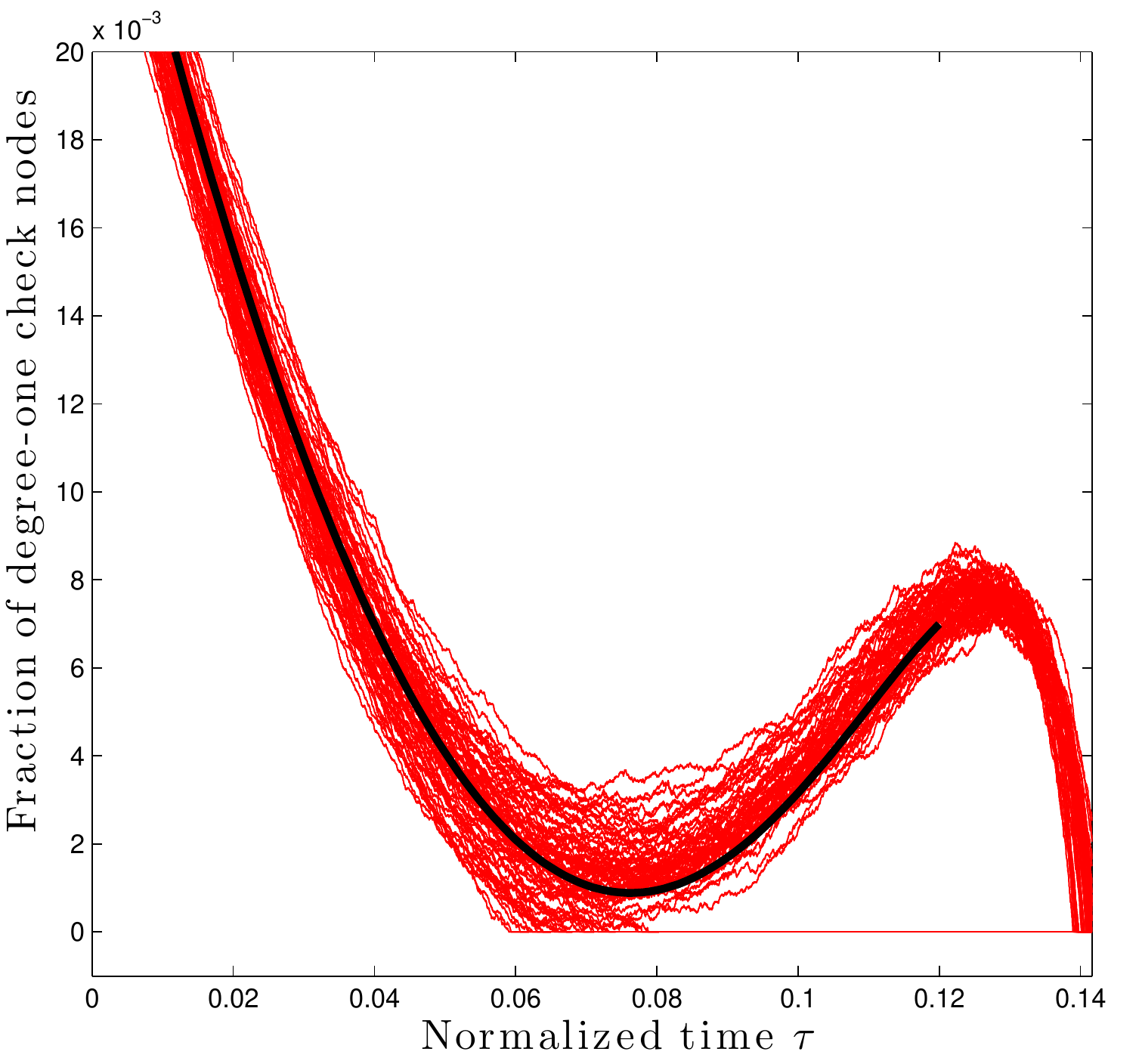}
\end{tabular}
\caption{In (a) we show a set of 100 realizations of $X_{r_{1}}^{\text{BP}}(\tau,\n)$ for $\n=2^{15}$ and $\epsilon=0.425$ along with the DE mean prediction in \EQ{R1PDLuby}, thick line. 
}\LABFIG{FigSCA1}
\end{figure}

Let us focus for simplicity on LDPC ensembles with a single critical point at $\tau^*$\footnote{This includes any regular ensemble and typically codes with small degree of irregularity, for instance the code defined in \EQ{IrrUrb} and \EQ{IrrUrb2}.}. In  \cite{Urbanke09, Amraoui05}, it is shown that the BP decoder at $\pe=\peBP+\Delta\pe$ successes with very high probability as long as $X_{r_{1}}^{\text{BP}}(\tau^*,\n)>0$, i.e., if there exists a positive fraction of degree-one check nodes at the BP-critical point. In Fig. \FIG{FigSCA1}, we include a random set of 100 realizations of $X_{r_{1}}^{\text{BP}}(\tau,\n)$ for the $(3,6)$ LDPC ensemble, $\n=2^{15}$ and $\epsilon=0.425$ along with the Density Evolution (DE) mean prediction in \EQ{R1PDLuby}, in thick line. The ensemble has a single critical point at $\tau^*=0.07$. Around this point, we can see that the DE curve reaches a relative minimum and the global error probability is clearly dominated by the failure probability at this point.  Thus, the finite-length performance can be roughly estimated by just evaluating the cumulative density function (cdf) of $X_{r_{1}}^{\text{BP}}(\tau,\n)$ at the unique critical point $\tau^{*}$:
\begin{align}\LABEQ{FINLENG}
1-\E_{(\lambda(x),\rho(x),\n)}\left[P ^{\text{BP}}_{\text{W}}(\code,\pe)\right]\leq p\left(X_{r_{1}}^{\text{BP}}(\tau^{*},\n)\geq0\right),
\end{align}
where $P ^{\text{BP}}_{\text{W}}(\code,\pe)$ is the average block error probability for the code $\code\in\LDPC$ under BP decoding. To evaluate \EQ{FINLENG} assuming  $X_{r_{1}}^{\text{BP}}(\tau^*,\n)$ is Gaussianly distributed, we need its mean and variance. The mean can be computed for each $\pe$ from \EQ{R1PDLuby}, but in \cite{Urbanke09} the authors show that the first-order Taylor expansion around the BP-critical point, i.e: 
\begin{align}\LABEQ{meanGauss}
r_{1}^{\text{BP}}(\tau^*)=\frac{\partial r_{1}^{\text{BP}}(\tau)}{\partial \pe}\bigg|_{\substack{\tau=\tau^{*}\\\;\pe=\peBP}}\Delta\pe,
\end{align}
is as precise and more convenient, as it allows computing a SL that measures the performance as a function of the distance to the threshold. 

Closed-form expression for the variance, $\delta^{\text{BP}}_{r_{1},r_{1}}(\tau^{*})$, for any LDPC ensemble can be found in \cite{Takayuki10} and, for large enough $\n$ at the critical point, we can ignore the dependency of $\delta^{\text{BP}}_{r_{1},r_{1}}(\tau^{*})$ with respect to $\epsilon$ \cite{Urbanke09}.The error probability can be approximated by \cite{Urbanke09}:
\begin{align}
\E_{(\lambda(x),\rho(x),\n)}\left[P ^{\text{BP}}_{\text{W}}(\code,\pe)\right]\geq\mathcal{Q}\left(\frac{\sqrt{\n}(\peBP-\pe)}{\alpha_{\text{BP}}}\right),\LABEQ{BASICSLBP}
\end{align}
where 
\begin{align}\LABEQ{alphaBP}
\alpha_{\text{BP}}=\frac{\sqrt{\delta^{\text{BP}}_{r_{1},r_{1}}(\tau^*)}}{\left|\frac{\partial r_{1}^{\text{BP}}(\tau)}{\partial \pe}\bigg|_{\substack{\tau=\tau^{*}\\\;\pe=\peBP}}\right|}.
\end{align}
For sufficiently large code lengths, this scaling function provides an accurate estimation of the BP error probability. A comparison between \EQ{BASICSLBP} (dashed lines) and empirical BP performance curves (solid lines) can be found, respectively, in Fig. \FIG{SLTEPFIGS}(a) and (b) for the regular $(3,6)$ LDPC ensemble and for the irregular LDPC ensemble in \EQ{IrrUrb} and \EQ{IrrUrb2}. 

\subsection{TEP decoding in the finite-length regime}\LABSEC{TEPFINIT}

For describing the TEP decoder finite-length performance, we follow a similar approach and study the random process $X_{r_{1}}^{\text{TEP}}(\tau,\n)$, which represents the evolution of the fraction of edges with right degree $1$ along the TEP decoding process for a given ensemble of code length $\n$, around its local minima\footnote{We could have also studied the evolution of
\begin{align}
X_{r_{1,2}}^{\text{TEP}}(\tau,\n)\doteq X_{r_{1}}^{\text{TEP}}(\tau,\n)+X_{r_{2}}^{\text{TEP}}(\tau,\n),
\end{align}
where $X_{r_{j}}^{\text{TEP}}(\tau,\n)$ represents the evolution of the fraction of edges with right degree $j$ along the decoding process for a given ensemble of code length $\n$. But the processing of degree two does not decode any additional variable unless degree-one check nodes are create and hence we only focus on the evolution of the random process $X_{r_{1}}^{\text{TEP}}(\tau,\n)$ that tell us how many additional variables we can decode.}. As for the BP, the analysis centers on the points during the decoding processes for which the presence of degree-one check nodes vanishes. As for the BP analysis, we define similarly the critical points for the TEP decoder and we also prove that these critical points are identical for both decoders.

\definition{TEP-critical point of an LDPC ensemble. For a given LDPC ensemble with TEP threshold $\peTEP=\peBP$, let $r_{1}^{\text{TEP}}(\tau)$ be the expected evolution of the fraction of degree-one check nodes under TEP decoding at $\pe$ in the limit $\n\rightarrow\infty$. We say that $\tau'$ is a TEP-critical point of the LDPC ensemble if 
\begin{align}\LABEQ{}
\lim_{\pe\rightarrow\peTEP^{-}}r_{1}^{\text{TEP}}(\tau')=0.
\end{align}
}

\lemma{For a given ensemble LDPC$[\lambda(x),\rho(x)]$, the number of TEP critical points is equal to the number of BP critical points.}\LABLEM{criticals}
\begin{proof}
In the regime $\n\rightarrow\infty$, we have proven that, with very high probability, the TEP decoder is not able to create any additional check nodes of degree one with respect to the  BP solution and, as a consequence, $\peTEP=\peBP$. Besides, given the TEP independence of the processing order proved in Appendix\SEC{A1}, we can implement the TEP decoder by first removing all the degree-one check nodes, i.e. a BP stage, and then process degree-two check nodes if the BP does not succeed. Hence, for $\pe\rightarrow\peBP^{-}$ and $\n\rightarrow\infty$, $r_{1}^{\text{BP}}(\tau)$ and $r_{1}^{\text{TEP}}(\tau)$ match during the whole decoding process and, therefore, the ensemble presents the same number of critical points for both decoding algorithms. 
\end{proof}

The process $X_{r_{1}}^{\text{TEP}}(\tau,\n)$ behaves as $X_{r_{1}}^{\text{BP}}(\tau,\n)$ and present the same statistical properties, namely:
\begin{enumerate}
\item $\E[X_{r_{1}}^{\text{TEP}}(\tau,\n)]$ closely follows $r_{1}^{\text{TEP}}(\tau)$, which is computed as solution of the TEP decoder differential equations in \EQ{System1}-\EQ{System3}. 
\item The variance decays as $\order(\n^{-1})$. We denote it as  $\delta^{\text{TEP}}_{r_{1},r_{1}}(\tau^{})/\n$. 
\item For any $\tau$, the distribution of $X_{r_{1}}^{\text{TEP}}(\tau,\n)$ tends with $\n$ to a (truncated) Gaussian pdf.
\end{enumerate}

\begin{figure}[]
\centering
\begin{tabular}{c}
\includegraphics[width=9cm,height=8.5cm]{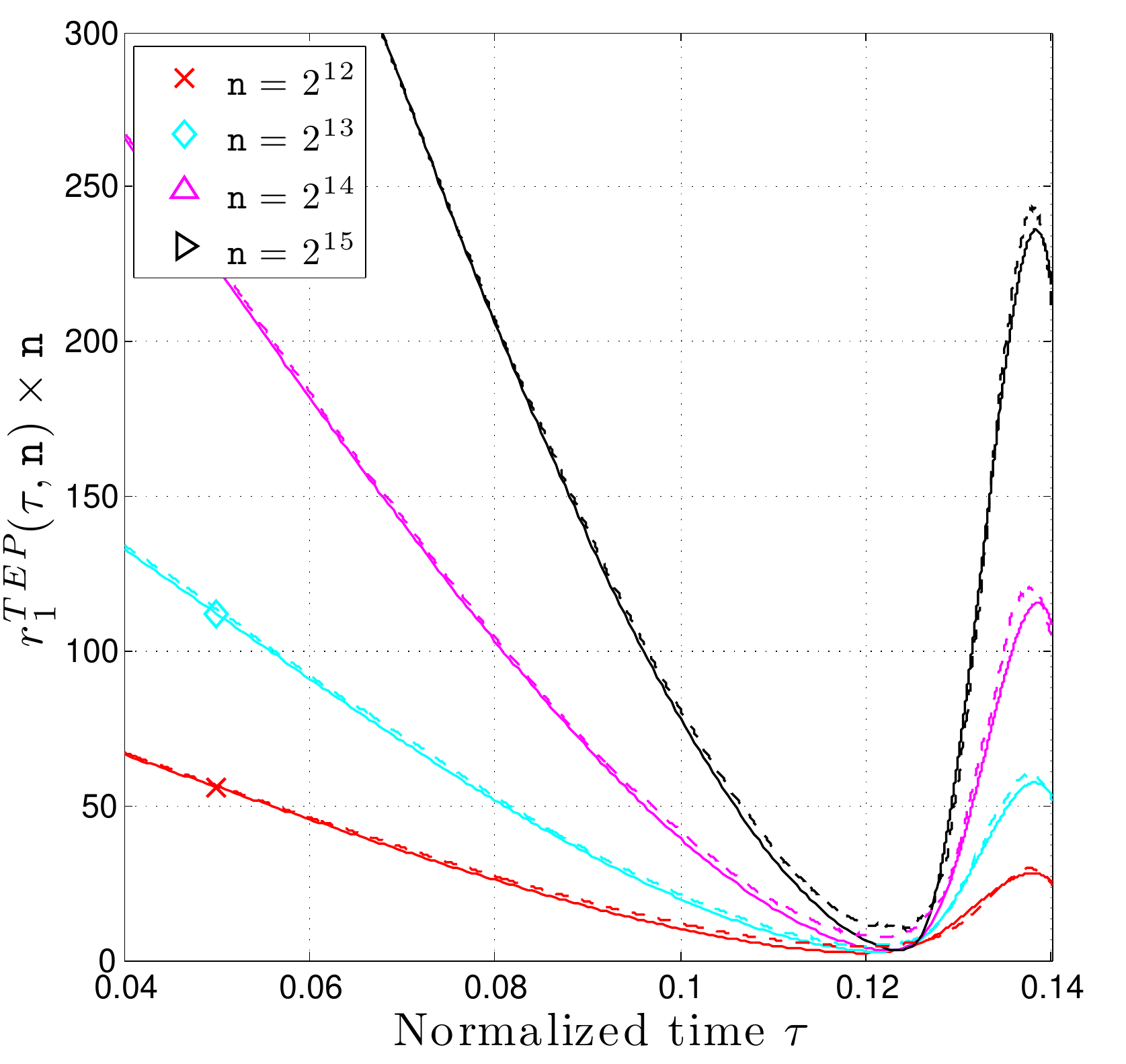} 
\end{tabular}\caption{ For the $(3,6)$ ensemble, we plot $\ri{1}{\tau',\n,\peTEP}\times \n$ (solid lines) computed from \EQ{System1}-\EQ{System3} for different code lengths, where $\peTEP=0.4294$. We include in dashed lines sample average curves obtained by simulation. We have averaged them over 500 realizations.}\LABFIG{AP6FIG1}
\end{figure} 

As discussed before, for codes with a single critical point $\tau'$, we estimate the TEP finite-length performance by computing the cdf of the process $X_{r_{1}}^{\text{TEP}}(\tau,\n)$ at $\tau'$:
\begin{align}\LABEQ{FINLENGTEP}
1-\E_{(\lambda(x),\rho(x),\n)}\left[P ^{\text{TEP}}_{\text{W}}(\code,\pe)\right]\approx p\left(X_{r_{1}}^{\text{TEP}}(\tau',\n)\geq0\right)
\end{align}
where $P ^{\text{TEP}}_{\text{W}}(\code,\pe)$ is the average block error probability for the code $\code\in\LDPC$ under TEP decoding. To evaluate \EQ{FINLENGTEP} assuming  $X_{r_{1}}^{\text{TEP}}(\tau',\n)$ is Gaussianly distributed, we need its mean and variance, which can be computed analogously to the BP case. The mean can be computed using the first-order Taylor expansion around the TEP-critical point, i.e.:
\begin{align}\LABEQ{meanTEP0}
r_{1}^{\text{TEP}}(\tau',\n,\pe)&=r_{1}^{\text{TEP}}(\tau',\n,\peTEP)+\frac{\partial r_{1}^{}(\tau,\n,\pe)}{\partial \pe}\bigg|_{\substack{\tau=\tau'\\\;\pe=\peTEP}}\Delta\pe
\end{align}

In this case, there exists a nonzero value for the zero-order Taylor expansion that indeed explains the improvement of the TEP decoder over the BP for finite-length codes. For the BP decoder, the mean process evolution in \EQ{R1PDLuby}  is independent of $\n$. Therefore, at $\peBP$ and the corresponding critical point, $r_{1}^{\text{BP}}(\tau^*,\n,\peBP)=0$\footnote{The BP gets stuck because it has run out of check nodes of degree one.} and hence the expression in \EQ{meanGauss}. For the TEP decoder, the mean evolution depends on the code length $\n$ and the mean at the critical point is nonzero. We can understand why $r_{1}^{\text{TEP}}(\tau',\n,\peTEP)$ is nonzero by assuming that we first process all degree-one check nodes, and then all degree-two check nodes. After we process all the degree-one check nodes for $\epsilon=\peBP$ and, at $\tau=\tau^*$ and large $\n$, we should get stuck at the BP critical point with high probability and hence $r_{1}^{\text{BP}}(\tau^*,\n,\peBP)\approx0$. Now, when we process all the degree-two check nodes, we move from $\tau^*$ to $\tau'$ and we can generate degree-one check nodes for finite length codes, because $\pA{\tau}$ in \EQ{pA} is nonzero. This value is captured by a nonzero $r_{1}^{\text{TEP}}(\tau',\n,\peTEP)$ in \EQ{meanTEP0}. The evaluation of
$r_{1}^{\text{TEP}}(\tau',\n,\peTEP)$ for different code lengths using the TEP differential equations  shows that it decays as $1/\n$, namely:
\begin{align}\LABEQ{r1avg}
r_{1}^{\text{TEP}}(\tau',\n,\peTEP)=\gamma_{\text{TEP}}\n^{-1},
\end{align}
because the probability that two variables nodes share the same two check nodes decays as $\n^2$ and we process a fraction of $\n$ of degree two check nodes after we have process all degree-one check nodes. For instance, in Fig. \FIG{AP6FIG1} we plot $\ri{1}{\tau',\n,\peTEP}\times \n$ (solid lines) computed from \EQ{System1}-\EQ{System3} for different code lengths, where $\peTEP=0.4294$. We include in dashed lines sample average curves obtained by simulation. Note that all curves converge at the critical point. The exact computation of $\gamma_{\text{TEP}}$ is obtained by solving \EQ{System1}-\EQ{System3} and it is shown in Appendix \SEC{A6} how it can be computed for each DD. For instance, the $(3,6)$ LDPC ensemble has a $\gamma_{\text{TEP}}^{-1}=0.3194$ and the irregular LDPC ensemble in \EQ{IrrUrb} and \EQ{IrrUrb2} presents a $\gamma_{\text{TEP}}^{-1}=0.2925$.

\begin{figure}[h]
\centering
\begin{tabular}{c}
\includegraphics[width=8cm]{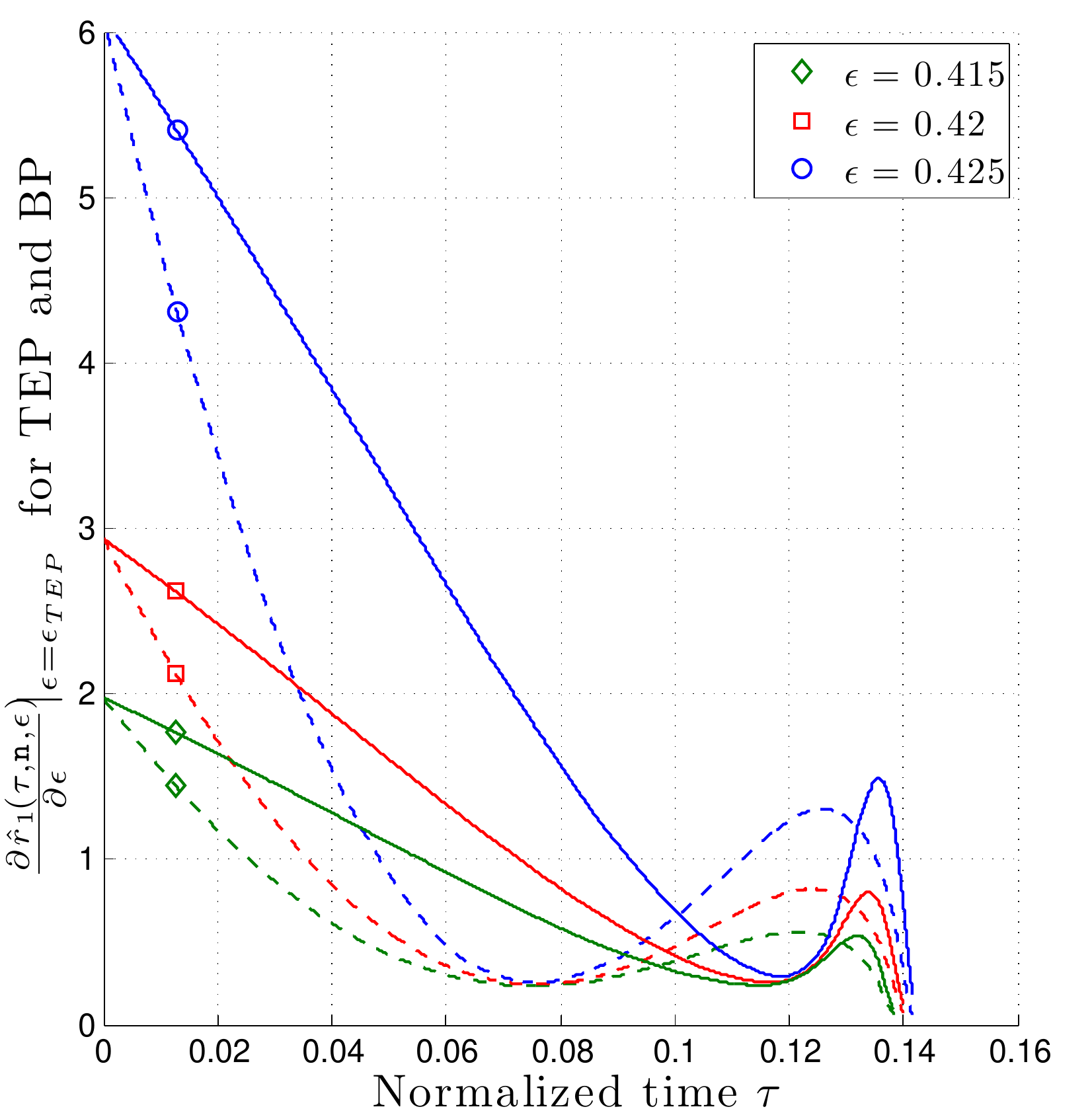}
\end{tabular}
\caption{We include the TEP sample mean evolution $(r_{1}^{\text{TEP}}(\tau)-\gamma_{\text{TEP}}\n^{-1})/\Delta\pe$ (in solid lines) computed using a set of 500 realizations of  $X_{r_{1}}^{\text{BP}}(\tau,\n)$ for the $(3,6)$ ensemble with $\n=2^{15}$ and $\epsilon=0.415$ $(\diamond)$, $\epsilon=0.42$ $(\square)$ and $\epsilon=0.425$ $(\circ)$. We also include the BP sample mean evolution in dashed lines for comparison.}\LABFIG{FigSCAepsilon}
\end{figure}






\begin{figure*}[htbp]
\centering
\begin{tabular}{cc}
\includegraphics[width=9cm,height=9cm]{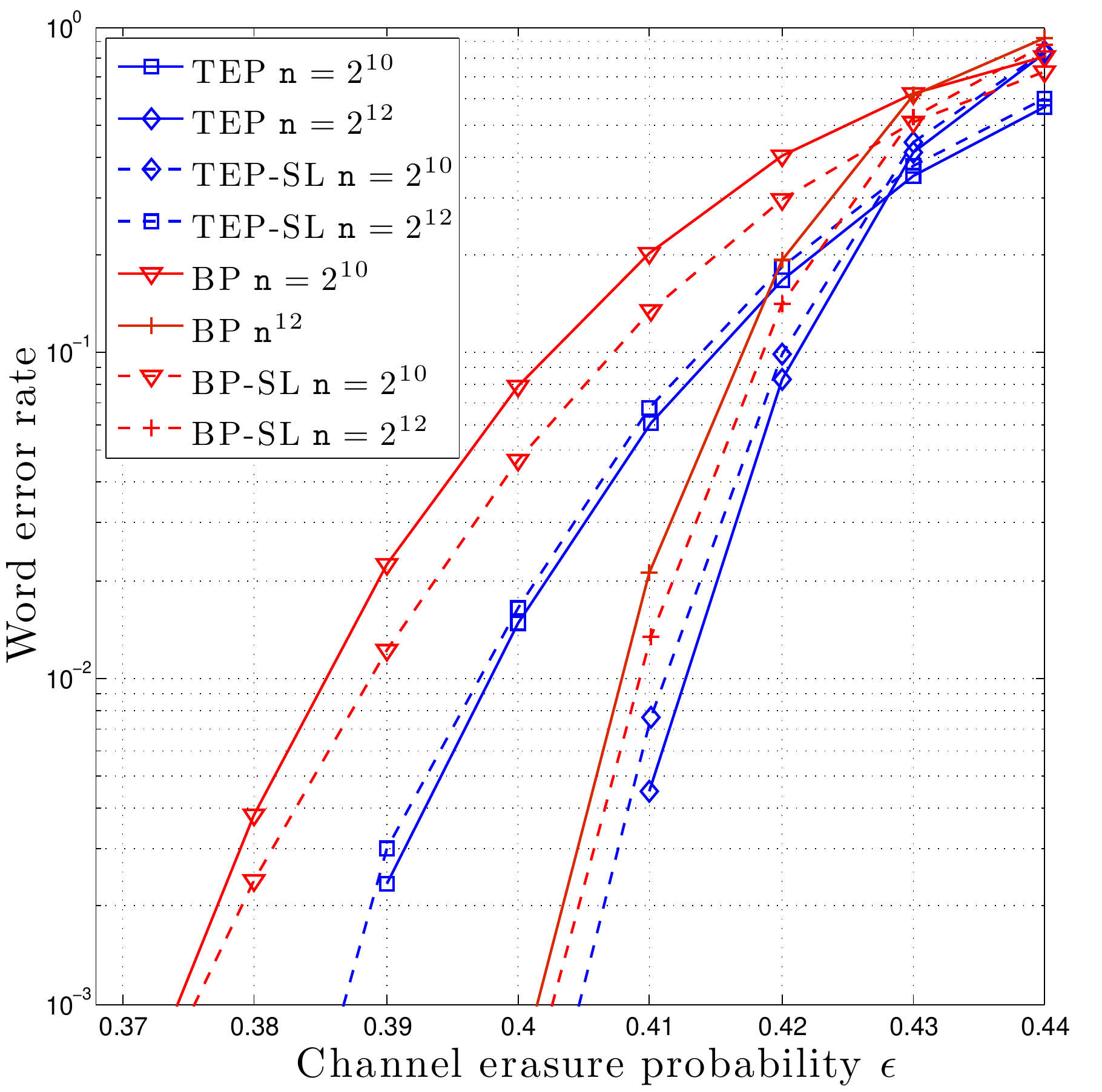} & \includegraphics[width=9cm,height=9cm]{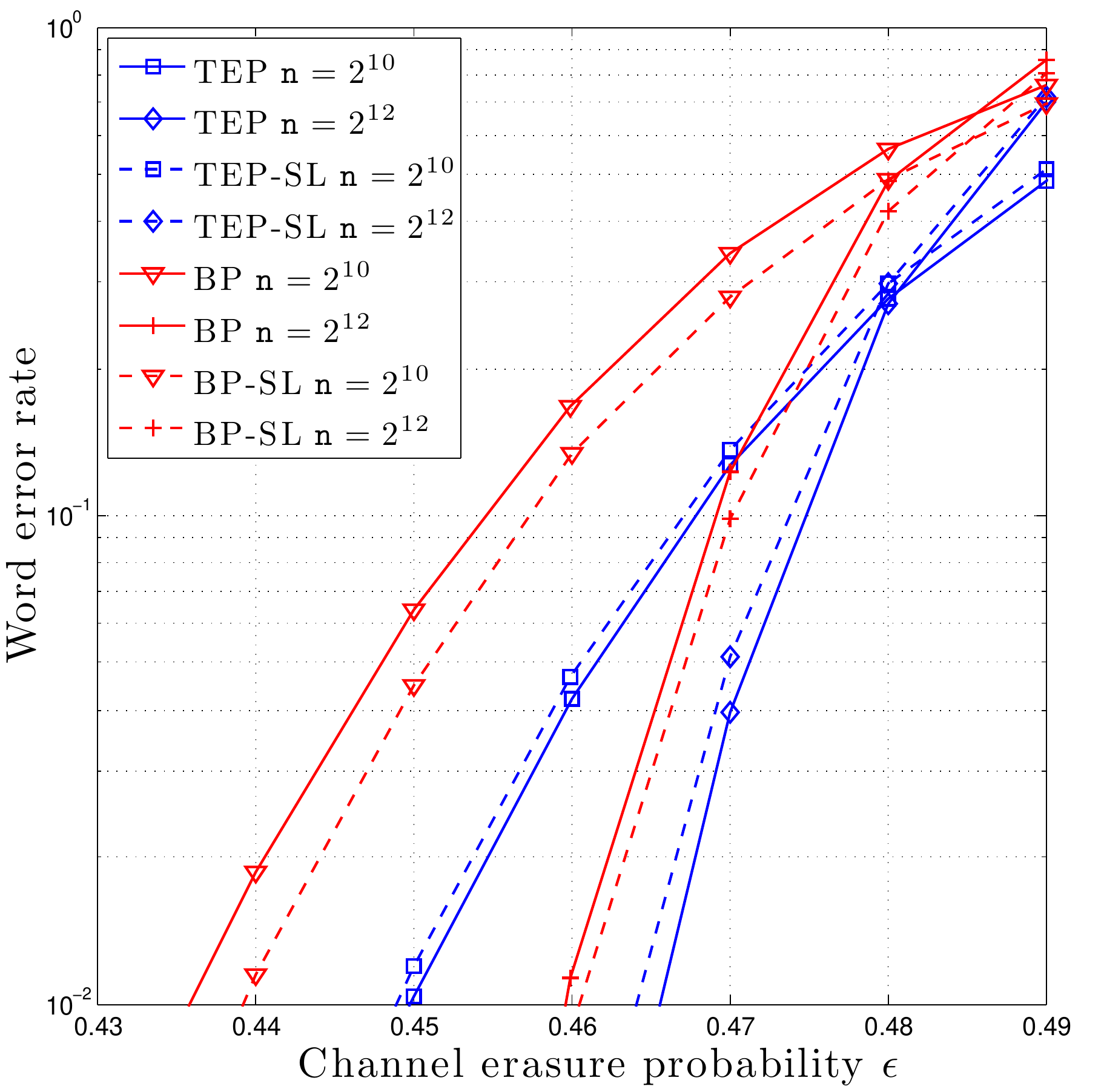}\\
(a) & (b)
\end{tabular}
\caption{In (a), we include the BP/TEP performance for the regular $(3,6)$ ensemble (solid lines) along with their respective SL approximations in \EQ{BASICSLBP}/\EQ{BASICSLTEP} (dashed lines) for $\n=2^{10}$ and $(+)$. The SL parameters are $\alpha_{\text{BP}}=\alpha_{\text{TEP}}\approx0.5603$, $\delta^{\text{TEP}}_{r_{1},r_{1}}(\tau^{*})\approx0.0526$ and $\gamma_{\text{TEP}}\approx0.3194$. In (b), we reproduce the same figure for the irregular LDPC code defined in \EQ{IrrUrb} and \EQ{IrrUrb2}. The SL parameter is $\alpha_{\text{BP}}=\alpha_{\text{TEP}}\approx0.5791$, $\delta^{\text{TEP}}_{r_{1},r_{1}}(\tau^{*})\approx0.0593$ and $\gamma_{\text{TEP}}\approx0.2925$.}
\LABFIG{SLTEPFIGS}
\end{figure*}

By \EQ{FINLENGTEP}, we estimate the TEP performance as follows:
\begin{align}\LABEQ{BASICSLTEP}
&\E_{(\lambda(x),\rho(x),\n)}\left[P ^{\text{TEP}}_{\text{W}}(\code,\pe)\right]\nonumber\\
&\approx1-\mathcal{Q}\left(-\left(\frac{\frac{\partial r_{1}^{\text{TEP}}(\tau,\n,\pe)}{\partial \pe}\bigg|_{\substack{\tau=\tau'\\\;\pe=\peTEP}}\Delta\pe+\gamma_{\text{TEP}}\n^{-1}}{\sqrt{\delta^{\text{TEP}}_{r_{1},r_{1}}(\tau')/\n}}\right)\right)\nonumber\\&=\mathcal{Q}\left(\frac{\sqrt{\n}(\peBP-\pe)}{\alpha_{\text{BP}}}+\frac{\gamma_{\text{TEP}}}{\sqrt{\n\;\;\delta^{\text{BP}}_{r_{1},r_{1}}(\tau^*)}}\right),
\end{align}
where $P ^{\text{TEP}}_{\text{W}}(\code,\pe)$ is the TEP block error rate for the code $\code\in\LDPC$ and we have used the values computed in \cite{Urbanke09, Takayuki10, Ezri07, Ezri08} for the BP (namely $\alpha_\text{BP}$ and $\delta^{\text{BP}}_{r_{1},r_{1}}(\tau^*)$) for the TEP performance. These values measure the mean and variance of the degree-one check nodes and mainly depend on the DD ensemble and the channel dispersion \cite{Verdu10}. We could expect the value of $\frac{\partial r_{1}^{}(\tau,\n,\pe)}{\partial \pe}\big|_{\substack{\tau=\tau'\\\;\pe=\peTEP}}$ and $\delta^{\text{TEP}}_{r_{1},r_{1}}(\tau')$ to slightly grow with respect to the BP estimates, because we are additionally processing degree-two check nodes. Empirically, we observe that $\frac{\partial r_{1}^{}(\tau,\n,\pe)}{\partial \pe}\big|_{\substack{\tau=\tau'\\\;\pe=\peTEP}}$ is slightly larger than $\frac{\partial r_{1}^{}(\tau)}{\partial \pe}\big|_{\substack{\tau=\tau^*\\\;\pe=\peBP}}$, as shown in Fig. \FIG{FigSCAepsilon}, and we observe an insignificant reduction in the variance.

In Fig. \FIG{SLTEPFIGS}, we compare the SL solution in \EQ{BASICSLTEP} in dashed lines with real performance data obtained through simulation in solid lines for the regular $(3,6)$ LDPC ensemble (a) and for the the irregular ensembel defined in \EQ{IrrUrb} and \EQ{IrrUrb2} (b). 
The match between dashed and solid lines is as good as for the BP estimates as for the TEP, showing the accuracy of the model for the TEP performance and the proposed parameter estimation. 
In both plots, due to the parameter overestimation when assuming the BP values for $\delta^{\text{TEP}}_{r_{1},r_{1}}(\tau')/\n$ and $\alpha_{\text{TEP}}$, we slightly overestimate the TEP error probability. The overestimation is also augmented because  in our estimate of $r_1(\tau,\n,\pe)$ under TEP decoding we do not take into account scenarios where two variables that share a degree-two check node, also share more than one additional check node. For the code lengths considered, these scenarios can be eventually found and the mean can be slightly higher. In addition, there exists $X_{r_{1}}^{\text{TEP}}(\tau,\n)$ trajectories that go to zero at some point but eventually can be positive again by removing degree-two check nodes. 


\section{Conclusions}\LABSEC{Conclusions}

In this paper, we present the expectation propagation algorithm to address the LDPC decoding process over any DMC. The posterior distribution of the variables is approximated with a Markov-tree probability distribution, over which the marginal estimate can be efficiently performed. By construction, the solution  improves the BP estimate. For the binary erasure channel, we show that the tree-EP algorithm reduces to a peeling-type algorithm, i.e. the TEP decoder, that outperforms the BP solution by additionally processing degree-two check nodes and not only degree-one check nodes. The TEP decoder improvement in performance is significant for practical finite-length LDPC codes. We prove that the complexity of this additional step is bounded even in the limiting case and therefore, the TEP complexity is of the same order than the BP algorithm. In the asymptotic regime, we have shows the conditions to be fulfilled by an LDPC ensemble to improve the BP threshold. The TEP decoder differential equations provide the graph mean evolution under TEP decoding. Using Wormald's theorem, we proved that the dispersion around the mean evolution is bounded. This results  are used to explain and predict the TEP decoder improvement. Along with empirical support, we have developed a scaling law to predict the performance. The analysis of the BEC case  developed in detail in this paper for both the asymptotic and finite-length regime is a guideline to construct efficient implementations of the Tree-EP algorithm to outperform the BP solution in other channels like the binary symmetry channel and the binary additive Gaussian noise channel. 

\bibliography{LDPC,LDPC_Conv,digwircom,Gmodels,gp,macler,PredisNLEq,UMTS,Others}

\begin{thebibliography}{10}
\providecommand{\url}[1]{#1}
\csname url@samestyle\endcsname
\providecommand{\newblock}{\relax}
\providecommand{\bibinfo}[2]{#2}
\providecommand{\BIBentrySTDinterwordspacing}{\spaceskip=0pt\relax}
\providecommand{\BIBentryALTinterwordstretchfactor}{4}
\providecommand{\BIBentryALTinterwordspacing}{\spaceskip=\fontdimen2\font plus
\BIBentryALTinterwordstretchfactor\fontdimen3\font minus
  \fontdimen4\font\relax}
\providecommand{\BIBforeignlanguage}[2]{{%
\expandafter\ifx\csname l@#1\endcsname\relax
\typeout{** WARNING: IEEEtran.bst: No hyphenation pattern has been}%
\typeout{** loaded for the language `#1'. Using the pattern for}%
\typeout{** the default language instead.}%
\else
\language=\csname l@#1\endcsname
\fi
#2}}
\providecommand{\BIBdecl}{\relax}
\BIBdecl

\bibitem{Gallager63}
R.~G. Gallager, \emph{Low Density Parity Check Codes}.\hskip 1em plus 0.5em
  minus 0.4em\relax MIT Press, 1963.

\bibitem{pearl88}
J.~Pearl, \emph{{Probabilistic reasoning in intelligent systems: networks of
  plausible Inference}}.\hskip 1em plus 0.5em minus 0.4em\relax Morgan
  Kaufmann, 1988.

\bibitem{Mackay96}
D.~J.~C. MacKay and R.~M. Neal, ``Near {S}hannon limit performance of low
  density parity check codes,'' \emph{Electronics Letters}, vol.~32, pp.
  1645--1646, 1996.

\bibitem{Mackay99}
D.~J.~C. MacKay, ``Good error-correcting codes based on very sparse matrices,''
  \emph{IEEE Transactions on Information Theory}, vol.~45, no.~2, pp. 399--431,
  1999.

\bibitem{Loeliger04}
H.~A. Loeliger, ``An introduction to factor graphs,'' \emph{IEEE Signal
  Processing Magazine}, vol.~21, no.~1, pp. 28--41, Feb. 2004.

\bibitem{Frey01}
F.~R. Kschischang, B.~I. Frey, and H.~A. Loeliger, ``Factor graphs and the
  sum-product algorithm,'' \emph{IEEE Transactions on Information Theory},
  vol.~47, no.~2, pp. 498--519, Feb. 2001.

\bibitem{Wiberg96}
N.~Wiberg, ``Codes and decoding on general graphs,'' Ph.D. dissertation,
  Department of Electrical Engineering Link{ö}ping University, 1996.

\bibitem{Aji00}
S.~M. Aji and R.~J. McEliece, ``The generalized distributive law,'' \emph{IEEE
  Transactions on Information Theory}, vol.~46, no.~2, pp. 325--343, Mar. 2000.

\bibitem{Tanner81}
R.~M. Tanner, ``A recursive approach to low complexity codes,'' \emph{IEEE
  Transactions on Information Theory}, vol.~27, no.~5, pp. 533--547, Sept.
  1981.

\bibitem{Urbanke01-2}
T.~Richardson and R.~Urbanke, ``The capacity of low-density parity check codes
  under message-passing decoding,'' \emph{{IEEE} {T}ransactions on
  {I}nformation {T}heory}, vol.~47, no.~2, pp. 599--618, Feb. 2001.

\bibitem{Urbanke08-2}
------, \emph{Modern Coding Theory}.\hskip 1em plus 0.5em minus 0.4em\relax
  Cambridge University Press, Mar. 2008.

\bibitem{Urbanke08}
C.~Measson, A.~Montanari, and R.~Urbanke, ``Maxwell construction: the hidden
  bridge between iterative and maximum a posteriori decoding,'' \emph{{IEEE}
  {T}ransactions on {I}nformation {T}heory}, vol.~54, no.~12, pp. 5277--5307,
  Dec. 2008.

\bibitem{Kudekar10}
S.~Kudekar, T.~Richardson, and R.~Urbanke, ``{T}hreshold {S}aturation via
  {S}patial {C}oupling: {W}hy {C}onvolutional {LDPC} {E}nsembles {P}erform {S}o
  {W}ell over the {BEC},'' \emph{IEEE Trans. on Information Theory}, vol.~57,
  no.~2, pp. 803 --834, feb. 2011.

\bibitem{Urbanke02}
C.~Di, D.~Proietti, T.~Richardson, E.~Telatar, and R.~Urbanke, ``Finite length
  analysis of low-density parity-check codes on the binary erasure channel,''
  \emph{{IEEE} {T}ransactions on {I}nformation {T}heory}, vol.~48, no.~6, pp.
  1570--1579, Jun. 2002.

\bibitem{Oswald02}
P.~Oswald and A.~Shokrollahi, ``Capacity-achieving sequences for the erasure
  channel,'' \emph{{IEEE} {T}ransactions on {I}nformation {T}heory}, vol.~48,
  no.~12, pp. 3017 -- 3028, Dec. 2002.

\bibitem{Luby01}
M.~Luby, M.~Mitzenmacher, A.~Shokrollahi, D.~Spielman, and V.~Stemann,
  ``Efficient erasure correcting codes,'' \emph{IEEE Transactions on
  Information Theory}, vol.~47, no.~2, pp. 569--584, Feb. 2001.

\bibitem{Brink04}
A.~E. Ashikhmin, G.~Kramer, and S.~ten Brink, ``Extrinsic information transfer
  functions: model and erasure channel properties,'' \emph{IEEE Transactions on
  Information Theory}, vol.~50, no.~11, pp. 2657--2673, Nov. 2004.

\bibitem{Orlitsky02}
J.~Zhang and A.~Orlitsky, ``Finite-length analysis of {LDPC} codes with large
  left degrees,'' in \emph{2002 {IEEE} {I}nternational {S}ymposium on
  {I}nformation {T}heory, {ISIT}}, 2002.

\bibitem{Urbanke09}
A.~Amraoui, A.~Montanari, T.~Richardson, and R.~Urbanke, ``Finite-length
  scaling for iteratively decoded {LDPC} ensembles,'' \emph{IEEE Transactions
  on Information Theory.}, vol.~55, no.~2, pp. 473--498, 2009.

\bibitem{Amraoui05}
A.~Amraoui, R.~Urbanke, and A.~Montanari, ``Finite-length scaling of irregular
  {LDPC} code ensembles,'' in \emph{2005 {IEEE} {I}nformation {T}heory
  {W}orkshop}, Aug. 2005.

\bibitem{Takayuki10}
\BIBentryALTinterwordspacing
N.~{Takayuki}, K.~{Kasai}, and S.~{Kohichi}, ``{Analytical solution of
  covariance evolution for irregular {LDPC} codes},'' \emph{e-prints}, Nov.
  2010. [Online]. Available:
  \url{http://adsabs.harvard.edu/abs/2010arXiv1011.1701T}
\BIBentrySTDinterwordspacing

\bibitem{Di06}
C.~Di, T.~Richardson, and R.~Urbanke, ``Weight distribution of low-density
  parity-check codes,'' \emph{IEEE Transactions on Information Theory},
  vol.~52, no.~11, pp. 4839 --4855, 2006.

\bibitem{Orlitsky05}
A.~Orlitsky, K.~Viswanathan, J.~Zhang, and S.~Member, ``Stopping set
  distribution of {LDPC} code ensembles,'' \emph{IEEE Transactions Information
  Theory}, vol.~51, pp. 929--953, 2005.

\bibitem{Minka01}
T.~P. Minka, ``{Expectation Propagation for approximate Bayesian inference},''
  in \emph{Proceedings of the 17th Conference in Uncertainty in Artificial
  Intelligence ({UAI} 2001)}.\hskip 1em plus 0.5em minus 0.4em\relax Morgan
  Kaufmann Publishers Inc., 2001, pp. 362--369.

\bibitem{Cover05}
T.~M. Cover and J.~A. Thomas, \emph{Elements of information theory}.\hskip 1em
  plus 0.5em minus 0.4em\relax New York, NY, USA: Wiley-Interscience, 2005.

\bibitem{Yedidia00}
J.~S. Yedidia, W.~T. Freeman, and Y.~Weiss, ``Generalized belief propagation,''
  in \emph{Proceedings of the Neural Information Processing Systems Conference,
  (NIPS)}, vol.~13, 2001, pp. 689--695.

\bibitem{Minka03}
T.~Minka and Y.~Qi, ``Tree-structured approximations by expectation
  propagation,'' in \emph{Proceedings of the Neural Information Processing
  Systems Conference, (NIPS)}, 2003.

\bibitem{Ghahramani97}
Z.~Ghahramani and M.~I. Jordan, ``Factorial hidden markov models,''
  \emph{Machine Learning}, vol.~29, pp. 245--273, Nov. 1997.

\bibitem{Olmos10-2}
P.~M. Olmos, J.~J. Murillo-Fuentes, and F.~P{\'e}rez-Cruz, ``{T}ree-structure
  expectation propagation for decoding {LDPC} codes over binary erasure
  channels,'' in \emph{2010 {IEEE} {I}nternational {S}ymposium on {I}nformation
  {T}heory, {ISIT}, {A}ustin, {T}exas}, 2010.

\bibitem{Olmos11}
P.~Olmos, J.~Murillo-Fuentes, and F.~P\'erez-Cruz, ``Tree-structured
  expectation propagation for decoding finite-length {LDPC} codes,'' \emph{IEEE
  Communications Letters}, vol.~15, no.~2, pp. 235 --237, Feb. 2011.

\bibitem{Pishro04}
H.~Pishro-Nik and F.~Fekri, ``On decoding of low-density parity-check codes
  over the binary erasure channel,'' \emph{{IEEE} {T}ransactions on
  {I}nformation {T}heory}, vol.~50, no.~3, pp. 439--454, Mar. 2004.

\bibitem{Burshtein04}
D.~Burshtein and G.~Miller, ``Efficient maximum-likelihood decoding of {LDPC}
  codes over the binary erasure channel,'' \emph{IEEE Transactions on
  Information Theory}, vol.~50, no.~11, pp. 2837 -- 2844, nov. 2004.

\bibitem{Liva09}
G.~Liva, B.~Matuz, E.~Paolini, and M.~Chiani, ``Pivoting {A}lgorithms for
  {M}aximum {L}ikelihood {D}ecoding of {LDPC} {C}odes over {E}rasure
  {C}hannels,'' in \emph{IEEE Global Telecommunications Conference, 2009.
  GLOBECOM 2009.}, dec. 2009, pp. 1 --6.

\bibitem{Saejoon08}
S.~Kim, S.~Lee, and S.-Y. Chung, ``An efficient algorithm for {ML} decoding of
  raptor codes over the binary erasure channel,'' \emph{IEEE Communications
  Letters}, vol.~12, no.~8, pp. 578 --580, aug. 2008.

\bibitem{Moon05}
T.~K. Moon, \emph{Error Correction Coding: Mathematical Methods and
  Algorithms}.\hskip 1em plus 0.5em minus 0.4em\relax Wiley-Interscience, 2005.

\bibitem{Vardy99}
T.~Etzion, A.~Trachtenberg, and A.~Vardy, ``Which codes have cycle-free tanner
  graphs?'' \emph{IEEE Transactions on Information Theory,}, vol.~45, no.~6,
  pp. 2173 --2181, sep 1999.

\bibitem{Wainwright08}
M.~J. Wainwright and M.~I. Jordan, \emph{Graphical Models, Exponential
  Families, and Variational Inference}.\hskip 1em plus 0.5em minus 0.4em\relax
  Foundations and Trends in Machine Learning, 2008.

\bibitem{Bishop06}
C.~M. Bishop, \emph{Pattern Recognition and Machine Learning (Information
  Science and Statistics)}.\hskip 1em plus 0.5em minus 0.4em\relax Secaucus,
  NJ, USA: Springer-Verlag New York, Inc., 2006.

\bibitem{Minka01thesis}
T.~P. Minka, ``A family of algorithms for approximate bayesian inference,''
  Ph.D. dissertation, Massachusetts Institute of Technology, 2001.

\bibitem{Lauritzen92}
S.~L. Lauritzen, ``Propagation of probabilities, means and variances in mixed
  graphical association models,'' \emph{Journal of the American Statistical
  Association}, vol.~87, no. 420, pp. 1098--1108, 1992.

\bibitem{Boyen98}
X.~Boyen and D.~Koller, ``{Tractable Inference for Complex Stochastic
  Processes},'' in \emph{Uncertainty in Artificial Intelligence.}\hskip 1em
  plus 0.5em minus 0.4em\relax Morgan Kaufmann, 1998, pp. 33--42.

\bibitem{Becker99}
A.~Becker, R.~Bar-Yehuda, and D.~Geiger, ``Randomized algorithms for the loop
  cutset problem,'' \emph{Journal of Artificial Intelligence Research},
  vol.~12, no.~1, pp. 219--234, 2000.

\bibitem{Luby97}
M.~Luby, M.~Mitzenmacher, A.~Shokrollahi, D.~Spielman, and V.~Stemann,
  ``Practical loss-resilient codes,'' in \emph{Proceedings of the 29th annual
  {ACM} Symposium on Theory of Computing}, 1997, pp. 150--159.

\bibitem{Abbasfar07}
A.~Abbasfar, D.~Divsalar, and K.~Yao, ``{A}ccumulate-{R}epeat-{A}ccumulate
  {C}odes,'' \emph{IEEE Transactions on Communications}, vol.~55, no.~4, pp.
  692 --702, april 2007.

\bibitem{Sason07}
H.~Pfister and I.~Sason, ``Accumulate-repeat-accumulate codes:
  Capacity-achieving ensembles of systematic codes for the erasure channel with
  bounded complexity,'' \emph{IEEE Transactions on Information Theory},
  vol.~53, no.~6, pp. 2088 --2115, june 2007.

\bibitem{Pfister05}
H.~Pfister, ``Finite-length analysis of a capacity-achieving ensemble for the
  binary erasure channel,'' in \emph{IEEE Information Theory Workshop, 2005},
  sept. 2005.

\bibitem{Wormald}
N.~C. Wormald, ``Differential equations for random processes and random
  graphs,'' \emph{{A}nnals of {A}pplied {P}robability}, vol.~5, no.~4, pp.
  1217--1235, 1995.

\bibitem{Montanari09}
M.~Mèzard and A.~Montanari, \emph{Information, Physics, and Computation},
  1st~ed.\hskip 1em plus 0.5em minus 0.4em\relax Oxford University Press, 2009.

\bibitem{Verdu10}
Y.~Polyanskiy, H.~Poor, and S.~Verdu, ``Channel coding rate in the finite
  blocklength regime,'' \emph{IEEE Transactions on Information Theory},
  vol.~56, no.~5, pp. 2307 --2359, may 2010.

\bibitem{Ezri07}
J.~Ezri, A.~Montanari, and R.~Urbanke, ``A generalization of the finite-length
  scaling approach beyond the {BEC},'' in \emph{2007 {IEEE} {I}nternational
  {S}ymposium on {I}nformation {T}heory, {ISIT}}, june 2007, pp. 1011 --1015.

\bibitem{Ezri08}
J.~Ezri, A.~Montanari, S.~Oh, and R.~Urbanke, ``The slope scaling parameter for
  general channels, decoders, and ensembles,'' in \emph{2008 {IEEE}
  {I}nternational {S}ymposium on {I}nformation {T}heory, {ISIT}}, jul. 2008,
  pp. 1443 --1447.

\bibitem{Papoulis02}
A.~Papoulis, \emph{Probability, Random Variables, and Stochastic Processes},
  4th~ed.\hskip 1em plus 0.5em minus 0.4em\relax {M}cgraw-{h}ill, 2002.

\bibitem{bollobas01}
B.~Bollobas, \emph{Random Graphs}, W.~Fulton, A.~Katok, F.~Kirwan, P.~Sarnak,
  B.~Simon, and B.~Totaro, Eds.\hskip 1em plus 0.5em minus 0.4em\relax
  Cambridge University Press, 2001.

\bibitem{Measson07}
C.~Measson, A.~Montanari, and R.~Urbanke, ``{A}symptotic {R}ate versus {D}esign
  {R}ate,'' in \emph{2007 {I}nternational {S}ymposium on {I}nformation
  {T}heory}, june 2007, pp. 1541 --1545.

\bibitem{Wad08}
T.~Wadayama, ``Average stopping set weight distributions of redundant random
  ensembles,'' \emph{IEEE Transactions on Information Theory}, vol.~54, no.~11,
  pp. 4991 --5004, nov. 2008.

\bibitem{Vardy06}
M.~Schwartz and A.~Vardy, ``On the stopping distance and the stopping
  redundancy of codes,'' \emph{Information Theory, IEEE Transactions on},
  vol.~52, no.~3, pp. 922 --932, march 2006.

\end{thebibliography}
\bibliographystyle{IEEEtran}

\appendices

\section{\TEP solution and processing order}\LABSEC{A1}

We prove that the \TEP decoder solution is independent of the processing order, i.e., the order in which check nodes of degree-one and two are removed. Given the parity check matrix of a linear block code $\Hs$, we first show that any operation of the \TEP decoder has an associated linear operator over $\Hs$.  Finally, we prove that these operations commute, proving the independence on the processing order.

Consider the $\k\times\n$ parity check matrix of the code, $\Hs$. 
The \TEP algorithm is initialized by removing from the graph all the known variables. 
The removal of any of these variables from the graph is equivalent to applying a binary linear transformation over the matrix $\Hs$ to get a new one, $\hat{\Hs}$, where the variable $\Vi{s}$ is completely disconnected, i.e.
\begin{align}\LABEQ{A1-1}
\hat{\Hs}=\Hs\mathbf{A}_{s},
\end{align}
where $\mathbf{A}_{s}$ is an $\n$-dimensional identity matrix where the $s$-th element is zero. 

To remove a check node of degree two from $\Hs$, e.g., $\CN{j}$ connected to variables $\Vi{o}$ and $\Vi{r}$, the resulting matrix, $\hat{\Hs}$, is obtained as follows:
\begin{align}\LABEQ{A1-3}
\hat{\Hs}&=\Hs\mathbf{B}_{o,r},\\\LABEQ{A1-32}
\mathbf{B}_{o,r}&=\mathbf{A}_{o}+\mathbf{Z}_{o,r},
\end{align}
where $o\neq r$ and $\mathbf{Z}_{o,r}$ is an $\n\times\n$ zero-matrix with$(\mathbf{Z}_{o,r})_{(o,r)}=1$. We have assumed that variable $\Vi{o}$ has been removed and $\Vi{r}$ inherits its connections. The symmetric transformation, i.e. remove $\Vi{r}$ instead of $\Vi{o}$, is performed by applying the $\mathbf{B}_{r,o}$ operator. Since the final result is one variable with all connections, $\mathbf{B}_{r,o}$ is equivalent to $\mathbf{B}_{o,r}$. With no loss of generality we assume that the removed variable is the leftmost variable in Tanner graph or, equivalently, the leftmost column in the parity check matrix.

When applying $\mathbf{A}_s$ after a sequence of operations (transformations), variable $s$ must be connected to a degree-one check node. If it was not in the original graph, in $\Hs$, some operations are needed first. Similarly, for $\mathbf{B}_{o,r}$ we need variables $o$ and $r$ to share a degree-two check node. Therefore, not every sequence of operations is valid. 

We study the commutativity of two valid sequences $I$ and $II$. If we prove that every pair of operations commutes, we may reorder the operations in the second sequence as in the first one. Particularly, we have to show the conditions for which following matrices commute:
\begin{enumerate}
\item $\mathbf{A}_{s}$ and $\mathbf{A}_{p}$ commute for all $s, p$,
\item $\mathbf{B}_{o,r}$  and $\mathbf{B}_{v,z}$ commute for all possible pairs $\{o,r\}$ and $\{v,z\}$ for $o\neq r$ and $v\neq z$,
\item $\mathbf{A}_{s}$ and $\mathbf{B}_{o,r}$ commute for all the possible triples $\{s,o,r\}$ for $o\neq r$,
\end{enumerate}
where $s$, $o$, $r$, $v$ and $z$ belong to $\{1,\ldots,\n\}$.
%
The first case is straightforward since diagonal matrices always commute. 

For the other two cases, we need to prove the conditions for which $\mathbf{B}_{o,r}\mathbf{B}_{v,z}=\mathbf{B}_{v,z}\mathbf{B}_{o,r}$, which can be express as follows using \EQ{A1-32}:
\begin{align}\LABEQ{}
\mathbf{B}_{o,r}\mathbf{B}_{v,z}&=\mathbf{A}_{o}\mathbf{A}_{v}+\mathbf{A}_{o}\mathbf{Z}_{v,z}+\mathbf{Z}_{o,r}\mathbf{A}_{v}+\mathbf{Z}_{o,r}\mathbf{Z}_{v,z}, \\
\mathbf{B}_{v,z}\mathbf{B}_{o,r}&=\mathbf{A}_{v}\mathbf{A}_{o}+\mathbf{A}_{v}\mathbf{Z}_{o,r}+\mathbf{Z}_{v,z}\mathbf{A}_{o}+\mathbf{Z}_{v,z}\mathbf{Z}_{o,r}.
\end{align}

Note that, to prove both 2) and 3), we have to show that the matrices $\mathbf{A}_{o}$  and $\mathbf{Z}_{v,z}$ commute for all the possible triples $\{o,v,z\}$ and that $\mathbf{Z}_{o,r}$ and $\mathbf{Z}_{v,z}$ commute as well. Regarding the first case, we have: 
\begin{align}\LABEQ{COMM1}
\mathbf{Z}_{v,z}\mathbf{A}_{o}&=\begin{cases}\mathbf{Z}_{v,z},& z\neq o\\  \mathbf{0}_{\n\times\n},& z=o\end{cases}\\
\mathbf{A}_{o}\mathbf{Z}_{v,z}&=\begin{cases}\mathbf{Z}_{v,z},& v\neq o\\  \mathbf{0}_{\n\times\n},& v=o\end{cases}\LABEQ{COMM3},
\end{align}
where $\mathbf{0}_{\n\times\n}$ is the $\n$-square zero matrix. Hence, we can conclude that $\mathbf{A}_{o}$ and $\mathbf{Z}_{v,z}$ commute for every triplet, as long as $o\neq z$ and $o\neq v$. If $o=v$, $\mathbf{Z}_{o,z}$ and $\mathbf{A}_{o}$ do not commute, but neither $\mathbf{Z}_{o,z}\mathbf{A}_{o}$ nor  $\mathbf{A}_{o}\mathbf{Z}_{o,z}$ are valid sequences, because once we have removed variable $o$, we cannot remove it again. If $o=z$, $\mathbf{Z}_{v,o}$ and $\mathbf{A}_{o}$ do not commute, but in this case $\mathbf{Z}_{v,o}\mathbf{A}_{o}$ is a valid sequence, while the reverse is not valid. But, if we examine the operation $\mathbf{B}_{v,o}\mathbf{A}_{o}$ closely, it means that $v$ and $o$ share a degree-two check node and $o$ is connected to degree-one check node. Hence, we can find in Sequence $I$ the operations $\mathbf{B}_{v,o}\mathbf{A}_{o}$ and in Sequence $II$ the operations $\mathbf{A}_{o}\mathbf{A}_{v}$, because once we have remove variable $o$, the check node that variables $v$ and $o$ shared is now a degree-one check node, and these two operations are equivalent, as it can be seen in in Fig. \FIG{FigA1}.

\begin{figure}[h]
\centering
\begin{tabular}{c}
\includegraphics[width=3 cm]{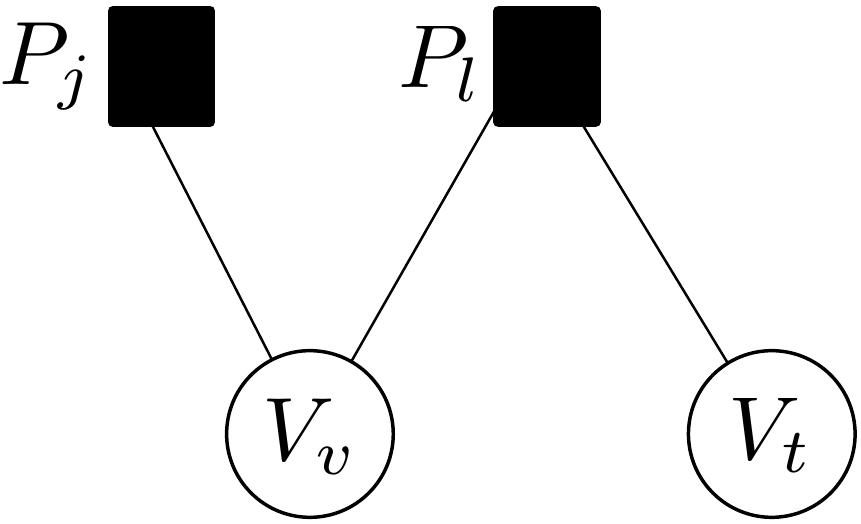}
\end{tabular}
\caption{The TEP decoder can process the graph by either removing the degree-one check node $\CN{j}$ and then process the degree-one check node $\CN{l}$ (once $\CN{j}$ has been processed $\CN{l}$ becomes degree one) or, indistinctly, remove first the degree-two check node $\CN{l}$ and then process the degree-one check node $\CN{j}$.}\LABFIG{FigA1}
\end{figure}

Finally, to conclude the proof, we have to show that the matrices $\mathbf{Z}_{o,r}$ and $\mathbf{Z}_{v,z}$ commute. It is easy to check that:
\begin{align}\LABEQ{}
\mathbf{Z}_{o,r}\mathbf{Z}_{v,z}=\begin{cases}\mathbf{Z}_{o,z},& r= v\\  \mathbf{0}_{\n\times\n},& \text{otherwise} \end{cases}.
\end{align}
If the four indices are different the matrices commute. If one of the indices is repeated, we have three possible scenarios:
\begin{enumerate}
\item $v=o$, the matrices commute and $\mathbf{B}_{o,r}\mathbf{B}_{o,z}$ and $\mathbf{B}_{o,z}\mathbf{B}_{o,r}$ are invalid sequences, because we would remove variable $o$ twice in both cases. 
\item $z=r$, the matrices commute and both sequence are valid. 
\item $v=r$ (or equivalently $o=z$), the matrices do not commute and one order gives a valid sequence, i.e. $\mathbf{B}_{o,r}\mathbf{B}_{r,z}$, and the other does not, i.e. $\mathbf{B}_{r,z}\mathbf{B}_{o,r}$, because once we have removed the variable $r$ we cannot use it again. But we could find in the Sequence $I$ the matrices in this order $\mathbf{B}_{o,r}\mathbf{B}_{r,z}$ and in the Sequence $II$ the matrices in this order $\mathbf{B}_{r,z}\mathbf{B}_{o,z}$, because after we process $\mathbf{B}_{r,z}$, $z$ inherits all the check nodes of $r$, hence if $r$ had a degree-two check node with $o$ after we have processed $\mathbf{B}_{r,z}$, now this degree-two check node is between $o$ and $z$. 
\end{enumerate}
Hence, we have proven that any two matrices in the first sequence commute or have a valid alternative in the second.

 \section{Probability of sharing two check nodes}\LABSEC{A3}
 
 In this appendix we compute the probability $\pA{t}$ in \EQ{pA}. Recall that it is defined as the probability for scenario $\ctwoshare$, in which two variables share a check node of degree two and, at least, another check node of arbitrary degree, as illustrated in Fig. \FIG{Cases}(b). 
Let $\ctwoScheck{m}$ the particular case where both variables share $m+1$ check nodes, one of them at least of degree two. If we compute the probability 
\begin{equation}
\P(\ctwoScheck{m}|\ctwo),
\end{equation}
then the probability $\pA{t}$ is computed by summing over the parameter $m$:
\begin{equation}\LABEQ{PBu}
\pA{t}=\P(\ctwoshare|\ctwo)=\sum_{m}\P(\ctwoScheck{m}|\ctwo).
\end{equation}

In what follows, we proceed by evaluating the probability of scenarios $\ctwoScheck{1}$ and $\ctwoScheck{2}$ to finally conclude that  
\begin{equation}
\P(\ctwoScheck{1}|\ctwo)\gg\P(\ctwoScheck{2}|\ctwo)\gg\P(\ctwoScheck{3}|\ctwo)\cdots
\end{equation}
and, for large enough graphs, we have
\begin{equation}\LABEQ{pBS1}
\P(\ctwoshare|\ctwo)=\P(\ctwoScheck{1}|\ctwo)+\order(1/\Edges^{2}),
\end{equation}
which means that, in practice, we only have to consider the scenario $\ctwoScheck{1}$ to study the TEP decoder for large code lengths.
 
\subsubsection{Probability of scenario $\ctwoScheck{1}$}

We first focus on the scenario $\ctwoScheck{1}$, illustrated in Fig. \FIG{Cases}(b). Our  goal is to compute the probability $\P\left(\ctwosharem|\ctwo\right)$, where $\ctwosharem$ corresponds to the case \emph{where $\Vi{p}$ and $\Vi{q}$ share just a check node of degree two and a check node of degree $j$}. Then, the probability \EQ{pBS1} is computed by summing over the degree $j$:
  \begin{equation}\LABEQ{PA1}
 \P\left(\ctwoScheck{1}|\ctwo\right)=\sum_{j=1}^{\rhimax}\P\left(\ctwosharem|\ctwo\right).
 \end{equation}
 
 

If $\vd{o}(j)$ denotes the number of edges connected to $\Vi{o}$ that have right degree $j$, we propose to obtain the probability $\P\left(\ctwosharem|\ctwo\right)$ by marginalizing over the following probability function:
\begin{equation}\LABEQ{Extend}
\P(\ctwosharem, \vd{o},\vd{r},\vd{o}(j)=\proofpa,\vd{r}(j)=\proofpb | \ctwo),
\end{equation}
for $\proofpa\in\{0,\ldots,\vd{o}-1\}$, $\proofpb\in\{0,\ldots\vd{r}-1\}$. The mass probability function in \EQ{Extend} represents the probability that $\Vi{o}$ and $\Vi{r}$ share another check node of degree $j$, the degree of $\Vi{o}$ and $\Vi{r}$ is, respectively, $\vd{o}$ and $\vd{r}$, $\Vi{o}$ has $\proofpa$ edges of right degree $j$ and $\Vi{r}$ has $\proofpb$ edges of right degree $j$. The probability density function in \EQ{Extend} can be rewritten applying the properties of conditional probability:
\begin{align}\LABEQ{Massf}
\P(&\ctwosharem,\vd{o},\vd{r},\vd{o}(j)=\proofpa,\vd{r}(j)=\proofpb |\ctwo)\nonumber\\
&=\P(\ctwosharem | \vd{o}(j)=\proofpa,\vd{r}(j)=\proofpb, \vd{o},\vd{r},\ctwo)\nonumber\\
&\;\;\cdot\P(\vd{o}(j)=\proofpa,\vd{r}(j)=\proofpb | \vd{o},\vd{r},\ctwo)\nonumber\\&\;\;\cdot\P(\vd{o},\vd{r}|\ctwo).
\end{align}
The degrees of two given nodes in the graph are (asymptotically) pairwise independent \cite{Montanari09} and, hence, for sufficiently large graphs we can assume
\begin{equation}\LABEQ{Last}
\P(\vd{o},\vd{r}|\ctwo)=\frac{\li{\vd{o}}{t}}{\eremain{t}}\frac{\li{\vd{r}}{t}}{\eremain{t}}+\order(\Edges^{-1}),
\end{equation}
for $\vd{o},\vd{o}\in\{2,\ldots,\Edges\}$. Each edge connected to a variable node has right degree $j$ with probability $\ri{j}{t}/\eremain{t}$. The number of edges connected to either $\Vi{o}$ or $\Vi{r}$ with right degree $j$ are asymptotically distributed according to binomial distributions 
$B(\proofpa,\vd{o}-1,\ri{j}{t}/\eremain{t})$ and $B(\proofpb,\vd{1}-1,\ri{j}{t}/\eremain{t})$ respectively, where $B(x,N,P)$ denotes a binomial distribution over $x\in\mathbb{N}$ with parameters $N$ and $P$ \cite{Papoulis02}. Note that we subtract $1$ to $\vd{o}$ and $\vd{r}$ since we know that one edge is connected to a check node of degree two. Therefore, the second term in \EQ{Massf} is
\begin{align}\LABEQ{Second}
&\P(\vd{o}(j)=\proofpa,\vd{r}(j)=\proofpb | \vd{o},\vd{r},\ctwo)=\\&=B\left(\proofpa,\vd{o}-1,\frac{\ri{j}{t}}{\eremain{t}}\right)\cdot B\left(\proofpb,\vd{r}-1,\frac{\ri{j}{t}}{\eremain{t}}\right)+\order(\Edges^{-1})\nonumber.
\end{align}
Finally, we have to compute the first term in \EQ{Massf}, i.e., the probability that variables $\Vi{o}$ and $\Vi{r}$ share a check node of degree $j$ when the variables have, respectively, $\proofpa$ and $\proofpb$ edges with right degree $j$. Note first that the graph has $\ri{j}{t}\Edges$ edges with right degree $j$. Let us assume that the edges of one of the variables, $\Vi{o}$ for example, are fixed and that the edges of $\Vi{r}$ are now randomly set.
%

If the graph is large enough, with probability 
\begin{align}\LABEQ{}
\frac{\proofpa(j-1)}{\Edges\ri{j}{t}},
\end{align}
each edge of $\Vi{r}$ shares a check node of degree $j$ with an edge of $\Vi{o}$. For large graphs, the number of checks shared between both variables is asymptotically described by a binomial distribution
$B(\proofpb,\proofpa(j-1)/\Edges\ri{j}{t})$. The probability that they share al least one check node of degree $j$ in \EQ{Massf} yields
\begin{align}\LABEQ{Allfixed}
&\P(\ctwosharem | \vd{o}(j)=\proofpa,\vd{r}(j)=\proofpb,\vd{o},\vd{r}, \ctwo)\nonumber\\&=\proofpb\frac{\proofpa(j-1)}{E\ri{j}{t}}\left(1-\frac{\proofpa(j-1)}{E\ri{j}{t}}\right)^{\proofpb-1}\nonumber\\&\approx\frac{\proofpb\proofpa(j-1)}{E\ri{j}{t}}+\order(\Edges^{-2}).
\end{align}
We have already computed all the factors in the joint mass function in \EQ{Massf}. In \EQ{marginalization} (see next page), we marginalize over $ \vd{o},\vd{r},\proofpa$ and $\proofpb$ to obtain $\P\left(\ctwosharem|\ctwo\right)$:
\begin{align}\LABEQ{}
&\P\left(\ctwosharem|\ctwo\right)&=\frac{(j-1)\ri{j}{t}}{\Edges e^{2}(t)}(\lav{t}-1)^{2}.
\end{align}

\begin{figure*}[]
\par\noindent\rule{\dimexpr(1\textwidth-0\columnsep-0pt)}{0.2pt}
\begin{align}\LABEQ{marginalization}
&\P\left(\ctwosharem|\ctwo\right)=\sum_{\substack{\vd{o},\vd{r}\\\proofpa,\proofpb}}
\P(\ctwosharem, \vd{o},\vd{r},\vd{o}(j)=\proofpa,\vd{r}(j)=\proofpb |\ctwo)\nonumber\\
&=\sum_{\vd{o},\vd{r}}\Bigg[\frac{\li{\vd{o}}{t}}{\eremain{t}}\frac{\li{\vd{r}}{t}}{\eremain{t}}\sum_{\proofpa}\binom{\vd{o}-1}{\proofpa}\left(\frac{\ri{j}{t}}{\eremain{t}}\right)^{\proofpa}\left(1-\frac{\ri{j}{t}}{\eremain{t}}\right)^{\vd{o}-1-\proofpa}\nonumber\\
&\;\;\;\;\;\qquad\qquad\sum_{\proofpb}\binom{\vd{r}-1}{\proofpb}\left(\frac{\ri{j}{t}}{\eremain{t}}\right)^{\proofpb}\left(1-\frac{\ri{j}{t}}{\eremain{t}}\right)^{\vd{r}-1-\proofpb}\frac{\proofpb\proofpa(j-1)}{E\ri{j}{t}}\Bigg]\nonumber\\
&=\sum_{\vd{o},\vd{r}}\frac{\li{\vd{o}}{t}}{\eremain{t}}\frac{\li{\vd{r}}{t}}{\eremain{t}}\frac{(j-1)}{E\ri{j}{t}}\E[\vd{o}(j)]\E[\vd{r}(j)]\nonumber\\&=\sum_{\vd{o},\vd{r}}\frac{\li{\vd{o}}{t}}{\eremain{t}}\frac{\li{\vd{r}}{t}}{\eremain{t}}\frac{(j-1)}{E\ri{j}{t}}\left(\vd{o}-1\right)\left(\vd{r}-1\right)\left(\frac{\ri{j}{t}}{\eremain{t}}\right)^{2}\nonumber\\
&=\frac{(j-1)\ri{j}{t}}{\Edges e^{2}(t)}\left(\sum_{\vd{o}}\frac{\li{\vd{o}}{t}}{\eremain{t}}\left(\vd{o}-1\right)\right)\left(\sum_{\vd{r}}\frac{\li{\vd{r}}{t}}{\eremain{t}}\left(\vd{r}-1\right)\right)=\frac{(j-1)\ri{j}{t}}{\Edges e^{2}(t)}(\lav{t}-1)^{2}.
\end{align}
\par\noindent\rule{\dimexpr(1\textwidth-0\columnsep-0pt)}{0.2pt}
\end{figure*}

%
Finally, the probability of scenario $\ctwoScheck{1}$ is computed by summing over the degree $j$ as follows
\begin{align}\LABEQ{PA2}
\P(\ctwoScheck{1}|\ctwo)=&\sum_{j=1}^{\rhimax}\P\left(\ctwosharem|\ctwo\right)=\frac{(\lav{t}-1)^{2}(\rav{t}-1)}{\Edges\eremain{t}}.
\end{align}

As expected, the probability of two variables sharing a check node in a random graph (aside from the check node of degree two that we know they are sharing) is $\order(\Edges^{-1}\eremain{t}^{-1})$, i.e., inverse to the total number of edges in the graph. This result is consistent with the theory of random graphs \cite{bollobas01}.


\subsubsection{Probability of scenario $\ctwoScheck{u}$ for $u>1$}

A similar analysis can be extended for the case $\ctwoScheck{2}$. If we define the subscenario $\cthreesharem$, where both variables share two check nodes of degrees $j$ and $\ell$, the probability $\P(\ctwoScheck{2}|\ctwo)$ is obtained by counting over all possible cases:
\begin{equation}\LABEQ{Marg2}
\P(\ctwoScheck{2}|\ctwo)=\sum_{j,\ell}\P(\cthreesharem|\ctwo).
\end{equation}

The probability $\P(\cthreesharem|\ctwo)$ can be obtained with a similar procedure than the one used to compute $\P(\ctwosharem|\ctwo)$. In this case, we marginalize over the joint mass probability function of the degrees of both variables $\left(\vd{o},\vd{r}\right)$ the number of edges in each variable with right degree $j$ $\left(\vd{o}(j),\vd{r}(j)\right)$ and the number of edges in each variable with right degree $\ell$ $\left(\vd{o}(\ell),\vd{r}(\ell)\right)$. Now we factorize the joint mass function applying the conditionality properties, as we did in \EQ{Massf}:

\begin{align}\LABEQ{Massf2}
\P(&\cthreesharem, \vd{o},\vd{r},\vd{o}(j),\vd{r}(j),\vd{o}(\ell),\vd{r}(\ell)|\ctwo)\nonumber\\
&=\P(\cthreesharem | \vd{o}(\ell),\vd{r}(\ell),\vd{o}(j),\vd{r}(j), \vd{o},\vd{r},\ctwo)\nonumber\\
&\cdot\P(\vd{o}(\ell),\vd{r}(\ell),\vd{o}(j),\vd{r}(j)=\beta | \vd{o},\vd{r},\ctwo)\nonumber\\
&\cdot\P(\vd{o},\vd{r}|\ctwo),
\end{align}
where the last factor $\P(\vd{o},\vd{r}|\ctwo)$ does not change with respect to \EQ{Last} and the second one, similarly to \EQ{Second}, can be expressed as a product of binomial distributions. In the first term in \EQ{Massf2}, we have fixed the degrees of $\Vi{o}$ and $\Vi{r}$ and the number of edges of both variables with right degree $j$ and $\ell$. Following the same procedure considered to derive \EQ{Allfixed}, it can be shown that
\begin{align}
\P&(\cthreesharem | \vd{o}(\ell),\vd{r}(\ell),\vd{o}(j),\vd{r}(j), \vd{o},\vd{r},\ctwo)\nonumber\\&=\frac{\vd{o}(j)\vd{r}(j)(j-1)}{E\ri{j}{t}}\frac{\vd{o}(\ell)\vd{r}(\ell)(\ell-1)}{E\ri{\ell}{t}}\propto\Edges^{-2}
\end{align} 

The constant $\Edges^{-2}$ does not depend on the marginalization in \EQ{Marg2}. Hence,
\begin{equation}
\P(\ctwoScheck{2}|\ctwo)\propto\frac{1}{\Edges^{2}}
\end{equation}
and, in general we have:
\begin{equation}
\P(\ctwoScheck{m}|\ctwo)\propto\frac{1}{\Edges^{m}},
\end{equation}
for $m\in\mathbb{N}$. Therefore $\P(\ctwoScheck{m}|\ctwo)$ for $m\geq2$ are negligible compared to $\P(\ctwoScheck{1}|\ctwo)$ for sufficiently large graphs.

Probability $\pA{t}$ is crucial to evaluate the expected graph evolution under TEP decoding in Section\SEC{EG}. In addition, for asymptotically large graphs, to evaluate how the graph changes in one iteration of the TEP decoder, it is a good approximation to consider that two variables that are sharing a check node of degree two,  share at most one extra check node, as illustrated in Fig. \FIG{Cases}(b).

\section{Boundedness on the graph degrees. Proof of Lemma\LEM{finiteLAVG}}\LABSEC{A4}


The proof of this lemma is straightforward but it is convenient to review first some basic definitions. We make use of LDPC ensembles
and degree distribution polynomials, which are defined in Section\SEC{LDPC}.

\subsection{Design rate, code rate and redundant codes}
Consider an LDPC ensemble defined by $(\lambda(x),\rho(x))$ with finite maximum degrees. Let $\Hs_{\n}$ be the parity matrix of a code sample of length $\n$ chosen at random and let  $\text{row}(\Hs_{\n})$ and $\text{col}(\Hs_{\n})$denote respectively the number of rows and columns of the matrix. 
In the limit $\n\rightarrow\infty$ the expected presence of double edges in $\Hs_{\infty}$ is zero \cite{Urbanke08-2} and the matrix 
%
presents the exact DD defined by $(\lambda(x),\rho(x))$. The following expression is true in this case:
\begin{align}\LABEQ{relation}
1-\frac{\text{row}\left(\Hs_{\infty}\right)}{\text{col}\left(\Hs_{\infty}\right)}=1-\frac{\Lambda_{\text{avg}}|_{\Hs_{\infty}}}{\Theta_{\text{avg}}|_{\Hs_{\infty}}}=1-\frac{\Lambda_{\text{avg}}}{\Theta_{\text{avg}}}=\rate,
\end{align}
where $\rate$ is the design rate, $\Lambda_{\text{avg}}|_{\Hs_{\infty}}$ and $\Theta_{\text{avg}}|_{\Hs_{\infty}}$ are the average 
node and check degrees that we compute from $\Hs_{\infty}$ and $\Lambda_{\text{avg}}$ and $\Theta_{\text{avg}}$ are the same parameters computed directly from the pair $(\lambda(x),\rho(x))$ using \EQ{nodeavg} and \EQ{nodeavg2}. Because the ensemble has bounded maximum degrees, the design rate is finite.

Besides, the \emph{true rate} $\rate_{\text{true}}$ of the code $\Hs_{\infty}$ is defined as
\begin{align}
\rate_{\text{true}}=1-\frac{\text{rank}(\Hs_{\infty})}{\text{col}(\Hs_{\infty})}\in[0,1]
\end{align}
and it can be easily verified that $\rate\leq\rate_{\text{true}}$, where in the limit $\n\rightarrow\infty$ the equality only holds for a certain subset of ensembles \cite{Di06,Measson07}.

Although unusual, we can construct LDPC ensembles with negative design rate. For instance, the ensemble:
\begin{align}\LABEQ{resemble}
\lambda(x)=x^3 \qquad \rho(x)=x^2,
\end{align}
has a design rate $\rate=-1$ but, despite this,  the DD is well-defined and the construction of graph samples follows the standard procedure \cite{Urbanke08-2}. When we sample from such class of random ensembles, we obtain parity matrices that contain dependent rows, which are sometimes referred to as redundant LDPC parity matrices. By extension, ensembles such as \EQ{resemble} are called redundant ensembles \cite{Wad08,Vardy06}. 

Redundant ensembles also appear when we analyze the residual graph after transmission over the erasure channel. Let $(\lambda(x),\rho(x))$ define an LDPC ensemble of positive design rate and finite maximum degrees that is used for transmission over the erasure channel. As shown in \cite{Luby01}, the DD that defines the residual ensemble after removing the set of known bits from the Tanner graph has the form
\begin{align}\LABEQ{initialcondDD}
\widehat{\lambda}(x)&=\lambda(x),\\
\widehat{\rho}(x)&=\sum_{j=1}^{\rhimax}\widehat{\rho}_{j}x^{j-1},
\end{align}
where
\begin{align}\LABEQ{initialcondDD2}
\widehat{\rho}_{j}=\frac{1}{\epsilon}\sum_{m\geq j}\rhi{j}\binom{m-1}{j-1}\pe^{j}(1-\pe)^{m-j}.
\end{align}
The ensemble $\left(\widehat{\lambda}(x),\widehat{\rho}(x)\right)$ has, in most cases, a negative design rate\footnote{For instance, for the $(3,6)$ ensemble and $\epsilon=0.42$ we get
\begin{align}
\widehat{\lambda}(x)=x^2, \quad
\widehat{\rho}(x)&=0.0656+0.2376x+0.3441x^2+0.2492x^3\nonumber
\\&\quad+0.0902x^4+0.013x^5 \nonumber,
\end{align} 
for which the design rate is $-0.1448$.}. An important property of this ensemble is that its design rate is always finite as long as the original ensemble has finite maximum degrees. As a consequence, by \EQ{relation}, the number of variable and check nodes in the residual graph are of the same order.

\subsection{Proof of Lemma\LEM{finiteLAVG}}

Let $(\lambda(x),\rho(x))$ define an LDPC ensemble of positive design rate and finite maximum degrees. Let $\code$ be a sample from such ensemble that is used for transmission over a BEC($\epsilon$). Any channel realization $\y\in\{0,?,1\}^{\n}$ gives rise to a residual parity matrix $\hat{\Hs}$ sampled from $\left(\widehat{\lambda}(x),\widehat{\rho}(x)\right)$ in \EQ{initialcondDD}-\EQ{initialcondDD2}. 

As discussed above, the residual ensemble after removing the known bits has finite rate and, therefore, $\text{row}(\hat{\Hs})$ and $\text{col}(\hat{\Hs})$ are of the same order. The TEP decoder works over the initialized graph by removing one column and one row per iteration and, thus, the sequence of residual parity matrices $\hat{\Hs}_t$ for $t=1,2,\ldots,$  satisfies the same property. 

By removing degree-one or degree-two check nodes, on the one hand we know that the average check degree $\Theta_{\text{avg}}|_{\hat{\Hs}_t}$ does not grow. On the other hand, $\Lambda_{\text{avg}}|_{\hat{\Hs}_t}$ might grow as we remove degree-two check nodes. To show that the latter cannot grow unbounded, note that $\Theta_{\text{avg}}|_{\hat{\Hs}_t}$ and $\Lambda_{\text{avg}}|_{\hat{\Hs}_t}$ are linked by $E(t)$, i.e., the number of edges in the residual matrix $\hat{\Hs}_t$:
%
\begin{align}\LABEQ{EAVG}
\frac{E(t)}{\text{col}\left(\hat{\Hs}_t\right)}=\Lambda_{\text{avg}}|_{\hat{\Hs}_t}, \quad \frac{E(t)}{\text{row}\left(\hat{\Hs}_t\right)}=\Theta_{\text{avg}}|_{\hat{\Hs}_t}.
\end{align}

Equation \EQ{EAVG} proves that, if $E(t)>0$ and $\text{col}\left(\hat{\Hs}_t\right)$ and $\text{row}\left(\hat{\Hs}_t\right)$ are of the same order, then $\Lambda_{\text{avg}}|_{\hat{\Hs}_t}$ and $\Theta_{\text{avg}}|_{\hat{\Hs}_t}$ have to keep the same order  as well,   which proves the lemma.

\section{TEP complexity per iteration. Proof of Lemma\LEM{Complexity}}\LABSEC{A5}

A step of the PD algorithm, summarized in the linear transformation in \EQ{A1-1}, has a constant complexity and the PD overall complexity is $\order(\n)$. By crossing \EQ{A1-1} and \EQ{A1-32}, we observe that the removal of a degree-two check node in a step of the TEP is performed by a basic PD operation followed by the summation of two columns of the code parity check matrix $\Hs$. 
The cost of this operation is given by the number of ones in both columns. For LDPC codes, once we select a degree-two check node, the two associated variable nodes have in average degree $\lav{t}$, where $\lav{t}$ is the average edge left degree defined in \EQ{l_avg}.

\begin{figure}
\centering
\begin{tabular}{cc}
  \includegraphics[width=2.5 cm]{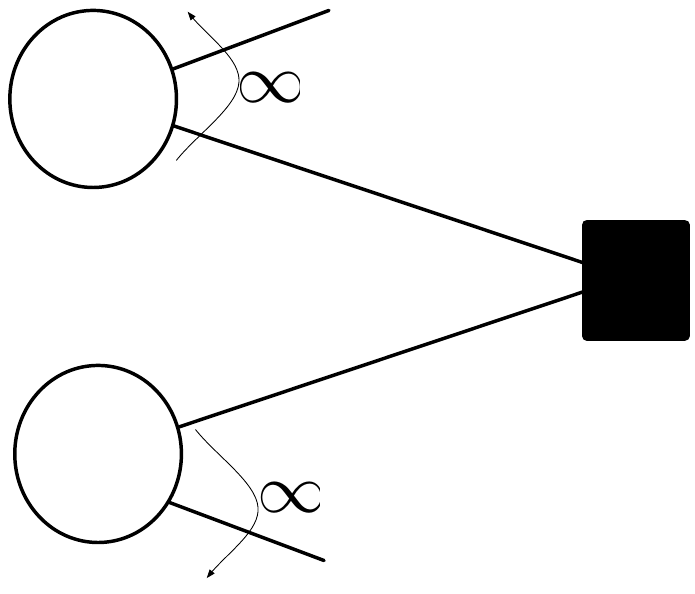} & \includegraphics[width=2.5 cm]{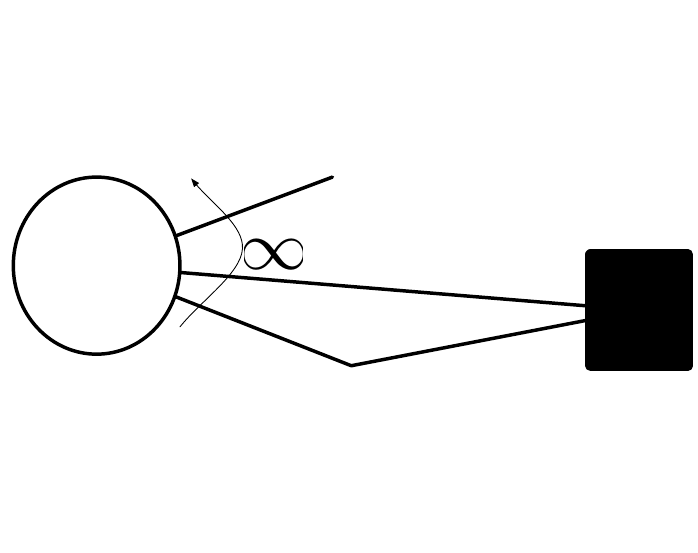} \\
  (a) & (b)
\end{tabular}
\caption{Most likely scenarios when $\lav{t}\rightarrow\infty$ and we are about to process a degree-two check node.}
\LABFIG{CasesAPEN}
\end{figure}


In Section\SEC{LAV}, we show that there might exist codes for which, in the limit $\n\rightarrow\infty$ the average edge left degree $\lav{t}$  diverges to $\infty$ at some point of the TEP decoding process. However, the conditions are too restrictive and we can say that, for  most of the ensembles,  $\lav{t}$ is bounded during the whole decoding process. In this case, TEP decoder iteration just adds a constant complexity to the PD cost per iteration.

For those ensembles for which $\lav{t}$ may diverge at some point, the complexity of the TEP decoder remains linear as long as we only focus on degree-one nodes. If there are not any of them left and $\lav{t}\rightarrow\infty$, an eventual iteration of quadratic cost (with $\n$) is required. However, we expect the fraction of such iterations throughout the decoding process to be low for several reasons. First, since the variable average degree $\Lambda_{\text{avg}}$ is always bounded, then the fraction of variable nodes with infinite degree is small compared to the rest of the code. And second, each time we remove a degree-two check node and $\lav{t}\rightarrow\infty$, we face very high probability one of the following two scenarios depicted in Fig. \FIG{CasesAPEN}. In the scenario (a), the probability that the two variables share an additional check nodes is close to one and then degree-one check nodes can be created. If this happens, $\lav{t}$ is of the same order and it is very likely that those recently created check nodes are connected to variables of infinite degree. By \EQ{RBP}, the removal of such nodes starts a process of massive creation of  degree-one check nodes that brings down $\lav{t}$. In Fig. \FIG{CasesAPEN} (b), we have a double connection that can be removed at no cost.

\section{Computation of the $\gamma_{\text{TEP}}$ parameter}\LABSEC{A6}


We compute $\gamma_{\text{TEP}}$, based on the TEP solution independency of the processing order proven in Appendix \SEC{A1}:
\begin{itemize}
\item Start the TEP algorithm by running a BP stage. Compute the BP residual graph expected DD at $\pe=\peBP$ \cite{Luby01}. Alternatively, we can obtain such DD by evaluating the differential equations for the TEP  in \EQ{System1}-\EQ{System3} if we set $p_{C}(\tau)=1$ in \EQ{PC} for any $\tau$ until the fraction of degree-one check nodes in the graph cancels.
\item Using this graph as input, evaluate the graph expected evolution when we only remove degree-two check nodes: solve the system \EQ{System1}-\EQ{System3} by setting $p_{C}(\tau)=0$ in \EQ{PC} for all $\tau$ until the graph runs out of degree-two check nodes. 
\item  $\gamma_{\text{TEP}}=\n\ri{1}{\tau',\n,\peTEP}$, once we have run out of degree two check nodes.
\end{itemize}

In Fig. \FIG{r136evol_method}, we represent the evolution of $\n r_1^{\text{TEP}}(\tau',\n,\peTEP)$ after following the previous steps for the regular $(3,6)$-LDPC code and $\n=2^{12}$ $(+)$, $\n=2^{14}$ $(\Box)$ and $\n=2^{16}$ $(\times)$. At $e(\tau)\approx0.22$, the BP decoder gets stuck and we begin to remove degree-two check nodes. In this second phase we can notice that $\n\ri{1}{\tau',\n,\peTEP}$ is independent of $\n$.


\begin{figure}[h]
\centering
\includegraphics[width=8cm,height=8cm]{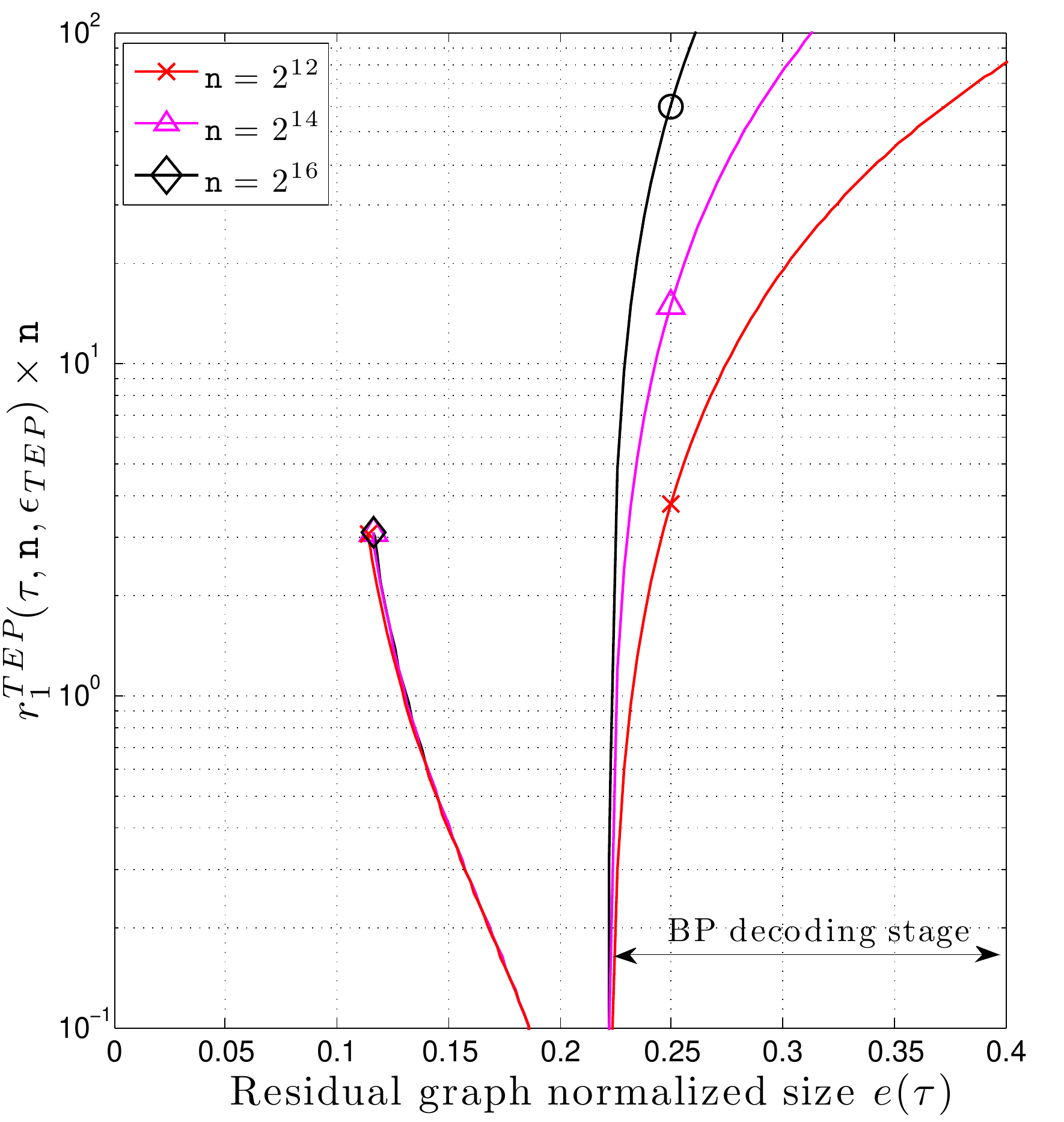}
\caption{We plot the solution for $\ri{1}{\tau}$ computed to estimate $\ri{1}{\tau',\n,\peTEP}$ for  a $(3,6)$ regular code at $\pe=\peBP=\peTEP$. The code lengths considered are $\n=2^{12}$ $(\times)$, $\n=2^{14}$ $(\triangle)$ and $\n=2^{16}$ $(\diamond)$. The BP was first run and then degree-two check nodes were processed.}\LABFIG{r136evol_method}
\end{figure}

\ifCLASSOPTIONcaptionsoff
  \newpage
\fi

\end{document}